\documentclass[a4paper,12pt,nofootinbib,notitlepage,preprintnumbers]{revtex4-1}

\usepackage{amsmath}           
\usepackage{amssymb}           
\usepackage{xcolor}             
\usepackage{bbm}               
\usepackage{braket}            
\usepackage{xspace}            
\usepackage{mathtools}         
\usepackage{graphicx}          
\usepackage{enumerate}         
\usepackage{hyperref}          
\usepackage{tabularx}  		   
\usepackage{multirow}		   
\usepackage{enumitem}
\usepackage{orcidlink}

\usepackage{etoolbox}
\makeatletter
\patchcmd{\subsubsection}{\itshape}{\itshape\bfseries}{}{}
\makeatother

\makeatletter
\newcommand*{\transpose}{%
  {\mathpalette\@transpose{}}%
}
\newcommand*{\@transpose}[2]{%
  \raisebox{\depth}{$\m@th#1\intercal$}%
}
\makeatother

\newcommand{\p}{\mathbf{p}}

\renewcommand{\P}{\mathbf{P}}

\newcommand{\0}{\mathbf{0}}
\newcommand{\1}{\mathbf{1}}
\newcommand{\2}{\mathbf{2}}
\newcommand{\3}{\mathbf{3}}
\newcommand{\n}{\mathbf{n}}
\newcommand{\m}{\mathbf{m}}

\newcommand{\Zbb}{\mathbbm{Z}}

\newcommand{\Cc}{\mathcal{C}}

\newcommand{\Ec}{\mathcal{E}}

\newcommand{\Gc}{\mathcal{G}}

\newcommand{\Ic}{\mathcal{I}}
\newcommand{\Jc}{\mathcal{J}}
\newcommand{\Kc}{\mathcal{K}}

\newcommand{\Mc}{\mathcal{M}}

\newcommand{\Pc}{\mathcal{P}}

\newcommand{\Rc}{\mathcal{R}}

\newcommand{\Tc}{\mathcal{T}}

\newcommand{\bs}[1]{\boldsymbol{ #1 }}

\newcommand{\bh}[1]{\mathbf{\hat{ #1 }}}

\newcommand{\cf}{cf.\xspace}
\newcommand{\eg}{e.g.\xspace}
\newcommand{\ie}{i.e.\xspace}

\newcommand{\nn}{\nonumber}
\newcommand{\diff}{\textrm{d}}

\usepackage{mfirstuc} 
\newcommand{\addReviewer}[2]{
  \expandafter\newcommand\csname #1\endcsname[1]{{\sf \color{#2} {#1}:\,##1}}
  \expandafter\newcommand\csname #1cor\endcsname[2]{{\color{#2} {#1}:\,\st{##1}{\sf ##2}}}
  \expandafter\newcommand\csname #1color\endcsname{#2}
}

\usepackage{soul,color}
\definecolor{chromeyellow}{rgb}{1.0, 0.65, 0.0}
\definecolor{DodgeBlue}{rgb}{0.118, 0.565,1.000}
\definecolor{asparagus}{rgb}{0.53, 0.66, 0.42}
\definecolor{cadmiumgreen}{rgb}{0.0, 0.42, 0.24}

\definecolor{jlab_red}{RGB}{192,39,45}
\definecolor{jlab_orange}{RGB}{249,102,0}
\definecolor{jlab_blue}{RGB}{47,122,121}
\definecolor{jlab_green}{RGB}{65,125,10}
\definecolor{jlab_gray}{RGB}{141,141,141}
\definecolor{wm_green}{HTML}{115740}
\definecolor{wm_gold}{HTML}{B9975B}

\addReviewer{rb}{jlab_blue}
\addReviewer{aj}{jlab_orange}
\addReviewer{nc}{jlab_red}

\hypersetup{
pdftitle     = {Symmetrizing relativistic three-body partial wave amplitudes},
pdfsubject   = {Scattering Theory},
pdfkeywords  = {Scattering, Partial Waves, QCD, Hadron, Physics, Lattice},
pdfauthor    = {Andrew W. Jackura},
pdfnewwindow = {true},
colorlinks   = {true},
linkcolor    = {wm_green},
citecolor    = {wm_green},
filecolor    = {wm_green},
urlcolor     = {wm_green}
} 

\newcommand{\wm}{Department of Physics, 
William \& Mary, 
Williamsburg, VA 23187, USA}
\newcommand{\ucb}{Department of Physics, 
University of California, 
Berkeley, CA 94720, USA}   
\newcommand{\lbnl}{Nuclear Science Division, 
Lawrence Berkeley National Laboratory, Berkeley, 
CA 94720, USA}

\begin{document}

\title{Symmetrizing relativistic three-body partial wave amplitudes}


\author{Andrew W. Jackura
\orcidlink{0000-0002-3249-5410} }
\email[e-mail: ]{awjackura@wm.edu}
\affiliation{\wm}

\author{Nicholas C. Chambers
\orcidlink{0009-0007-4891-9963} }
\email[e-mail: ]{ncchambers@wm.edu}
\affiliation{\wm}

\author{Ra\'ul A. Brice\~no
\orcidlink{0000-0003-1109-1473} }
\email[e-mail: ]{rbriceno@berkeley.edu}
\affiliation{\ucb}
\affiliation{\lbnl}

\begin{abstract}
S matrix principles and symmetries impose constraints on three-particle scattering amplitudes, which can be formulated as a class of integral equations for their partial wave projections. However, these amplitudes are typically expressed in an asymmetric basis, where one of the initial and final state particles is singled out, and all quantum numbers are defined relative to this spectator. In this work, we show how to construct symmetric partial wave amplitudes, which have been symmetrized over all possible spectator combinations, using their asymmetric counterparts and sets of recoupling coefficients. We derive these recoupling coefficients for arbitrary angular momentum and isospin for arbitrary systems of spinless particles with SU(2) flavor symmetry. We propose a simple intensity observable suitable for visualizing the structure of three-body dynamics in Dalitz distributions. Finally, we provide some numerical examples of Dalitz distributions relevant for future lattice QCD calculations of $3\pi$ systems and provide numerical evidence that the symmetrization procedure is consistent with expected symmetries of Dalitz plots.
\end{abstract}
\date{\today}
\maketitle

\section{Introduction}
\label{sec:intro}

One of the major challenges of modern-day nuclear and particle physics is finding a systematically improvable procedure for studying few-hadron dynamics directly from the Standard Model. This is motivated by intertwined goals. On the theoretical side, many features of the strong interaction remain unexplained, and only a handful of the hadrons seen in experiments—including all light nuclei beyond the proton—have been directly reproduced from quantum chromodynamics (QCD). The theoretical challenges with reproducing the spectrum of QCD include the fact that they appear as either bound states or unstable resonances coupling to few-hadron states~\cite{ParticleDataGroup:2024cfk}. On the experimental side, searches for physics beyond the Standard Model, whether they involve nuclear targets as in Deep Underground Neutrino
Experiment (DUNE)~\cite{DUNE:2016hlj} or rare heavy-hadron decays~\cite{LHCb:2019xmb,LHCb:2022fpg,LHCb:2014mir,LHCb:2019jta,LHCb:2013lcl, Suzuki:1999uc,Wolfenstein:1990ks,Suzuki:2007je,AlvarengaNogueira:2015wpj,Bediaga:2013ela,Garrote:2022uub}, often run into the same problem. Namely that it is hard to interpret new signals without a precise understanding of the QCD ``background". The non-perturbative nature of QCD has, until recently, prevented the community from achieving such a goal. A program to connect hadronic and nuclear processes directly to QCD is being built on two complementary and powerful tools: scattering theory and lattice QCD~\cite{Briceno:2017max,Hansen:2019nir,Mai:2021lwb}. 

Scattering theory principles, such as unitarity and analyticity, provide strict constraints on the properties of scattering amplitudes. Motivated by progress in lattice QCD, there has been a resurgence of studies aimed at finding such constraints for three-hadron systems~\cite{Jackura:2018xnx, Hansen:2015zga, Dawid:2023jrj,Dawid:2023kxu, Jackura:2020bsk, Jackura:2023qtp, Jackura:2022gib, Mai:2017vot, Mikhasenko:2019vhk, Dawid:2020uhn, Feng:2024wyg}. Generally, one can show that S matrix unitarity imposes that the scattering amplitude satisfies a set of linear integral equations in the physical scattering region. These integral equations involve a class of real-meromorphic functions in the physical region for which they are defined, commonly referred to as K matrices, which are otherwise unconstrained by unitarity. Such functions can, in principle, be determined experimentally, but as first pointed out in Refs.~\cite{Polejaeva:2012ut, Briceno:2012rv}, can also be determined via lattice QCD. 

In practice, lattice QCD calculations are performed in a finite, discretized spacetime, where scattering observables can not rigorously be defined. As has now been established by a large body of work~\cite{Luscher:1985dn, Luscher:1986n2, Luscher:1990ux, Rummukainen:1995vs, Kim:2005gf, He:2005ey, Hansen:2012tf, Briceno:2012yi, Briceno:2013lba, Briceno:2014oea, Polejaeva:2012ut, Hansen:2014eka, Hansen:2015zga, Briceno:2017tce, Briceno:2018mlh, Briceno:2018aml, Briceno:2019muc, Blanton:2019igq, Hansen:2020zhy, Blanton:2020gha, Hammer:2017uqm, Hammer:2017kms, Meng:2017jgx, Pang:2019dfe, Muller:2021uur, Blanton:2020gmf, Blanton:2021mih, Jackura:2022gib,Hansen:2024ffk,Raposo:2023oru,Raposo:2025dkb}, one can develop non-perturbative relations between the finite-volume energy spectrum as computed from lattice QCD  and the infinite-volume K matrices, which are used to reconstruct the scattering amplitudes of interest. By combining high-precision lattice calculations~\cite{Detmold:2008fn, Culver:2019vvu, Alexandru:2020xqf, Hansen:2020otl, Draper:2023boj, Dudek:2010ew, Pelissier:2012pi, Dudek:2012xn, Liu:2012zya, Wilson:2014cna, Dudek:2014qha, Lang:2015hza, Wilson:2015dqa, Dudek:2016cru, Briceno:2016mjc, Moir:2016srx, Bulava:2016mks, Hu:2016shf, Alexandrou:2017mpi, Bali:2017pdv, Wagman:2017tmp, Andersen:2017una, Briceno:2017qmb, Woss:2018irj, Brett:2018jqw, Mai:2019pqr, Woss:2019hse, Wilson:2019wfr, Cheung:2020mql, Rendon:2020rtw, Woss:2020ayi, Horz:2020zvv} with these finite-volume formalisms, it is indeed possible to extract two- and now also three-hadron scattering information from first principles, bridging the gap between QCD and experimental observables for at least some kinematic regions.

\begin{figure}[t]
	\centering
	\includegraphics[width=\textwidth]{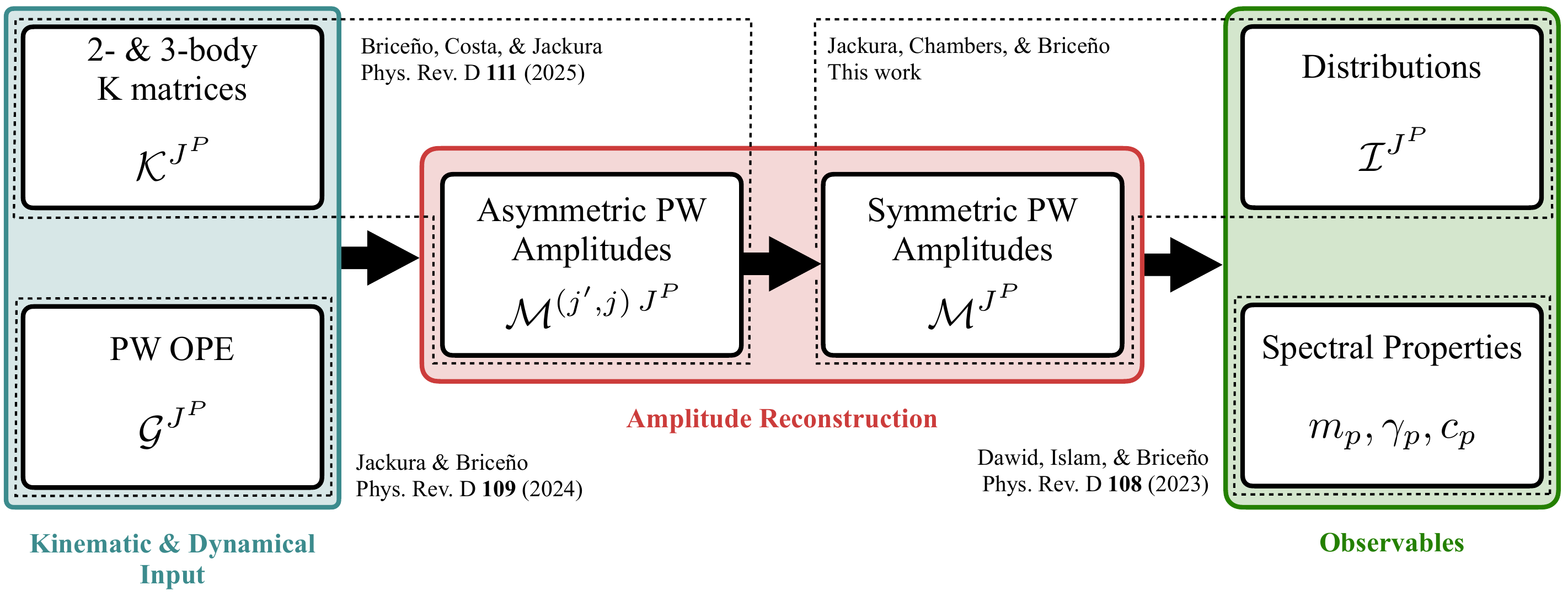}
	\caption{Shown is the workflow for computing physical observables from the key inputs ($\Kc^{J^P},\Gc^{J^P}$) to the integral equations that three-body scattering amplitudes, $\Mc^{J^P}$, must satisfy. A detailed description of each step is described in the text, and the main references include Refs.~\cite{Jackura:2023qtp, Briceno:2024ehy, Dawid:2023jrj,Dawid:2023kxu, Jackura:2020bsk}.}
	\label{fig:workflow}
\end{figure}

This work falls into this program by providing the final components of the now-understood workflow for determining three-body observables from lattice QCD. Our focus here is on $\3\to\3$ scattering amplitudes of spinless particles of any species,~\footnote{Here $\n\to\m$ refers to reactions of QCD stable hadrons, $\n$ incoming and $\m$ outgoing.} including the scenario where the hadrons respect strong isospin symmetry. The first step, which we do not discuss in any detail, is constraining K matrices from three-body finite-volume observables~\cite{Horz:2020zvv, Hansen:2020otl, Draper:2023boj,Dawid:2025doq}. The rest of the workflow is depicted in Fig.~\ref{fig:workflow}, which we describe here. These K matrices, denoted by $\Kc^{J^P}$ when projected to definite spin-parity, serve as input to the integral equations governing the three-body scattering amplitudes, $\Mc^{J^P}$. If the hadrons respect isospin symmetry, as one often assumes or imposes in lattice QCD calculations, then the K matrices and scattering amplitudes are projected to total isospin and spin-parity quantum numbers, $I(J^P)$. The other key input to the integral equations is the one-particle exchange (OPE) amplitude, $\Gc$. The OPE, required by unitarity, depends entirely on kinematics associated with an exchange of a particle between two-body scattering sub-processes.~\footnote{Constraining the two-body scattering amplitude from lattice QCD spectra is well-known, see Ref.~\cite{Briceno:2017max} for a review.} In Ref.~\cite{Jackura:2023qtp}, we showed how the OPE can be partial-wave projected for arbitrary three-particle systems with no intrinsic spin, which we labeled as $\Gc^{J^P}$. 

Given these kinematic and dynamical inputs, the partial wave amplitudes can be reconstructed. In Ref.~\cite{Briceno:2024ehy}, we derived the integral equations that the asymmetric analog of $\Mc^{J^P}$ must satisfy using $\Kc^{J^P}$ and $\Gc^{J^P}$ as input. The resultant amplitude, labeled as $\Mc^{(j',j)\,J^P}$, is asymmetric because we have singled out one particle, called the \emph{spectator}, and have defined all angular momenta with respect to this particle. Specifically, the remaining two particles, called the \emph{pair}, are projected to definite angular momentum $S$, and this resulting state is subsequently coupled with the spectator to some orbital angular momentum $L$. Amplitudes of definite $J^P$ are then formed by the usual rules of angular momentum addition. The $j$ and $j'$ indices in $\Mc^{(j',j)\,J^P}$ represent the spectator choice in the initial and final state, respectively. In Sec.~\ref{sec:amps}, we review the kinematics and definition of three-body scattering amplitudes and give a detailed description of their partial wave projection.~\footnote{Note, since this work builds from Refs.~\cite{Jackura:2023qtp, Briceno:2024ehy}, we use notation introduced in those works. We recommend the reader first familiarize themselves with those works.}

In this work, we complete an outstanding key step in this workflow. We show in Sec.~\ref{sec:symm} how to go from $\Mc^{(j',j)\,J^P}$ to an amplitude that is independent of the choice of the spectator, but still projected to a definite $J^P$. 
Heuristically, one can do this by introducing \emph{recoupling coefficients}, $\Rc$, defined such that 
\begin{align}
    \Mc^{\,J^P} = \sum_{j',j} \Rc_{j'} \, \Mc^{(j',j)\,J^P} \, \Rc_{j} . 
    \label{eq:main_result}
\end{align}
These recoupling coefficients are defined for a particular partial wave channel and can include isospin recoupling coefficients if the system is symmetric under SU(2) flavor symmetry. In Sec.~\ref{sec:symm_ang_mom}, we derive a detailed and explicit form of Eq.~\eqref{eq:main_result} when only considering angular momentum recoupling. We generalize this result for systems with isospin quantum numbers in Sec.~\ref{sec:isospin} to form \emph{symmetrized} amplitudes of definite $I(J^P)$.~\footnote{Recent work has defined symmetrized amplitudes of definite $J^P$, but that still carry angular dependence~\cite{Dawid:2025doq}. Those amplitudes are the $J^P$-th term in a particular partial wave expansion, which includes the angular functions. In contrast, the amplitudes derived here are independent of the angular degrees of freedom.}

%

To aid in the analysis of three-body systems from lattice QCD, we provide a set of simple observables closely related to experimental cross-sections in order to study features of the partial wave amplitudes. Previously, Refs.~\cite{Jackura:2020bsk,Dawid:2023jrj,Dawid:2023kxu} had explained how the asymmetric amplitudes can be used to study spectral properties, e.g., bound state and resonances poles and couplings. Here we provide a definition for intensity distributions, $\Ic^{J^P}$, which can be qualitatively understood as the helicity/channel-average modulus squared of the symmetrized amplitudes. These distributions can be used to visualize kinematic and dynamic enhancements of the amplitude using, among others, \emph{Dalitz} plots as those shown in Ref.~\cite{Hansen:2020otl}. For a fixed initial state, the Dalitz plots for $\3\to\3$ amplitudes satisfy the same symmetry properties of Dalitz regions known in the literature for three-body decays. This provides a robust check on numerical investigations performed in Sec.~\ref{sec:application}, where we provide detailed examples of relevance for $3\pi$ systems.

We provide two appendices, \ref{sec:ls_recoupling} and \ref{sec:asym_lsi}. In Appendix~\ref{sec:ls_recoupling},  we present the recoupling coefficients to construct symmetric amplitudes in the spin-orbit basis. In Appendix~\ref{sec:asym_lsi}, we summarize some key expressions from Ref.~\cite{Jackura:2023qtp} needed to construct the asymmetric amplitudes in the spin-orbit basis that appear in our numerical examples.

\section{Kinematics \& Amplitudes}
\label{sec:amps}

Consider a system of three spinless particles, denoted by $\varphi_j$ where $j\in \Zbb_3 =\{1,2,3\}$ labels the particle. Let $m_j$ and $\eta_j$ be the mass and intrinsic parity of particle $j$. Particle $j$ carries a momentum $\p_j$ and an energy $E_j = \sqrt{m_j^2 + p_j^2}$ where $p_j \equiv \lvert\p_j\rvert$.~\footnote{We work with natural units, $\hbar = c = 1$, in a spacetime with signature $\mathrm{diag}(+1,-1,-1,-1)$.} Let $\ket{\p_j}$ represent a single-particle state which is normalized as
\begin{align}
    \label{eq:state_norm}
    \braket{\p_{k}'|\p_j} = \delta_{kj}\, (2\pi)^3\,2E_j\,\delta^{(3)}(\p'_{k}-\p_j) \, ,
\end{align}
where $\delta_{kj}$ enforces that the species are the same. As is customary, we keep the particle mass and parity implicit in the particle state. We make no restriction on the species of the particles involved, and for the moment assume that the particles have no additional quantum numbers. In Sec.~\ref{sec:isospin}, we extend the results to systems of hadrons with flavor SU(2) symmetry, i.e. isospin.

A state of three particles is denoted $\ket{\{\p\}} \equiv \ket{\p_1,\p_2,\p_3}$.~\footnote{Such a state is defined as 
\begin{align}
	\ket{\{\p\}} \equiv \bigotimes_{j=1}^{N}\frac{1}{\sqrt{n_j!}} \left( \sum_\pi  \bigotimes_{i=1}^{n_j} \ket{\p_{\pi(i)}} \right) \, ,
\end{align}
where the $N$ is the number of sets of distinct particles for the scattering channel ($N = 1,2,$ or $3$) and $n_{j}$ is the number of identical particles in each set, with the sum over $\pi$ indicating permutations within that set. The three cases are then: (\emph{i}) $N=1$ with $n_1 = 3$, (\emph{ii}) $N=2$ with either $n_1 = 1, n_2 = 2$ or $n_1 = 2, n_2 = 1$, and (\emph{iii}) $N=3$ with $n_1 = n_2 = n_3 = 1$.
}
In general, a three-particle state could have (\emph{i}) all three particles identical, (\emph{ii}) two of the particles identical, and (\emph{iii}) all three particles distinguishable. We do not restrict the species of the particles involved, and the main results are valid for any of these cases. If any or all particles are identical, one has further symmetries of the system under interchange of any pair of particles, which imposes constraints on their allowed quantum numbers. Unless otherwise noted, we work in the three-particle CM frame, i.e., the frame with zero total momentum,
\begin{align}
	\sum_{j=1}^{3}\p_j = \0 \, .
\end{align}
Consequently, only two of the three momenta are independent, and the invariant mass of the three-particle system, $\sqrt{s}$, is its total energy,
\begin{align}
	\sum_{j = 1}^{3} E_j \equiv \sqrt{s} \, .
\end{align}
Thus for physical kinematics, $m_1+m_2+m_3 \le \sqrt{s} < \infty$.

%
\begin{figure}[t]
	\centering
	\includegraphics[width=0.3\textwidth]{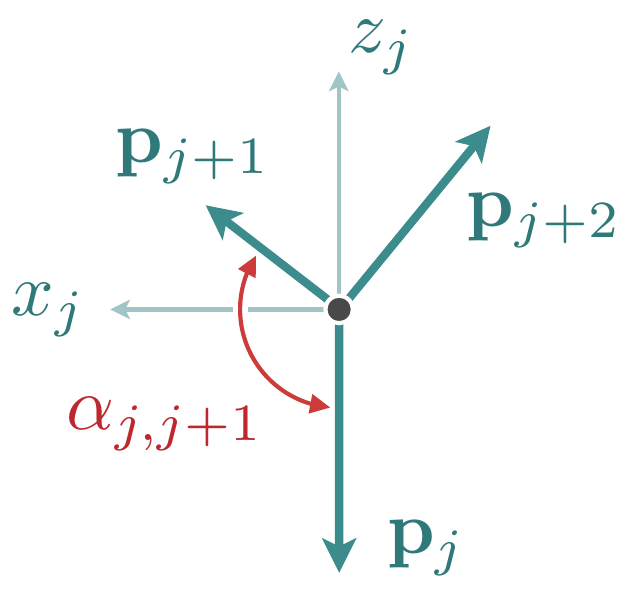}
	\caption{Three-particle plane formed by $\p_j + \p_{j+1} + \p_{j+2} = \0$ for any spectator choice $j\in \Zbb_3$. By convention, the body-fixed axes are specified by spectator particle $j$ and the cyclic notation as discussed in the text. The angle between momenta $\p_j$ and $\p_{k}$ is denoted $\alpha_{j,k}$, here shown for the case where $k = j+1$.}
	\label{fig:three_particle_plane}
\end{figure}

In the CM frame, the momenta of the three-particle system are constrained to lie in a plane.
The orientation of the three-particle plane with respect to some \emph{space-fixed} Cartesian coordinate system ($xyz$) can be specified by the polar and azimuthal angles of a unit normal, $\bh{n} \equiv \n / n$, and some angle specifying the rotation of the plane about the normal. While $\bh{n}$ can be fixed unambiguously, the rotation about the unit normal is dependent on the choice of the origin. The arbitrariness of the origin illustrates the need for a convention in defining the three angles specifying the three-particle plane. To specify the orientation, we define a \emph{body-fixed} coordinate system ($x_jy_jz_j$), where the subscript $j$ indicates that this system is defined with respect to the $j$th particle. The choice of particle $j$ is arbitrary, but one must choose to fix the orientation. We call the chosen $j$th particle the \emph{spectator}, while the other two form a \emph{pair}. To specify the ordering of the pair, we adopt a \emph{cyclic} convention when labeling particles and their associated quantities. Therefore, we denote the first and second particle of the pair associated with the $j$th spectator as $j+1$ and $j+2$, respectively, where we implicitly assume arithmetic modulo 3. For example, if $j = 2$ is the spectator, then $j+1 = 3$ and $j+2 = 1$.

\begin{figure}[t]
	\centering
	\includegraphics[width=\textwidth]{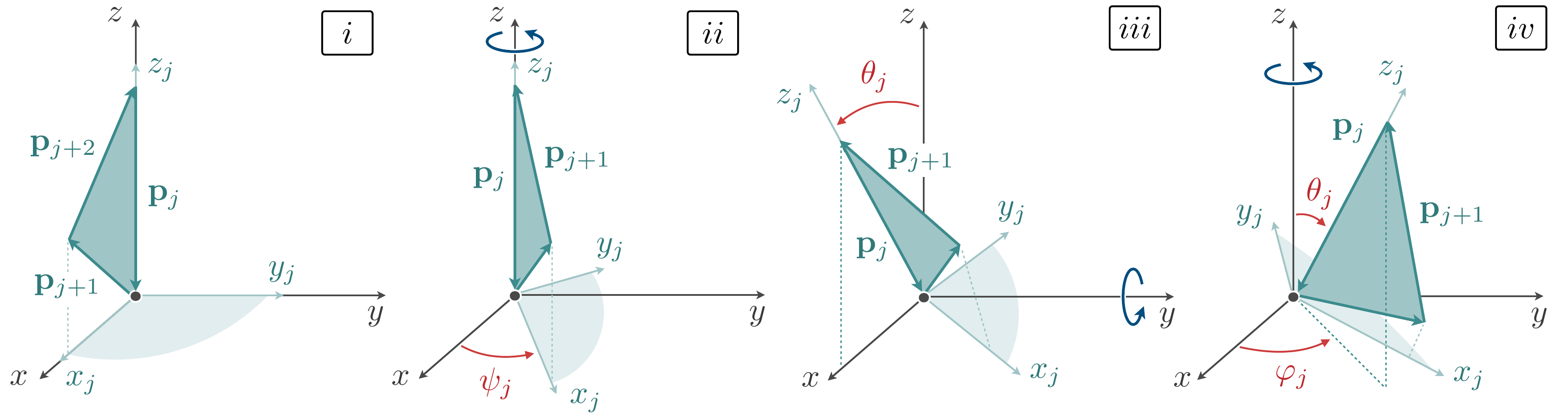}
	\caption{Euler angles $\Ec_j\equiv \{\varphi_j,\theta_j,\psi_j\}$ defining the three-particle plane. Euler angles are defined by the sequence: (\textit{i}) Begin with the standard configuration $j$, (\textit{ii}) rotate the plane about the space-fixed $z$-axis by $\psi_j$, (\textit{iii}) rotate the plane about the space-fixed $y$-axis by $\theta_j$, and (\textit{iv}) rotate the plane about the space-fixed $z$-axis by $\varphi_j$.}
	\label{fig:euler}
\end{figure}

Choosing $j$ as the spectator, the body-fixed ($x_jy_jz_j$) coordinate system is then defined such that the $z_j$ axis is aligned along the $-\bh{p}_j$ direction, the $x_j$ axis is such that $\bh{p}_{j+1}\cdot \bh{x}_j > 0$, and $y_j$ is perpendicular to the plane formed by $\p_j \times \p_{j+2}$ (see Fig.~\ref{fig:three_particle_plane}). We then specify the orientation by a set of Euler angles, $\Ec_j\equiv \{\varphi_j,\theta_j,\psi_j\}$, defined with respect to the space-fixed coordinate system and dependent on the choice of body-fixed system. These angles are defined within the usual $zyz$ convention~\cite{VMK}, with the plane initially lying in the $xz$ plane with $z_j\parallel z$ and $x_j \parallel x$. We call this configuration the \emph{standard configuration} with respect to the $j$th particle, or $\mathrm{SC}_j$. The standard configuration is then rotated about the $z$ axis by an angle $\psi_j \in [0,2\pi)$, followed by a rotation about the $y$ axis by $\theta_j \in [0,\pi)$, and finally another rotation about $z$ by an angle $\varphi_j \in [0,2\pi)$. This sequence is illustrated in Fig.~\ref{fig:euler}.
%

A three-particle state can then be described by the three Euler angles, the total invariant mass $\sqrt{s}$, and the three invariant masses of each pair $\sqrt{\sigma_j}$, which are defined in the CM frame as~\cite{byckling1973particle,Jackura:2022gib}
\begin{align}
    \label{eq:pair_inv_mass}
    \sigma_j = (\sqrt{s}-E_j)^2 - p_j^2 \, ,
\end{align}
which is bounded as $m_{j+1} + m_{j+2} \le \sqrt{\sigma_j} \le \sqrt{s} - m_j$ for physical kinematics. Therefore, we specify $\ket{\{\p\}} = \ket{\{\sigma\},\Ec_j}$, where $\{\sigma\} = \{\sigma_1,\sigma_2,\sigma_3\}$, and we suppress the $s$ dependence for notational convenience. Notice that the pair invariant mass is labeled by its corresponding spectator. For a given $s$, we may trade the pair invariant masses for the corresponding spectator momenta by solving Eq.~\eqref{eq:pair_inv_mass} for $p_j$,
\begin{align}
    \label{eq:momentum}
    p_j = \frac{1}{2\sqrt{s}}\,\lambda^{1/2}(s,m_j^2,\sigma_j) \, , 
\end{align}
where $\lambda(x,y,z) \equiv x^2 + y^2 + z^2 - 2(xy + yz + zx)$~\cite{byckling1973particle}. This in turn relates the energy of particle $j$ to Mandelstam invariants,
\begin{align}
    \label{eq:energy}
    E_j = \frac{1}{2\sqrt{s}}\,(s + m_j^2 - \sigma_j) \,.
\end{align}

The relative orientations of the three momenta within the three-particle plane are entirely fixed by kinematics. We denote the CM frame separation angle between particle $j$ and $k$ by $\alpha_{j,k}$, which is defined for $0\le\alpha_{j,k} <\pi$. The cosine of this angle is defined by
\begin{align}
    \label{eq:alpha_angle}
    \cos\alpha_{j,k} = \frac{\p_j\cdot\p_k}{p_jp_k} \, . 
\end{align}
For $j \ne k$, we express the cosine of the separation angle as
\begin{align}
    \label{eq:alpha_angle_a_ne_j}
    \cos\alpha_{j,k} = \frac{p_{3-j-k}^2 - p_j^2 - p_k^2}{2p_jp_k} \, ,
\end{align}
which can be found by squaring $\p_j + \p_k = -\p_{3-j-k}$, where $\p_{3-j-k}$ represents the momentum of the particle that is neither $j$ nor $k$, reminding the reader that we implicitly use arithmetic modulo 3. For example, if $j = 3$ and $k = 1$, then $\p_{3 - 3 - 1} = \p_{-1} = \p_2$. When $j=k$, we get the trivial identity $\alpha_{j,j} = 0$. One may further express $\alpha_{j,k}$ in terms of invariants by using Eq.~\eqref{eq:momentum}.

For a fixed $s$, energy-momentum conservation imposes the Mandelstam relation
\begin{align}
    \label{eq:mandelstam}
	\sum_{j=1}^3\sigma_j = s + \sum_{j=1}^3 m_j^2  \equiv \Delta(s) \, ,
\end{align}
which can be verified by direct evaluation using Eq.~\eqref{eq:pair_inv_mass}. The Mandelstam relation imposes a constraint that one of the pair invariant masses is fixed for a given $s$, leaving only two independent. The physical region, or Dalitz region, at fixed $s$ is defined by $\Phi(\{\sigma\}) \ge 0$~\cite{byckling1973particle,Kibble:1960zz}, where $\Phi$ is the Kibble cubic defined as 
\begin{align}
    \label{eq:kibble}
	\Phi(\{\sigma\}) & = \prod_{j=1}^3\sigma_j - \frac{1}{\Delta(s)}\sum_{j=1}^{3}c_j\sigma_j \, ,
\end{align}
where the $c_j$ coefficients are given by
\begin{align}
    \label{eq:kibble_coef}
    c_j = (m_{j+1}^2m_{j+2}^2 - m_j^2 s)(m_{j+1}^2 + m_{j+2}^2 - m_j^2 - s) \, .
\end{align}

We consider a generic $\3\to\3$ reaction of spinless particles, where the initial and final states need not be the same. The corresponding scattering amplitude $\Mc$ is defined as the fully connected piece of the S matrix element~\cite{Jackura:2022gib},
\begin{align}
    \label{eq:amp_def}
	\bra{\{\p'\}}\mathsf{S}\ket{\{\p\}}_{\mathrm{f.c.}} \equiv (2\pi)^4\,\delta^{(4)}(P_{\mathrm{out}} - P_{\mathrm{in}}) \, i\Mc(\{\p'\};\{\p\}) \, ,
\end{align}
where the Dirac delta distribution enforces total energy-momentum conservation and ``f.c.'' emphasizes that only the fully connected piece is considered.~\footnote{Fully connected means there are no interactions where one or more of the particles remain unscattered, \cf Ref.~\cite{Jackura:2022gib}.} The amplitude is a function of eight kinematic variables formed from the initial and final sets of momenta, $\{\p\} = \{\p_1,\p_2,\p_3\}$ and $\{\p'\} = \{\p_{1'}',\p_{2'}',\p_{3'}'\}$, respectively. Note that primes denote final state variables throughout this work, and we emphasize that the final particle species $j'\in \{1', 2', 3'\}$ need not be the same as the initial set. The first variable we choose is $s$, which we leave implicit in all expressions throughout this work. The sets of initial and final pair invariant masses, $\{\sigma\} = \{\sigma_1,\sigma_2,\sigma_3\}$ and $\{\sigma'\} = \{\sigma_{1'}',\sigma_{2'}',\sigma_{3'}'\}$, respectively, serve as four additional variables given that two are fixed by their respective Mandelstam relations, Eq.~\eqref{eq:mandelstam}. The remaining three variables are chosen to be the set of Euler angles that describe the orientation of the final state plane with respect to the initial state plane. We denote these Euler angles as $\Ec_{j'j} = \{\varphi_{j'j},\theta_{j'j},\psi_{j'j}\}$, which are defined as follows. If $\mathsf{R}(\Ec_j) = \mathsf{R}_z(\varphi_j)\cdot \mathsf{R}_y(\theta_j)\cdot \mathsf{R}_z(\psi_j)$~\footnote{Here $\mathsf{R}_n(\theta)$ represents a Euclidean rotation matrix of a positive rotation $\theta$ about the $n$-axis.} represents the Euler rotation matrix for the initial state with the $j$th chosen spectator, and $\mathsf{R}(\Ec_{j'}')$ is the Euler rotation matrix for the final state with $j'$th spectator, then $\mathsf{R}^{\transpose}(\Ec_{j'j}) = \mathsf{R}^\transpose(\Ec_{j'}')\cdot \mathsf{R}(\Ec_j)$ is the combined Euler rotation matrix with Euler angles $\Ec_{j'j}$. The combined angles are given by the usual relations from spherical trigonometry~\cite{VMK},
\begin{align}
	\label{eq:euler_combine}
	\begin{gathered}
		\cot(\varphi_{j'j}+\psi_j) =  \cos\theta_{j} \, \cot(\varphi_{j'}' - \varphi_j) - \cot\theta_{j'}' \, \frac{\sin\theta_j}{\sin(\varphi_{j'}' - \varphi_j)}  \, , \\[5pt]
		\cot(\psi_{j'j}-\psi_{j'}') = \cos\theta_{j'}' \, \cot(\varphi_{j'}' - \varphi_j) - \cot\theta_j \,\frac{\sin\theta_{j'}'}{\sin(\varphi_{j'}' - \varphi_j)} \, , \\[5pt]
        \cos\theta_{j'j} = \cos\theta_{j'}'\cos\theta_j + \sin\theta_{j'}'\sin\theta_j\cos(\varphi_{j'}' - \varphi_j)\, , \\[5pt]
		\frac{\sin(\varphi_{j'j}+\psi_j)}{\sin\theta_{j'}'} = -\frac{\sin(\psi_{j'j}-\psi_{j'}')}{\sin\theta_{j}} = \frac{\sin(\varphi_{j'}' - \varphi_j)}{\sin\theta_{j'j}} \, .
	\end{gathered}	
\end{align}
Note that in the special case where the initial state plane is aligned with the space-fixed coordinate system, $\Ec_j = \{0,0,0\}$, then $\Ec_{j'j} = \Ec_{j'}'$ identically. 

\subsection{Partial wave expansion}
\label{sec:pwe}

We expand the three-particle system in terms of definite total angular momentum, $J$, states through the formalism introduced in Ref.~\cite{Berman:1965gi},
\begin{align}
    \label{eq:dalitz_pw}
	\ket{\{\p\}} \propto \sum_{J = 0}^{\infty} \sum_{m_J = -J}^{J}\sum_{\lambda_j = -J}^{J}\ket{\{\sigma\},Jm_J\lambda_j}\,\sqrt{2J+1}\,D_{m_J\lambda_j}^{(J)}(\Ec_j) \, ,
\end{align}
where $m_J$ is the projection of $J$ along the space-fixed $z$ axis, while $\lambda_j$ is the projection of $J$ along the body-fixed $z_j$ axis.~\footnote{We define the state up to some irrelevant normalization which is absorbed into the definition of the amplitudes.} Since we consider spinless particles, $J$ is restricted to non-negative, integer values. For any $J$, the projections lie within the range $-J \le m_J , \lambda_j \le J$. Notice that the definition of $\lambda_j$ depends on the chosen spectator, while $m_J$ is independent of this choice. The state $\ket{\{\sigma\},Jm_J\lambda_j}$ then describes a three-particle state with total angular momentum $J$. The orientation of the three-body system is encoded in the Wigner $D$ matrix elements,  $D_{m_J\lambda_j}^{(J)}(\Ec_j) = D_{m_J\lambda_j}^{(J)}(\varphi_j,\theta_j,\psi_j)$.~\footnote{The Wigner $D$ matrices are defined as $ D_{m'm}^{(j)}(\varphi,\theta,\psi) \equiv e^{-im'\varphi} d_{m'm}^{(j)}(\theta) e^{-im\psi}$ where $d_{m'm}^{(j)}(\theta) = \bra{jm'}e^{-i\Jc_y\theta}\ket{jm}$ is the Wigner little $d$ matrix with $\Jc_y$ the $y$ component of the angular momentum operator and $\ket{jm}$ are the usual SU(2) eigenstates~\cite{VMK}.} As this expansion makes explicit the variables used in Dalitz plots for $\1\to\3$ decays, we refer to Eq.~\eqref{eq:dalitz_pw} as the Dalitz partial wave expansion~\cite{JPAC:2019ufm}.

The partial wave expansion of the amplitude is then given by expanding the states in Eq.~\eqref{eq:amp_def} with~\eqref{eq:dalitz_pw},
\begin{align}
	\label{eq:Dalitz_expand}
	\Mc(\{\p'\};\{\p\}) & = \sum_{J,m_J}\sum_{\lambda_{j'}',\lambda_j}(2J+1)D_{m_J\lambda_{j'}'}^{(J)\,*}(\Ec_{j'}') \, \Mc_{\lambda_{j'}'\lambda_j}^{J}(\{\sigma'\};\{\sigma\})\, D_{m_J\lambda_j}^{(J)}(\Ec_j) \, ,
\end{align}
where $\bra{\{\sigma'\},J'm_{J'} \lambda_{j'}'}\mathsf{S}\ket{\{\sigma\},Jm_J\lambda_j}_{\mathrm{f.c.}} \propto \delta_{J'J}\delta_{m_{J'}m_J} \, \Mc_{\lambda_{j'}'\lambda_j}^{J}(\{\sigma'\};\{\sigma\})$ defines the partial wave amplitude. We have used the fact that rotational invariance ensures that $J$ and $m_J$ is conserved and the amplitude is independent of $m_J$. However, the body-fixed projection quantum numbers $\lambda_j$ and $\lambda_{j'}'$ are not generally conserved as they are not invariant under global rotations. The inverse relation of Eq.~\eqref{eq:Dalitz_expand} is 
\begin{align}
    \label{eq:dalitz_pw_from_full}
    \Mc_{\lambda_{j'}'\lambda_j}^{J}(\{\sigma'\};\{\sigma\}) = \frac{1}{(8\pi^2)^2}\sum_{m_J}\int\!\diff\Ec_{j'}' \, \int\!\diff \Ec_j \, D_{m_J \lambda_{j'}'}^{(J)}(\Ec_{j'}') \, \Mc(\{\p'\};\{\p\}) \, D_{m_J\lambda_j}^{(J)\,*}(\Ec_j) \, ,
\end{align}
where we have used the orthogonality of the Wigner $D$ matrices, 
\begin{align}
    \label{eq:wigner_d_ortho}
    \frac{2J+1} {8\pi^2}\,\int\!\diff\Ec \, D_{m'\lambda'}^{(J')\,*}(\Ec) D_{m\lambda}^{(J)}(\Ec) =  \delta_{J'J}\delta_{m'm}\delta_{\lambda'\lambda}\, ,
\end{align}
with the integration measure being defined as $\diff \Ec = \diff \varphi \, \diff\!\cos\theta\,\diff\psi$ over the domain $\varphi \in [0,2\pi)$, $\theta \in [0,\pi)$, and $\psi\in[0,2\pi)$.

While the Dalitz partial wave amplitude depends on all two-body invariant masses, there remains residual spectator dependence due to the choice of the body-fixed angular momentum projections, $\lambda_j$ and $\lambda_{j'}'$, and the orientations from their associated standard configurations. This residual dependence on spectator choice cannot be removed, as it reflects the freedom in how a plane is oriented with respect to another. That is, the angle between two planes can be defined, \eg, by the angle between their normal vectors. However, any arbitrary rotation about each normal vector will have the same orientation. This contrasts the case in $\2\to\2$ scattering, where the orientation is completely fixed by the scattering angle between initial and final states. Therefore, we cannot have partial wave amplitudes independent of the body-fixed coordinates and must choose some spectators as references for the standard configurations. We can see this more explicitly by adding the two rotations in Eq.~\eqref{eq:Dalitz_expand},
\begin{align}
	\sum_{m_J} D_{m_J\lambda_{j'}}^{(J)\,*}(\Ec_{j'}') D_{m_J\lambda_j}^{(J)}(\Ec_j) = D_{\lambda_j\lambda_{j'}'}^{(J)\,*}(\Ec_{j'j}) \, ,
\end{align}
with $\Ec_{j'j} = \{\varphi_{j'j}, \theta_{j'j}, \psi_{j'j}\}$ as defined in Eq.~\eqref{eq:euler_combine}. Using this in Eq.~\eqref{eq:Dalitz_expand}, we find
\begin{align}
    \label{eq:dalitz_exp_one_wigner}
	\Mc(\{\p'\};\{\p\}) & = \sum_{J} (2J+1)\sum_{\lambda_{j'}',\lambda_j} \Mc_{\lambda_{j'}' \lambda_j}^{J}(\{\sigma'\};\{\sigma\}) \, D_{\lambda_j\lambda_{j'}'}^{(J)\,*}(\Ec_{j'j})  \, ,
\end{align}
with an inverse relation,
\begin{align}
    \Mc_{\lambda_{j'}'\lambda_j}^{J}(\{\sigma'\};\{\sigma\}) = \frac{1}{8\pi^2}\int\!\diff\Ec_{j'j} \, D_{\lambda_j \lambda_{j'}'}^{(J)}(\Ec_{j'j}) \, \Mc(\{\p'\};\{\p\}) \,  \, .
\end{align}
Again, we find dependence on the initial and final state spectators as they define the origins for the respective body-fixed coordinate systems. The resulting Dalitz partial wave amplitudes depend on five independent kinematic variables, which are $s$, two initial state invariant masses, and two final state invariant masses.

The partial wave states in Eq.~\eqref{eq:dalitz_pw} do not possess definite parity, which is a conserved quantum number in strongly interacting processes. To form definite spin-parity, $J^P$, states, we take the linear combination~\cite{Berman:1965gi},
\begin{align}
    \label{eq:dalitz_JP_state}
	\ket{\{\sigma\},J^P m_J\lambda_j} = \frac{1}{2} \Big[ \ket{\{\sigma\},J m_J\lambda_j} + \eta P (-1)^{J+\lambda_j} \ket{\{\sigma\},J m_J(-\lambda_j)} \Big] \, ,
\end{align}
where $P$ is the target parity and $\eta$ is the product of the intrinsic parities of the three particles, $\eta \equiv \eta_1\eta_2\eta_3$. Partial wave amplitudes of definite $J^P$ are then given by
\begin{align}
    \label{eq:dalitz_JP_amp}
	\Mc^{J^P}_{\lambda_{j'}'\lambda_j} & = \frac{1}{4}\Bigg[\Mc^{J}_{\lambda_{j'}'\lambda_j} + \eta'\eta (-1)^{\lambda_{j'}' + \lambda_j}\Mc^{J}_{-\lambda_{j'}'\,-\lambda_j} \nn\\[5pt]
    & \qquad \qquad +  P(-1)^{J} \left(\eta'(-1)^{\lambda_{j'}'}\Mc_{-\lambda_{j'}'\lambda_j}^{J} + \eta(-1)^{\lambda_j} \Mc_{\lambda_{j'}'\,-\lambda_j}^{J}\right) \Bigg] \, ,
\end{align}
where $\eta'$ is the product of the intrinsic parities of the final state particles. Notice that for any $J$ with $\lambda_j = \lambda_{j'}' = 0$, the amplitude $\Mc^{J^P}_{00}$ is non-zero only when $\eta'=\eta$ and the target parity is $P = \eta(-1)^J$.

\subsection{Connecting to spin-orbit amplitudes}
\label{sec:so_to_dalitz}

In Refs.~\cite{Jackura:2023qtp,Briceno:2024ehy}, an alternative partial wave expansion was used, which was useful for reconstructing amplitudes via sets of integral equations involving the three-particle K matrix. The amplitudes were projected to definite $J^P$ in the spin-orbit, or $LS$, basis. In this basis, the pair and spectator are treated as a quasi-two-body system, where the pair has an angular momentum $S$ and $L$ is the orbital angular momentum between the pair and the spectator. Here, we summarize the procedure presented in Refs.~\cite{Jackura:2023qtp,Briceno:2024ehy} and show the connection to the Dalitz partial waves as defined by Eq.~\eqref{eq:dalitz_pw}.

We consider a coupling scheme where, for a given spectator, the first and second particle of the pair is determined cyclically. That is, we specify a three-particle state $\ket{\{\p\}_j}$,~\footnote{Note that $\ket{\{\p\}_j} = \ket{\{\p\}}$ identically, the subscript merely points out a spectator choice.} where $\{\p\}_j = \{\p_j,\p_{j+1},\p_{j+2}\}$ where $\p_j$ is the spectator momentum and $\p_{j+1}$ ($\p_{j+2}$) is the momentum of the first (second) particle in the pair. We first expand the pair into definite angular momentum $S_j$,~\footnote{In Ref.~\cite{Jackura:2023qtp} we referred to $S_j$ as $\ell_j$.}
\begin{align}
    \label{eq:pw_pair}
    \ket{\{\p\}_j} \propto \sqrt{4\pi}\,\sum_{S_j = 0}^{\infty}\sum_{\lambda_j = -S_j}^{S_j} \ket{\p_j,S_j\lambda_j} \, Y_{S_j \lambda_j}^{*}(\bh{p}_{j+1}^\star) \, ,
\end{align}
where $\bh{\p}_{j+1}^\star$ is the orientation of the first particle in the pair, in the pair CM frame (as indicated by the $\star$ superscript on the momentum). We define the quantization axis $z_j$ coincident with $-\p_j$, so that $\lambda_j$ may be interpreted as the helicity of the pair when boosted back to the three-body CM frame. Notice that this quantization axis is identical to the body-fixed axis in the Dalitz decomposition, and we may identify the projection $\lambda_j$ in Eq.~\eqref{eq:dalitz_pw} with the helicity label in Eq.~\eqref{eq:pw_pair}.

\begin{figure}[t]
	\centering
	\includegraphics[width=0.8\textwidth]{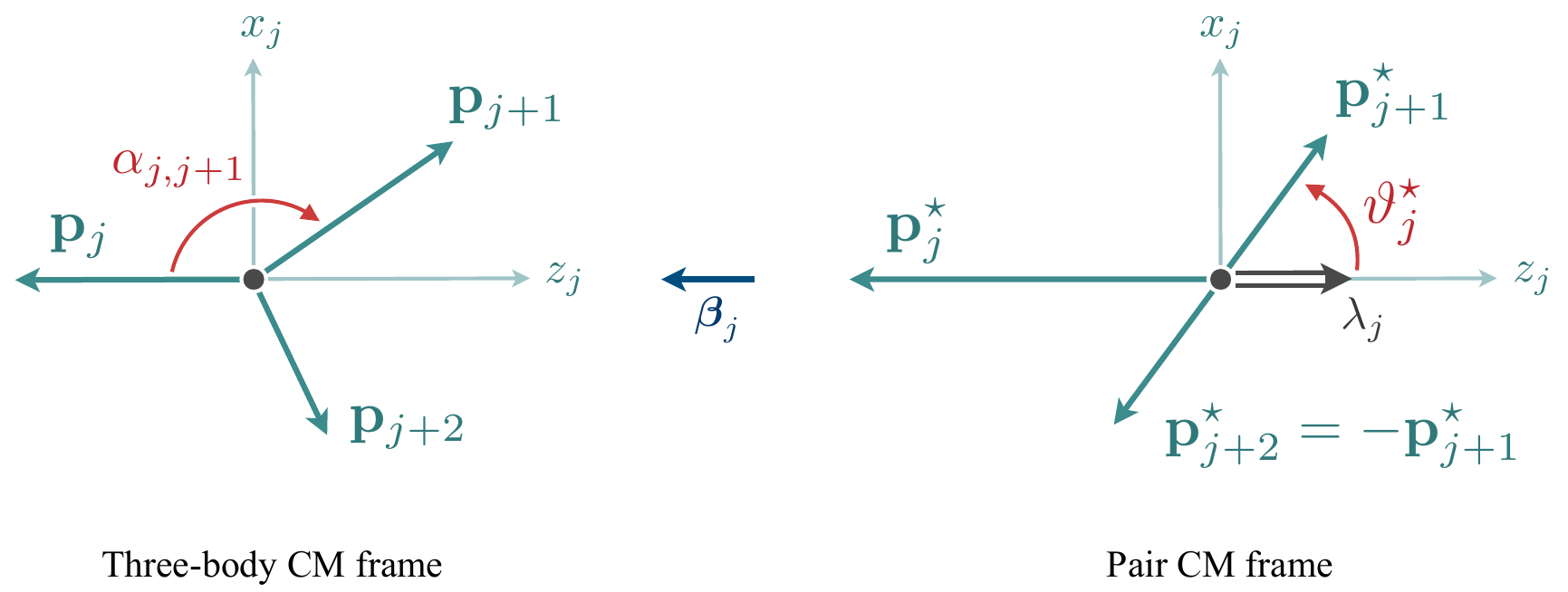}
	\caption{Kinematics in the pair CM frame.}
	\label{fig:pair_cm_frame}
\end{figure}

The orientation of the first particle in the pair, \ie, the $(j+1)$th particle, is specified by a polar angle $\vartheta_{j}^\star$ defined with respect to the $z_j$ axis (see Fig.~\ref{fig:pair_cm_frame}) and an azimuthal angle $\psi_{j}^\star$ that measures rotations about $z_j$. The remaining angles are specified by the orientation of the $z_j$ axis (equivalently, the negative of the spectator momentum) with respect to the space-fixed coordinate system. We define the polar and azimuthal angles of $\bh{z}_j \equiv -\bh{p}_j$ with respect to ($xyz$) as $\theta_j$ and $\varphi_j$, respectively. With these definitions, note that under a Lorentz boost along $z_j$ from the pair CM frame to the three-body CM frame, the angle $\psi_{j}^\star = \psi_j$ is invariant as well as the helicity $\lambda_j$, and we have identified this azimuthal angle as the first Euler angle for the three-particle plane, \cf Fig.~\ref{fig:euler}.
%

Next, we expand the quasi-two-body pair-spectator state into total angular momentum states using the helicity basis of Ref.~\cite{Jacob:1959at},
\begin{align}
    \label{eq:pw_quasi}
    \ket{\p_j,S_j,\lambda_j} \propto \sum_{J,m_J} \ket{p_j,Jm_J,S_j\lambda_j} \, \sqrt{2J+1} \, D_{m_J\lambda_j}^{(J)}(\varphi_j,\theta_j,0) \, ,
\end{align}
where $J$ and $m_J$ are as before. The combined expansions Eq.~\eqref{eq:pw_pair} and \eqref{eq:pw_quasi} can be expressed in terms of the Euler angles $\Ec_j$ as defined in Eq.~\eqref{eq:dalitz_pw} as follows. First, we use the known relation between spherical harmonics and Wigner $D$ matrices~\cite{VMK},
\begin{align}
    \label{eq:Y_to_D}
    \sqrt{4\pi}\,Y_{S_j\lambda_j}^{*}(\bh{\p}_{j+1}^\star) = \sqrt{2S_j+1} \, D_{\lambda_j 0}^{(S_j)}(\psi_j^\star,\vartheta_j^\star,0) = \sqrt{2S_j+1}\,d_{\lambda_j 0}^{(S_j)}(\vartheta_j^\star) \, e^{-i\lambda_j\psi_j} \, .
\end{align}
Next, we combine the phase $e^{-i\lambda_j\psi_j}$ from Eq.~\eqref{eq:Y_to_D} with $D_{m_j\lambda_j}^{(J)}(\varphi_j,\theta_j,0)$ of Eq.~\eqref{eq:pw_quasi} to form $D_{m_J\lambda_j}^{(J)}(\Ec_j)$. Therefore, combining both expansions we find that $\ket{\{\p\}_j}$ can be expanded as
\begin{align}
    \label{eq:pw_combined}
    \ket{\{\p\}_j} \propto \sum_{J,m_J} \sum_{S_j,\lambda_j} \ket{p_j,Jm_J,S_j\lambda_j}\, \sqrt{2S_j + 1} \, d_{\lambda_j 0 }^{(S_j)}(\vartheta_j^\star)  \, \sqrt{2J+1} \, D_{m_J \lambda_j}^{(J)}(\varphi_j,\theta_j,\psi_j) \, .
\end{align}

We make an identification of Eq.~\eqref{eq:pw_combined} with the Dalitz expansion~\eqref{eq:dalitz_pw} by first arguing that the sum of $\lambda_j$ bounded by $-S_j \le \lambda_j \le S_j$ can be extended to the range in Eq.~\eqref{eq:dalitz_pw} , if $S_j<J$. Notice that for a fixed $S_j$ that if $\lambda_j$ is chosen outside the boundary that $d_{\lambda_j 0}^{(S_j)}$ is identically zero. Thus, we may trivially extend the range knowing that these additional states will not contribute. Therefore, we have found that the Dalitz state is related to the quasi-two-body angular momentum state as
\begin{align}
    \label{eq:dalitz_to_pair}
    \ket{\{\sigma\},Jm_J\lambda_j} = \sum_{S_j} \ket{p_j,Jm_J,S_j\lambda_j}\, \sqrt{2S_j + 1} \, d_{\lambda_j 0 }^{(S_j)}(\vartheta_j^\star) \, .
\end{align}
It is important to note that $\vartheta_j^\star$ can be written in terms of invariant masses. This can be seen by first writing the energy of the $j+1$ particle in three-body CM frame in terms of the kinematics of the $j$ pair CM frame,
\begin{align}
    \label{eq:angle_vartheta_j}
    E_{j+1} & = \gamma_j(E_{j+1}^\star + p_{j+1}^\star \beta_j \cos\vartheta_j^\star) \, ,
\end{align}
where $\gamma_j = (\sqrt{s} - E_j)/\sqrt{\sigma_j}$ is the Lorentz factor and $\beta_j = p_j / (\sqrt{s} - E_j)$ is the boost velocity between the two frames. Furthermore, both $p_j$ and $E_j$ are defined in terms of Lorentz scalars in Eqs.~\eqref{eq:momentum} and~\eqref{eq:energy}, respectively. The energy of the $(j+1)$th particle in the three-body CM frame is given by Eq.~\eqref{eq:energy} with the replacement $j \to j+1$. 

Next, we express both $p_{j+1}^\star$ and $E_{j+1}^\star$ in terms of Lorentz scalars. This can be shown by observing that in the $j$ pair CM frame, the momentum $p_{j+1}^\star$ is defined by~\cite{byckling1973particle},
\begin{align}
    p_{j+1}^\star = \frac{1}{2\sqrt{\sigma_j}} \, \lambda^{1/2}(\sigma_j,m_{j+1}^2,m_{j+2}^2) \, .
\end{align}
Given this, one can then use the definition of the energy $E_{j+1}^\star=\sqrt{\left(p_{j+1}^\star\right)^2+m_{j+1}^2}$ to find
\begin{align}
    E_{j+1}^\star = \frac{1}{2\sqrt{\sigma_j}} \, (\sigma_j + m_{j+1}^2 - m_{j+2}^2) \, .
\end{align}
 Therefore, $\vartheta_j^\star$ is entirely fixed by the kinematics $s$, $\sigma_j$, and $\sigma_{j+1}$ through Eq.~\eqref{eq:angle_vartheta_j}.

This quasi-two-body angular momentum state does not possess definite parity. To form states of definite spin-parity $J^P$, we form linear combinations of the partial wave helicity states as
\begin{align}
	\label{eq:state_LS}
	\ket{p_j,Jm_J,S_j \lambda_j} = \sum_{L_j = 0}^{\infty}\ket{p_j,Jm_J,L_jS_j}\Pc_{\lambda_j}({}^{2S_j+1}({L_j})_J) \, ,
\end{align}
where the parity of the state is $\eta(-1)^{L_j+S_j}$ with $\eta = \eta_1\eta_2\eta_3$, $L_j$ is the orbital angular momentum of the pair-spectator system, and $S_j$ is now interpreted as the intrinsic spin of the pair-spectator system. We emphasize that the subscripts on the quantum numbers are defined with respect to the spectator $j$. The spin-orbit coupling coefficients~\footnote{In Ref.~\cite{Jackura:2023qtp} we denoted $\ell_j$ as the angular momentum of the pair, and $S_j$ as the intrinsic spin of the system, with an additional label of the spin-orbit couplings, \ie, $\Pc_\lambda^{(\ell)}({}^{2S+1}L_J)$. Since we exclusively work with spinless particles, we have identically $S_j = \ell_j$, so we drop this redundant notation for brevity throughout this article.} are defined as~\cite{Jacob:1959at,Jackura:2023qtp},
\begin{align}
    \label{eq:spin_orbit_coupling}
	\Pc_\lambda({}^{2S+1}L_J) = \sqrt{\frac{2L+1}{2J+1}}\,\Cc^{J\lambda}_{L0;S\lambda} \, ,
\end{align}
where $\Cc^{j m_j}_{j_1 m_1, j_2 m_2} \equiv \braket{j m_j | j_1 m_1,j_2 m_2}$ is an SU(2) Clebsch-Gordan coefficient~\cite{VMK}. Inserting Eq.~\eqref{eq:state_LS} into~\eqref{eq:dalitz_to_pair} gives the relation
\begin{align}
    \label{eq:dalitz_to_LS}
    \ket{\{\sigma\},Jm_J\lambda_j} = \sum_{L_j,S_j} \ket{p_j,Jm_J,L_jS_j}\, \sqrt{2S_j + 1} \, d_{\lambda_j 0 }^{(S_j)}(\vartheta_j^\star)\, \Pc_{\lambda_j}({}^{2S_j+1}({L_j})_J) \, . 
\end{align}
The abundance of subscripts is essential in discriminating which pair-spectator system is used to define quantum numbers and kinematic variables. We will see in Sec.~\ref{sec:symm} that this is necessary when we symmetrize over all possible pair-spectator combinations.

Finally, we take the linear combinations of Eq.~\eqref{eq:dalitz_to_LS} as shown in Eq.~\eqref{eq:dalitz_JP_state} to form definite parity states. Using the Wigner $d$ symmetry property $d_{-\lambda 0}^{(S)}(\vartheta^\star) = (-1)^\lambda d_{\lambda 0}^{(S)}(\vartheta^\star)$ as well as the symmetry relations for Clebsch-Gordon coefficients to write $\Pc_{-\lambda}({}^{2S+1}L_J) = (-1)^{J-L-S}\,\Pc_{\lambda}({}^{2S+1}L_J)$ (\cf Eq.~\eqref{eq:spin_orbit_coupling} and Refs.~\cite{VMK,Jackura:2023qtp}), one can show that Dalitz states of definite $J^P$ are related to spin-orbit states through
\begin{align}
    \label{eq:dalitz_JP_to_LS}
	\ket{\{\sigma\},J^P m_J \lambda_j} = \sum_{L_j,S_j} \ket{p_j,Jm_J,L_jS_j} \, \sqrt{2S_j + 1} \, d_{\lambda_j 0}^{(S_j)}(\vartheta_j^\star) \, \Pc_{\lambda_j}({}^{2S_j+1}(L_j)_J) \, \delta_{PP_j} \, .
\end{align}
Here, $\delta_{P P_j}$ restricts the sum to enforce the parity of the spin-orbit state, $P_j \equiv \eta (-1)^{L_j + S_j}$, to be equal to the target parity $P$.

Taking appropriate S matrix elements in the basis given by Eq.~\eqref{eq:dalitz_JP_to_LS}, we find the connection between the definite-parity Dalitz partial wave amplitudes and the spin-orbit amplitudes,
\begin{align}
    \label{eq:dalitz_amp_to_LS}
    \Mc_{\lambda_{j'}'\lambda_j}^{J^P}(\{\sigma'\};\{\sigma\}) & = \sum_{L_{j'}',S_{j'}'}\sum_{L_j,S_j}\sqrt{2S_{j'}'+1} \, d_{\lambda_{j'}' 0}^{(S_{j'}')}(\vartheta_{j'}'^\star)\,\Pc_{\lambda_{j'}'}({}^{2S_{j'}'+1}(L_{j'}')_{J}) \, \delta_{PP_{j'}'}  \nn \\[5pt]
	& \quad \times \, \Mc_{L_{j'}'S_{j'}';L_jS_j}^{J}(p_{j'}';p_j) \sqrt{2S_j+1} \,d_{\lambda_j 0}^{(S_j)} (\vartheta_j^\star) \, \Pc_{\lambda_j}({}^{2S_j+1}(L_j)_J) \, \delta_{PP_j} \, ,
\end{align}
where the final state parity is defined as $P_{j'}' = \eta' (-1)^{L_{j'}' + S_{j'}'}$ with $\eta' \equiv \eta_{1'}\eta_{2'}\eta_{3'}$. The $LS$ partial wave amplitude is defined, up to an overall momentum conserving delta function, through the S matrix element, 
\begin{align}
    \bra{p_{j'}',Jm_J, L_{j'}' S_{j'}'}\mathsf{S}\ket{p_{j},Jm_J, L_{j} S_{j}}_{\mathrm{f.c.}} \propto \Mc_{L_{j'}' S_{j'}';L_{j} S_{j}}^{J}(p_{j'}';p_j) \, .
\end{align}
%

\section{Symmetrizing partial wave amplitudes}
\label{sec:symm}

We have thus far concentrated on the total $\3\to\3$ amplitude, meaning that all possible dynamical processes are encoded in the resulting partial wave amplitudes given by Eq.~\eqref{eq:dalitz_JP_amp}. In applications to phenomenology or lattice QCD, it is convenient to work first with amplitudes of definite pair-spectator systems, ignoring all other possible combinations~\cite{Hansen:2015zga,Blanton:2020gha,Jackura:2022gib,Briceno:2024ehy}. We call these \emph{asymmetric} amplitudes and denote them by $\Mc^{({j'},j)}$, where the superscripts inform which spectators have been chosen. These amplitudes are the primary objects that are reconstructed by sets of integral equations given a three-body K matrix~\cite{Jackura:2023qtp,Briceno:2024txg}. The full $\3\to\3$ amplitude is then given by summing all pair-spectator combinations, 
\begin{align}
	\label{eq:amp_symm}
	\Mc(\{\p'\},\{\p\}) = \sum_{j,{j'}} \Mc^{({j'},j)}(\{\p'\}_{j'};\{\p\}_j) \, ,
\end{align}
where we remind the reader that $\{\p\}_j$ represents a set of momenta where we have singled out momentum $\p_j$ to be the spectator momentum, while the other two are the momenta of the pair. We call the processes given by Eq.~\eqref{eq:amp_symm} \emph{symmetrizing} the amplitude, and emphasize that it does not refer to constructing a state of identical particles.

\subsection{Angular momentum}
\label{sec:symm_ang_mom}

The full amplitude on the left-hand side of Eq.~\eqref{eq:amp_symm} is expanded into partial waves through Eq.~\eqref{eq:Dalitz_expand} for some choice of spectators. Like the full amplitude, the asymmetric amplitudes admit a partial wave expansion similar to Eq.~\eqref{eq:Dalitz_expand},
\begin{align}
	\label{eq:Dalitz_expand_pair_spectator}
	\Mc^{({j'},j)}(\{\p'\}_{j'},\{\p\}_j) & = \sum_{J,m_J}\sum_{\lambda_{j'}',\lambda_j}(2J+1)D_{m_J\lambda_{j'}'}^{(J)\,*}(\Ec_{j'}') \, \Mc_{\lambda_{j'}'\lambda_j}^{({j'},j)\,J}(\{\sigma'\};\{\sigma\})\, D_{m_J\lambda_j}^{(J)}(\Ec_j) \, ,
\end{align}
where now the Euler angles are defined with respect to the spectator involved in the symmetrization of Eq.~\eqref{eq:amp_symm}. While symmetrizing is a straightforward operation, the procedure for projecting to partial wave amplitudes requires care, as each combination is defined with its own set of Euler angles to define orientations. The key idea is that no matter which $\mathrm{SC}$ is chosen to define the Euler angles, the final orientations of the three-particle planes are identical. Therefore, we can always find a sequence of rotations relating Euler angles defined with respect to some $\mathrm{SC}$ to another. 

Explicitly, consider an initial state $\ket{\{\sigma\},J m_J,\lambda_a}$ for some chosen spectator $a$. We call $a$ the \emph{target} spectator. By inverting the expansion given by Eq.~\eqref{eq:dalitz_pw}, we find the projection
\begin{align}
    \ket{\{\sigma\},J m_J,\lambda_a} & \propto \frac{\sqrt{2J+1}}{8\pi^2} \int\!\diff\Ec_{a} \, D_{m_J \lambda_a}^{(J)\,*}(\Ec_a) \,\ket{\{\p\}} \, .
\end{align}
Since the orientation of the three particles described by $\ket{\{\p\}}$ is independent of the spectator choice, we now expand the state in basis with Euler angles defined by $\mathrm{SC}_j$, giving the relation
\begin{align}
    \label{eq:dalitz_basis_hel_rec}
    \ket{\{\sigma\},J m_J \lambda_a} & = \sum_{J',m_{J'}'\lambda_{j}'} \ket{\{\sigma\},J'm_{J'}'\lambda_j'} \nn \\[5pt]
    & \qquad \qquad \times \frac{\sqrt{(2J+1)(2J'+1)}}{8\pi^2}  \int\!\diff\Ec_{a} \, D_{m_{J'}'\lambda_j'}^{(J')}(\Ec_j)  \,  D_{m_J \lambda_a}^{(J)\,*}(\Ec_a) \, .
\end{align}
The integral relates how a three-particle state defined with Euler angles with respect to $\mathrm{SC}_j$ to those of the target from $\mathrm{SC}_a$. In the sub-space where the invariant masses are fixed, this integral defines the helicity-basis angular momentum recoupling coefficients for the three-particle system,
\begin{align}
    \braket{\{\sigma\}, J'm_{J'}' \lambda_j' | \{\sigma\}, Jm_J \lambda_a } = \delta_{J'J} \delta_{m_{J'}'m_J} \, \left(\Rc_{j\to a}^J\right)_{\lambda_j' \lambda_a} \, ,
\end{align}
where $\Rc_{j\to a}^J$ is defined by
\begin{align}
    \label{eq:hel_ang_mom_rec}
    \left(\Rc_{j\to a}^J\right)_{\lambda_j' \lambda_a} \equiv \frac{1}{8\pi^2} \sum_{m_J} \int\!\diff\Ec_{a} \, D_{m_{J}\lambda_j'}^{(J)}(\Ec_j) \,  D_{m_J \lambda_a}^{(J)\,*}(\Ec_a) \, .
\end{align}
In the following paragraphs we show that $\Rc_{j \to a}^{J}$ is diagonal in $J$, independent of $m_J$, and depends only on the kinematics of the particles, \ie, $\Rc_{j\to a}^J = \Rc_{j\to a}^{J}(\{\sigma\})$. Notice that if $a = j$, then the recoupling coefficient is the identity in helicity space, \ie,
\begin{align*}
    \left(\Rc_{j\to j}^J\right)_{\lambda_j'\lambda_j} = \delta_{\lambda_j'\lambda_j} \, ,
\end{align*}
which can be seen by the orthornormality of Wigner $D$ matrices, Eq.~\eqref{eq:wigner_d_ortho}.

%
%
\begin{figure}[t]
	\centering
	\includegraphics[width=\textwidth]{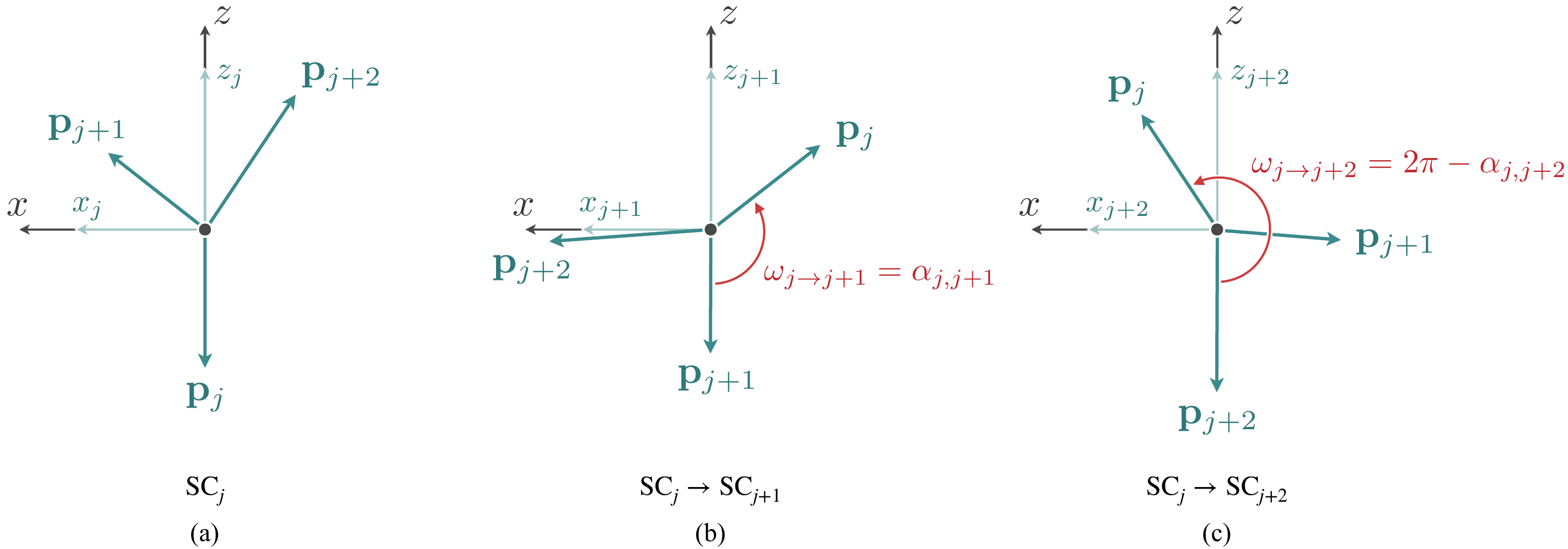}
	\caption{Three standard configurations for the three-particle plane.}
	\label{fig:stdcfg}
\end{figure}
%
We evaluate the integrals of Eq.~\eqref{eq:hel_ang_mom_rec} by connecting $\Ec_j$ to $\Ec_a$. We recognize that the Euler angles $\Ec_j$ depend on the three-particle plane being defined with respect to $\mathrm{SC}_j$, while $\Ec_a$ is defined with respect to $\mathrm{SC}_a$. As illustrated in Fig.~\ref{fig:stdcfg}, the three standard configurations are related by rotations about $y$, and we use this fact to reduce the integral over $\Ec_a$ in Eq. \ref{eq:dalitz_basis_hel_rec}. If $\mathsf{R}(\Ec_j)$ represents the Euler rotation matrix corresponding to $\mathrm{SC}_j$, and $\mathsf{R}(\Ec_a)$ to $\mathrm{SC}_a$, then we may use the composition of rotations to relate the two sets of angles as
\begin{align}
    \label{eq:rotate_stdcfg}
    \mathsf{R}(\Ec_j) = \mathsf{R}(\Ec_a)\cdot \mathsf{R}_y(\omega_{j \to a}) \, ,
\end{align}
where $\mathsf{R}_y(\omega_{j\to a})$ represents a positive rotation by an angle $\omega_{j\to a}$ about the $+y$-axis. This rotates the three-particle system from $\mathrm{SC}_j$ to $\mathrm{SC}_a$.

 First, notice that if $a = j$, then identically $\omega_{j\to j} = 0$, as expected. For $a\ne j$, one must consider if $a$ is cyclic or anti-cyclic with respect to $j$, as we have explicitly defined Eq.~\eqref{eq:rotate_stdcfg} by a positive rotation about the $+y$-axis. If $a = j+1$, \ie, the cyclic case, then $\omega_{j \to j+1} = \alpha_{j,j+1}$ with $\alpha_{j,j+1}$ being defined in Eq.~\eqref{eq:alpha_angle} (see Fig.~\ref{fig:stdcfg}). However, if $a = j+2$ (equivalently, $a = j-1$), which is the anti-cyclic case, then $\omega_{j \to j+2} = 2\pi - \alpha_{j,j+2}$, with $\alpha_{j,j+2}$ being the interior angle between $j$ and $j+2$ also defined by Eq.~\eqref{eq:alpha_angle}. One can also think of the rotation from $\mathrm{SC}_{j}$ to $\mathrm{SC}_{j+2}$ as two successive rotations: one from $j$ to $j+1$, followed by a rotation $j+1$ to $j+2$. Therefore, the angle $\omega_{j\to a}$ is given by
\begin{align}
    \label{eq:omega_aj_values}
    \omega_{j\to a} = \begin{cases}
        0 \, , &  a = j \, , \\
        \alpha_{j,j+1} \, ,&  a = j+1 \, ,\\
        2\pi - \alpha_{j,j+2} \, , & a = j+2 \, .
    \end{cases}
\end{align}

Representations of Eq.~\eqref{eq:rotate_stdcfg} in the angular momentum basis yield
\begin{align}
    D_{m_J\lambda_j'}^{(J)}(\Ec_j) = \sum_{\mu} D_{m_J\mu}^{(J)}(\Ec_a) \, d_{\mu\lambda_j'}^{(J)}(\omega_{j \to a}) \,.
\end{align}
Therefore, the integrals of Eq.~\eqref{eq:dalitz_basis_hel_rec} can be evaluated directly as
\begin{align}
    \label{eq:wigner_d_recoup}
    \frac{\sqrt{(2J+1)(2J'+1)}}{8\pi^2} & \int\!\diff\Ec_a \,  D_{m_{J'}'\lambda_j'}^{(J')}(\Ec_j) D_{m_J\lambda_a}^{(J)\,*}(\Ec_a) = \delta_{J'J} \delta_{m_{J'}' m_J} \, d_{\lambda_a\lambda_j'}^{(J)}(\omega_{j\to a}) \, .
\end{align}
We conclude that the $j\to a$ helicity recoupling coefficient, Eq.~\eqref{eq:hel_ang_mom_rec}, is simply
\begin{align}
    \left(\Rc_{j \to a}^J\right)_{\lambda_j' \lambda_a} = d_{\lambda_a\lambda_j'}^{(J)}(\omega_{j\to a}) \, .
\end{align}
Noticing the order of the helicities, we emphasize that this is the recoupling coefficient \emph{from} $\mathrm{SC}_j$ \emph{to} $\mathrm{SC}_a$. Moreover, since $\omega_{j\to a}$ is determined by $\alpha_{j,a}$, the recoupling coefficient is thus a function of the invariants $\{\sigma\}$. Given this result, Eq.~\eqref{eq:dalitz_basis_hel_rec} reduces to
\begin{align}
    \label{eq:dalitz_symm_state}
    \ket{\{\sigma\},Jm_J\lambda_a} = \sum_{\lambda_j'} \ket{\{\sigma\},Jm_J\lambda_j'}\,d_{\lambda_a\lambda_j'}^{(J)}(\omega_{j \to a}) \, .
\end{align}
As expected, the recoupling coefficient is the identity when $a = j$. In which case, $\omega_{j\to j} = 0$, and $d_{\lambda_j\lambda_j'}^{(J)}(0) = \delta_{\lambda_j\lambda_j'}$. We can also use the symmetry properties of the Wigner $d$ matrix elements~\cite{VMK} to write the recoupling coefficient in terms of $\alpha_{j,a}$ directly at the cost of an index-dependent phase that depends on the cyclicity of the recoupling. For $j = a+2$, we use $d_{\lambda_a \bar{\lambda}_j}^{(J)}(2\pi - \alpha_{j,a}) = (-1)^{2J + \lambda_a - \bar{\lambda}_j} d_{\lambda_a \bar{\lambda}_j}^{(J)}(\alpha_{j,a})$, so that the helicity recoupling coefficient is
\begin{align*}
    \left(\Rc_{j\to a}^J\right)_{\bar{\lambda}_j \lambda_a} = (-1)^{\varepsilon_{j\to a}} \, d_{\lambda_a\bar{\lambda}_j}^{(J)}(\alpha_{j,a}) \, ,
\end{align*}
where the phase $\varepsilon_{j \to a} = 0$ if $a = j$ or $a = j+1$ (cyclic coupling), and $\varepsilon_{j \to a} = 2J + \lambda_a - \bar{\lambda}_j$ if $a = j+2$ (anti-cyclic coupling).

With the helicity-basis recoupling coefficients identified, we now apply them to the partial wave expansion of the symmetrized amplitude. We project the left-hand side of Eq.~\eqref{eq:amp_symm} as given by Eq.~\eqref{eq:dalitz_pw_from_full} with initial and final target spectators $a$ and $a'$, respectively, and insert the expansion in Eq.~\eqref{eq:Dalitz_expand_pair_spectator} for each asymmetric amplitude, giving
\begin{align}
    \label{eq:amp_sym_intermediate}
	\Mc_{\lambda_{a'}'\lambda_a}^{J}(\{\sigma'\};\{\sigma\}) & = \frac{2J+1}{(8\pi^2)^2}\sum_{j,{j'}}\sum_{m_J}\sum_{J',m_J'} \sum_{\mu_{j'}',\mu_j} \int\!\diff\Ec_{a'} '\, D_{m_J'\mu_{j'}'}^{(J')\,*}(\Ec_{j'}') D_{m_J\lambda_{a'}'}^{(J)}(\Ec_{a'}')  \nn\\[5pt]
	& \qquad \times \Mc_{\mu_{j'}'\mu_j}^{({j'},j)\,J}(\{\sigma'\};\{\sigma\})\, \int\!\diff\Ec_a \, D_{m_J'\mu_j}^{(J')}(\Ec_j)  D_{m_J\lambda_a}^{(J)\,*}(\Ec_a)  \, .
\end{align}
Each of the integrals is a recoupling coefficient; thus, using the result in Eq.~\eqref{eq:wigner_d_recoup}, we find for the symmetrized partial wave amplitude
\begin{align}
    \label{eq:symm_procedure}
    \Mc_{\lambda_{a'}'\lambda_a}^{J}(\{\sigma'\};\{\sigma\}) = \sum_{j,{j'}} \sum_{\mu_{j'}',\mu_j} d_{\lambda_{a'}' \mu_{j'}'}^{(J)}(\omega_{j' \to a'}') \, \Mc_{\mu_{j'}'\mu_j}^{({j'},j)\,J}(\{\sigma'\};\{\sigma\}) \, d_{\lambda_a \mu_j}^{(J)}(\omega_{j\to a}) \, ,
\end{align}
where $\omega_{j' \to a'}'$ is the angle which brings $\mathrm{SC}_{j'}$ to $\mathrm{SC}_{a'}$. We emphasize that the angles $\omega_{j\to a}$ and $\omega_{j'\to a'}'$ are given entirely by $s$ and invariant masses via Eqs.~\eqref{eq:omega_aj_values} and ~\eqref{eq:alpha_angle}. 

In computing the final state recoupling coefficient in Eq.~\eqref{eq:symm_procedure}, we have
\begin{align}
    \braket{\{\sigma\}, Jm_J \lambda_a|\{\sigma\}, Jm_J \lambda_j'} = \left(\Rc_{a\to j}^J\right)_{\lambda_a \lambda_j'} \, ,
\end{align}
which is for transformations from $\mathrm{SC}_{a}$ to $\mathrm{SC}_{j}$. Following the steps outlined above, we find $(\Rc_{a\to j}^J)_{\lambda_a \lambda_j'} = d_{\lambda_j'\lambda_a}^{(J)}(\omega_{a \to j})$. However, note that $\omega_{a \to j} = -\omega_{j\to a}$, thus 
\begin{align}
    d_{\lambda_{j}'\lambda_{a}}^{(J)}(-\omega_{j\to a}) = d_{\lambda_{a}\lambda_{j}'}^{(J)}(\omega_{j \to a}) \, ,
\end{align}
which follows from the symmetry properties of the Wigner $d$ matrices~\cite{VMK}. From this, we find that $\bs{\Rc}_{a\to j}^J = (\bs{\Rc}_{j\to a}^J)^{\transpose}$ as a matrix in helicity space.

We form definite parity relations by inserting Eq.~\eqref{eq:symm_procedure} into~\eqref{eq:dalitz_JP_amp}. Using states of definite parity given by Eq.~\eqref{eq:dalitz_JP_state} and substituting Eq.~\eqref{eq:dalitz_symm_state}, we find
\begin{align}
    \label{eq:dalitz_state_parity_symm}
    \ket{\{\sigma\},J^Pm_J\lambda_a} & = \frac{1}{2}\bigg[ \sum_{\lambda_j} \ket{\{\sigma\},Jm_J\lambda_j}\,d_{\lambda_a\lambda_j}^{(J)}(\omega_{j\to a}) \nn \\[5pt]
    & \qquad\qquad  +\, \eta P (-1)^{J+\lambda_a} \sum_{\lambda_j} \ket{\{\sigma\},Jm_J\lambda_j}\,d_{-\lambda_a\lambda_j}^{(J)}(\omega_{j\to a})\bigg] \, , \nn \\[5pt]
    & = \sum_{\lambda_j} \ket{\{\sigma\},J^Pm_J \lambda_j} d_{\lambda_a\lambda_j}^{(J)}(\omega_{j\to a}) \, .
\end{align}
In going to the second line we take $\lambda_j \to -\lambda_j$ in the second term and used the symmetry relation $d_{-\lambda\,-\lambda'}^{(J)} = (-1)^{\lambda - \lambda'} d_{\lambda\lambda'}$ (see Ref.~\cite{VMK}) to form a Dalitz state of definite parity defined with respect to $\mathrm{SC}_j$. Therefore, the helicity-basis angular momentum recoupling coefficients for definite-parity states are identical to those for Eq.~\eqref{eq:dalitz_symm_state}. Applying this basis transformation to Eq.~\eqref{eq:symm_procedure}, the symmetrization procedure for definite $J^P$ amplitudes is then
\begin{align}
    \label{eq:symm_procedure_JP}
    \Mc_{\lambda_{a'}'\lambda_a}^{J^P}(\{\sigma'\};\{\sigma\}) = \sum_{j,{j'}} \sum_{\mu_{j'}',\mu_j} d_{\lambda_{a'}' \mu_{j'}'}^{(J)}(\omega_{j'\to a'}') \, \Mc_{\mu_{j'}'\mu_j}^{({j'},j)\,J^P}(\{\sigma'\};\{\sigma\}) \, d_{\lambda_a \mu_j}^{(J)}(\omega_{j\to a}) \, .
\end{align}

Our final step is to relate the symmetric $J^P$ amplitudes given by Eq.~\eqref{eq:symm_procedure_JP} to the spin-orbit amplitudes $\Mc_{L_{j'}'S_{j'}';L_jS_j}^{({j'},j)\,J}$, which are the primary objects analyzed in reaction phenomenological studies and lattice QCD. Since the asymmetric amplitudes admit identical partial wave expansions to the fully symmetrized amplitude, we can further express the asymmetric amplitudes of Eq.~\eqref{eq:symm_procedure_JP} in terms of spin-orbit amplitudes via Eq.~\eqref{eq:dalitz_amp_to_LS}, which gives the main result of this work,
\begin{align}
    \label{eq:amp_symm_JP}
        \Mc_{\lambda_{a'}'\lambda_a}^{J^P}(\{\sigma'\};\{\sigma\}) & = \sum_{j,{j'}} \sum_{L_{j'}',S_{j'}'}\sum_{L_j,S_j} \left(\Rc_{j'\to a'}^{J^P}\right)_{L_{j'}'S_{j'}' ; \lambda_{a'}' } \,  \nn\\[5pt]
    & \qquad \times \, \Mc_{L_{j'}'S_{j'}';L_jS_j}^{({j'},j)\,J}(p_{j'}';p_j) \, \left(\Rc_{j\to a}^{J^P}\right)_{L_jS_j ; \lambda_a }\, ,
\end{align}
where $(\Rc_{j\to a}^{J^P})_{LS;\lambda}$ are \emph{helicity-to-spin-orbit angular momentum recoupling coefficients},
\begin{align}
    \label{eq:R_LS_to_hel}
    \left(\Rc_{j\to a}^{J^P}\right)_{L_jS_j ; \lambda_a }\equiv \delta_{PP_j} \, \sqrt{2S_j+1}  \sum_{\mu_j} d_{\lambda_a \mu_j}^{(J)}(\omega_{j\to a}) \, d_{\mu_j 0}^{(S_j)} (\vartheta_j^\star) \, \Pc_{\mu_j}({}^{2S_j+1}(L_j)_J) \, . 
\end{align}

One can also form spin-orbit-to-spin-orbit recoupling coefficients to form a symmetric $\Mc_{L_{a'}'S_{a'}';L_aS_a}^{J^P}$ amplitude. These amplitudes are not necessary for the observables we examine in the following section. However, as these amplitudes are important for other studies, (\eg, analytic continuation), we provide their definition and the associated recoupling coefficients in App.~\ref{sec:ls_recoupling}.

\subsection{Flavor symmetry - SU(2) isospin}
\label{sec:isospin}

We extend the above symmetrization procedure to accommodate three particles with flavor symmetry. Specifically, we consider particles that respect SU(2) strong isospin symmetry, which is a common scenario in many lattice QCD calculations of the hadron spectrum. Other extensions, such as to systems with SU(3) flavor symmetry, \ie, degenerate light and strange quarks in QCD, follow a similar procedure. Let a single particle state with mass $m_j$, momentum $\p_j$, isospin $i_j$, and projection $\tau_j$ be denoted $\ket{\p_j,\tau_j} = \ket{\p_j}\otimes \ket{\tau_j}$, where the particle's isospin quantum number is implicit. The isospin states have unit normalization, $\braket{\tau'|\tau} = \delta_{i' i}\delta_{\tau' \tau}$. A three particle state is then denoted by
\begin{align}
    \ket{\{\p,\tau\}} \equiv \ket{\p_1\tau_1,\p_2\tau_2,\p_3\tau_3} = \ket{\{\p\}}\otimes \ket{\{\tau\}} \, ,
\end{align}
with $\ket{\{\tau\}}\equiv \ket{\tau_1,\tau_2,\tau_3}$ being defined similarly to $\ket{\{\p\}}$ as above. Since the Hilbert space factorizes, we focus on the flavor isospin states, as we have already formulated the relations for symmetric partial wave amplitudes. 

We form states of definite total isospin $I$ using a similar procedure as what is done for angular momentum states in Sec.~\ref{sec:so_to_dalitz}. For a given spectator $j$, we first couple the particles $j+1$ and $j+2$ in cyclic order to a total isospin $I_j$ and projection $m_{I_j}$. Then, the pair and spectator are coupled to a total isospin and projection $I$ and $m_I$, respectively. This coupling scheme is written as
\begin{align}
    \label{eq:iso_pair_spec_coupling}
    \ket{\{\tau\}} = \sum_{I, m_I}\sum_{I_j m_{I_j}} \ket{(i_{j+1} i_{j+2})\, I_j i_j, I m_I} \, \Cc^{I m_I}_{I_j m_{I_j}, i_j \tau_j} \Cc^{I_j m_{I_j}}_{i_{j+1} \tau_{j+1}, i_{j+2} \tau_{j+2}}  \, ,
\end{align}
where $\Cc$'s are Clebsch-Gordan coefficients. The sums are over all allowed quantum numbers. That is, for the pair, we have $\lvert i_{j+1} - i_{j+2} \rvert \le I_j \le i_{j+1} + i_{j+2}$ and $-I_j \le m_{I_j} \le I_j$, while for the total system we have $\lvert I_j - i_j \rvert \le I \le  I_j + i_j$ and $-I\le m_I \le I$. 

The scattering amplitude, defined similarly as in Eq.~\eqref{eq:amp_def},
\begin{align}
    \bra{\{\p',\tau'\}}\mathsf{S}\ket{\{\p,\tau\}}_{\mathrm{f.c.}} \equiv (2\pi)^4\,\delta^{(4)}(P_{\mathrm{out}} - P_{\mathrm{in}})\, i\Mc_{\{\tau'\};\{\tau\}}(\{\p'\};\{\p\}) \, ,
\end{align}
can then be block diagonalized in the space of total isospin as
\begin{align}
    \label{eq:amp_isospin_expand}
    \Mc_{\{\tau'\};\{\tau\}} = \sum_{I,m_I}\ \sum_{I_{j'}',I_{j}} \Cc^{I m_I}_{I_{j'}' m_{I_{j'}'}, i_{j'}' \tau_{j'}'} \Cc^{I_{j'}' m_{I_{j'}'}}_{i_{j'+1}' \tau_{j'+1}', i_{j'+2}' \tau_{j'+2}'} \, \Mc_{I_{j'}'  I_{j}}^{I} \, \Cc^{I m_I}_{I_j m_{I_j}, i_j \tau_j} \Cc^{I_j m_{I_j}}_{i_{j+1} \tau_{j+1}, i_{j+2} \tau_{j+2}},
\end{align}
where we have used the fact that total isospin is conserved in the reaction and the amplitude is independent of the projection $m_I$. Note that the projections of the pair are fixed such that $m_{I_{j'}'} = \tau_{j'+1}' + \tau_{j'+2}'$ and $m_{I_j} = \tau_{j+1} + \tau_{j+2}$ for the initial and final state, respectively. Thus, the amplitude depends on the individual particle isospins implicitly, and $I_j$, $I_{j'}'$, and $I$ explicitly.

Expressing the amplitude as a sum over pair-spectator combinations, 
\begin{align}
    \label{eq:amp_symm_iso}
    \Mc_{\{\tau'\};\{\tau\}}(\{\p'\};\{\p\}) = \sum_{j,j'} \Mc_{\{\tau'\}_{j'};\{\tau\}_j}^{(j',j)}(\{\p'\}_{j'};\{\p\}_j) \, ,
\end{align}
we can construct definite isospin amplitudes from individual pair-spectator amplitudes by using the expansion in Eq.~\eqref{eq:amp_isospin_expand} for each asymmetric amplitude and projecting the full amplitude to definite isospin $I$. This procedure leads to isospin recoupling coefficients, which transform states defined in one coupling scheme to another. Again, we denote the target spectators as $a$ and $a'$ for the initial and final states, respectively. 

To construct the recoupling coefficients, we focus on the initial state and consider the inverse of Eq.~\eqref{eq:iso_pair_spec_coupling} for some target spectator $a$ and subsequently expand the state $\ket{\{\tau\}}$ in terms of a spectator $j$. This leads to
\begin{align}
    \label{eq:project_iso}
    \ket{(i_{a+1}i_{a+2}) I_a i_a,I m_I} & = \sum_{m_{I_a}}\sum_{\{\tau\}} \ket{\{\tau\}} \, \Cc_{I_a m_{I_a},i_a\tau_a}^{I m_I} \Cc_{i_{a+1}\tau_{a+1},i_{a+2} \tau_{a+2}}^{I_a m_{I_a}} \, , \nn \\[5pt]
    & = \sum_{\bar{I}_j} \ket{(i_{j+1} i_{j+2})\, \bar{I}_j i_j, I m_I} (\Rc^I_{j\to a})_{\bar{I}_j I_a}  \, ,
\end{align}
where $(\Rc_{j\to a}^I)_{\bar{I}_j I_a}$ is the \emph{isospin recoupling coefficient} from the $j$ coupling scheme to the $a$ coupling scheme, defined via the matrix element
\begin{align}
    \braket{(i_{j+1} i_{j+2})\, \bar{I}_j i_j, \bar{I} m_{\bar{I}} |(i_{a+1}i_{a+2}) I_a i_a,I m_I} = \delta_{\bar{I} I} \delta_{m_{\bar{I}} m_I} \, (\Rc^I_{j\to a})_{\bar{I}_j I_a} \, .
\end{align}
Explicitly, following the procedure above, we find
\begin{align}
    \label{eq:isospin_recoup}
    (\Rc^I_{j\to a})_{\bar{I}_j I_a} = \frac{1}{2I+1} \sum_{m_I} \sum_{m_{\bar{I}_j}, m_{I_a}}\sum_{\{\tau\}} \Cc^{I m_I}_{\bar{I}_j m_{\bar{I}_j}, i_j \tau_j} \Cc^{\bar{I}_j m_{\bar{I}_j}}_{i_{j+1} \tau_{j+1}, i_{j+2} \tau_{j+2}} \Cc_{I_a m_{I_a},i_a\tau_a}^{I m_I} \Cc_{i_{a+1}\tau_{a+1},i_{a+2} \tau_{a+2}}^{I_a m_{I_a}} \, ,
\end{align}
where we have used the fact that the expression is independent of $m_I$, as it is fixed from the projections $\{\tau\}$. Thus, we averaged over its values, resulting in the $2I+1$ factor. Note that if $a = j$, we obtain the trivial case $(\Rc_{j\to j}^I)_{\bar{I}_j I_a} = \delta_{\bar{I}_j I_a}$ for all $I$. We can see this from Eq.~\eqref{eq:isospin_recoup} by evaluating $a = j$ and expressing it as
\begin{align}
    \label{eq:R_jj_ortho}
    (\Rc^{I}_{j\to j})_{\bar{I}_j I_j} 
    &= \frac{1}{2I+1}\sum_{m_{I}} \sum_{m_{I_j}, \tau_j} \Cc^{I m_{I}}_{\bar{I}_j m_{\bar{I}_j},\, i_{j} \tau_{j}}\Cc^{I m_{I}}_{I_j m_{I_j},\, i_j \tau_j}  
    \sum_{m_{\bar{I}_j}} \sum_{\tau_{j+1}, \tau_{j+2}} \Cc^{\bar{I}_j m_{\bar{I}_j}}_{i_{j+1} \tau_{j+1},\, i_{j+2} \tau_{j+2}} \Cc^{I_j m_{I_j}}_{i_{j+1} \tau_{j+1},\, i_{j+2} \tau_{j+2}} \, , \\[5pt]
    & = \frac{1}{2I+1}\sum_{m_{I}} \sum_{m_{I_j}, \tau_j} \Cc^{I m_{I}}_{\bar{I}_j m_{\bar{I}_j},\, i_{j} \tau_{j}}\Cc^{I m_{I}}_{I_j m_{I_j},\, i_j \tau_j}  \sum_{m_{\bar{I}_j}} \delta_{\bar{I}_j I_j} \delta_{m_{\bar{I}_j} m_{I_j}} \, , \\[5pt]
    &= \delta_{\bar{I}_j I_j} \, ,
\end{align}
where, in going to the second and last lines, we used the orthonormality of Clebsch-Gordan coefficients~\cite{VMK}.

A computationally efficient form for calculating the isospin recoupling coefficient is given by expressing them in terms of Wigner 6$j$ symbols~\cite{VMK}. To do so, let us examine the two non-trivial cases: $a = j+1$ and $a = j+2$. For $a = j+1$, that is, when we have a cyclic recoupling, we find from Eq.~\eqref{eq:isospin_recoup}
\begin{align} 
    \label{eq:isospin_recoup_cyclic}
    \left(\Rc^{I}_{j \to j+1} \right)_{\bar{I}_j I_{j+1}} & = \frac{1}{2I + 1} \sum_{m_I}\sum_{m_{I_{j+1}}, m_{\bar{I}_j}}\sum_{\{\tau\}} \Cc^{I m_{I}}_{\bar{I}_j m_{\bar{I}_j},\, i_j \tau_j} 
    \Cc^{\bar{I}_j m_{\bar{I}_j}}_{i_{j+1} \tau_{j+1},\, i_{j+2} \tau_{j+2}} \nn \\[5pt]
    & \qquad\qquad  \times \Cc^{I m_{I}}_{I_{j+1} m_{I_{j+1}},\, i_{j+1} \tau_{j+1}}
    \Cc^{I_{j+1} m_{I_{j+1}}}_{i_{j+2} \tau_{j+2},\, i_j \tau_j} \, .
\end{align}
Comparing Eq.~\eqref{eq:isospin_recoup_cyclic} to Eq. 9.1.8 in Ref.~\cite{VMK}, we find 
\begin{align}
    \label{eq:Rcyc}
    \left(\Rc^I_{j \to j+1} \right)_{\bar{I}_{j} I_{j+1}} = (-1)^{i_j + 2i_{j+1}+i_{j+2} + I_{j+1}}\sqrt{(2\bar{I}_j + 1)(2 I_{j+1} + 1)}\begin{Bmatrix}
        i_{j+1} & i_{j+2} & \bar{I}_j \\ i_j & I & I_{j+1}
    \end{Bmatrix} \, ,
\end{align}
where the $\{\cdots\}$ represents the Wigner 6$j$ symbol, see Ref.~\cite{VMK}. Similarly, for $a = j+2$, which is the anti-cyclic case as $j+2 = j-1\,\mathrm{mod}\, 3$, we have
\begin{align}
    \left(\Rc^{I}_{j \to j+2} \right)_{\bar{I}_{j}I_{j+2}} & = \frac{1}{2I + 1} \sum_{m_I}\sum_{m_{I_{j+2}}, m_{\bar{I}_j}}\sum_{\{\tau\}} \Cc^{I m_{I}}_{\bar{I}_j m_{\bar{I}_j},\, i_j \tau_j} 
    \Cc^{\bar{I}_j m_{\bar{I}_j}}_{i_{j+1} \tau_{j+1},\, i_{j+2} \tau_{j+2}} \nn \\[5pt]
    & \qquad \qquad \times \Cc^{I m_{I}}_{I_{j+2} m_{I_{j+2}},\, i_{j+2} \tau_{j+2}}
    \Cc^{I_{j+2} m_{I_{j+2}}}_{i_j \tau_j,\, i_{j+1} \tau_{j+1}} \,  \nn \\[5pt]
    \label{eq:Ranticyc}
    & = (-1)^{2 i_j + i_{j+1}+i_{j+2} + \bar{I}_{j}}\sqrt{(2\bar{I}_j + 1)(2 I_{j+2} + 1)}\begin{Bmatrix}
        i_{j} & i_{j+1} & I_{j+2} \\ i_{j+2} & I & \bar{I}_{j}
    \end{Bmatrix} \, .
\end{align}
Using symmetry properties of Wigner 6$j$ symbols~\cite{VMK}, all of these cases can be written as
\begin{align}
    \label{eq:R_iso_all}
    \left(\Rc_{j\to a}^{I}\right)_{\bar{I}_j I_a} = (-1)^{i_j + 2i_a + i_{3-j-a} + I_a + \delta_{j\to a}} \, \sqrt{(2\bar{I}_j+1)(2I_a + 1)} \, \begin{Bmatrix}
    i_a & i_{3-j-a} & \bar{I}_j \\
    i_j & I & I_a
    \end{Bmatrix} \, ,
\end{align}
where $\delta_{j\to a} = 0$ if $a = j$ or $a = j+1$ (cyclic coupling), and $\delta_{j\to a} = i_j + \bar{I}_j  - i_a - I_a$ if $a = j+2$ (anti-cyclic coupling).

Applying these definitions to the scattering amplitude, we find that projecting Eq.~\eqref{eq:amp_symm_iso} to definite isospin $I$ via Eq.~\eqref{eq:project_iso}, where the initial and final state target spectators are $a$ and $a'$, respectively, and expanding the pair-spectator amplitudes into definite isospin with Eq.~\eqref{eq:amp_isospin_expand}, the definite isospin symmetrized amplitude is given by
\begin{align}
    \label{eq:isospin_symm}
    \Mc_{I_{a'}' I_a}^I(\{\p'\};\{\p\}) = \sum_{j,{j'}} \sum_{\bar{I}_j,\bar{I}_{j'}'} \left(\Rc^I_{j' \to a'}\right)_{\bar{I}_{j'}' I_{a'}' } \,\Mc_{\bar{I}_{j'}' \bar{I}_j}^{({j'},j)\,I}(\{\p'\};\{\p\}) \, \left(\Rc^I_{j\to a}\right)_{\bar{I}_j I_a} \, .
\end{align}
For the final state, we used the fact that $a\to j$ recoupling coefficients are related to the $j\to a$ coefficients through a transposition, that is $\bs{\Rc}_{a\to j}^I = \left(\bs{\Rc}_{j\to a}^{I}\right)^{\transpose}$, or component-wise
\begin{align}
    \label{eq:isospin_transpose}
    \left(\Rc_{a \to j}^I\right)_{I_a\bar{I}_j} = \left(\Rc_{j \to a}^I\right)_{\bar{I}_j I_a}  \, .
\end{align}

To see that Eq.~\eqref{eq:isospin_transpose} is true, consider the recoupling coefficient $\left(\Rc_{a \to j}^I\right)_{I_a\bar{I}_j}$, written similarly to Eq.~\eqref{eq:R_iso_all},
\begin{align}
    \left(\Rc_{a \to j}^I\right)_{I_a\bar{I}_j} & = \braket{(i_{a+1}i_{a+2}) I_a i_a, I m_I | (i_{j+1} i_{j+2}) \bar{I}_j i_j, I m_I  } \, , \nn \\[5pt]
    & = (-1)^{i_a + 2i_j + i_{3-j-a} + \bar{I}_j + \delta_{a\to j}} \, \sqrt{(2\bar{I}_j+1)(2I_a + 1)} \, \begin{Bmatrix}
    i_j & i_{3-j-a} & I_a \\
    i_a & I & \bar{I}_j
    \end{Bmatrix} \, , \nn  \\[5pt]
    \label{eq:isospin_transpose_proof}
    & = (-1)^{i_j + \bar{I}_j - i_a - I_a + \delta_{a\to j} - \delta_{j\to a}} \, \left(\Rc_{j\to a}^I\right)_{\bar{I}_j I_a} \, .
\end{align}
In going to the last line, we used the Wigner 6$j$ symmetry relation,
\begin{align*}
    \begin{Bmatrix}
    a & b & c \\
    d & e & f
    \end{Bmatrix} = 
    \begin{Bmatrix}
    d & b & f \\
    a & e & c
    \end{Bmatrix} \, .
\end{align*}
Trivially, if $a = j$, we find $\bs{\Rc}_{a\to j}^I = \left(\bs{\Rc}_{j\to a}^{I}\right)^{\transpose} = \mathbf{I}$. Now, if the original $a\to j$ coupling is cyclic, then the $j\to a$ coupling is anti-cyclic. Therefore, $\delta_{a\to j} = 0$ but $\delta_{j\to a} = i_j + \bar{I}_j - i_a - I_a$. Thus, the phase of Eq.~\eqref{eq:isospin_transpose_proof} reduces to unity. If instead the coupling $a\to j$ is anti-cyclic and $j\to a$ cyclic, then $\delta_{a\to j} = i_a + I_a - i_j - \bar{I}_j$ and $\delta_{j\to a} = 0$, and again the phase of~\eqref{eq:isospin_transpose_proof} is unity. We have verified Eq.~\eqref{eq:isospin_transpose}, therefore concluding that $\bs{\Rc}_{a\to j}^I = \left(\bs{\Rc}_{j\to a}^{I}\right)^{\transpose}$ for every $j$ and $a$. Alternatively, we can see this by shifting $j\to j+1$ in $\bs{\Rc}_{j \to j+2}^I$ of Eq.~\eqref{eq:Ranticyc}, and recognizing that $j+3 = j$ mod 3, yielding $\bs{\Rc}_{j+1\to j}^I$, which is equivalent to the transposition above.

These recoupling coefficients can be interpreted as representations of the cyclic group of order 3. From $\bs{\Rc}_{j\to a}^I = (\bs{\Rc}_{a\to j}^I)^{\transpose}$, as well as recognizing that an anti-cyclic permutation is equivalent to permuting the system twice, we find $(\bs\Rc_{j \to j+1}^I)^2 = \bs\Rc_{j \to j+2}^I = (\bs\Rc_{j \to j+1}^I)^\transpose$ and $(\bs\Rc_{j \to j+1}^I)^3 = (\bs\Rc_{j \to j+2}^I)^3 = \mathbf{I}$. The coefficients presented here are a generalization of those presented for three pions, which have $i_j = 1$ for each $j$, as discussed in Ref.~\cite{Hansen:2020zhy}. However, Ref.~\cite{Hansen:2020zhy} primarily considers the cyclic group to generate the coefficients. Indeed, by explicit evaluation of Eq.~\eqref{eq:R_iso_all} (see the next section), we recover the results shown in Ref.~\cite{Hansen:2020zhy}.

We can now combine the results of Eqs.~\eqref{eq:amp_symm_JP} and~\eqref{eq:isospin_symm} in a straightforward manner. We find
\begin{align}
    \label{eq:amp_sym_final}
    \Mc_{I_{a'}'\lambda_{a'}';I_{a}\lambda_a}^{I(J^P)}(\{\sigma'\};\{\sigma\}) & = \sum_{j,{j'}} \sum_{L_{j'}',S_{j'}',\bar{I}_{j'}'}\sum_{L_j,S_j,\bar{I}_j} \left(\Rc_{j'\to a'}^{I(J^P)}\right)_{L_{j'}'S_{j'}'\bar{I}_{j'}' ; I_{a'}'\lambda_{a'}' } \,  \nn\\[5pt]
    & \qquad \times \, \Mc_{L_{j'}'S_{j'}'\bar{I}_{j'}';L_jS_j\bar{I}_j}^{({j'},j)\,I(J^P)}(p_{j'}';p_j) \, \left(\Rc_{j\to a}^{I(J^P)}\right)_{L_jS_j \bar{I}_j ; I_{a}\lambda_a }\, ,
\end{align}
where the $I(J^P)$ recoupling coefficients are defined as
\begin{align}
    \left(\Rc_{j\to a}^{I(J^P)}\right)_{L_jS_j \bar{I}_j ; I_{a}\lambda_a} \equiv 
    \left(\Rc_{j\to a}^I\right)_{\bar{I}_j I_a} \left( \Rc^{J^P}_{j\to a}\right)_{L_j S_j ; \lambda_a} \, ,
\end{align}
with the isospin coefficient being given by Eq.~\eqref{eq:R_iso_all} and the $J^P$ from the spin-orbit-to-helicity basis coefficient given by Eq.~\eqref{eq:R_LS_to_hel}. Of course, one could construct similar relations for symmetrizing helicity amplitudes or for projecting the symmetrized amplitude into the spin-orbit basis. See App.~\ref{sec:ls_recoupling} for this scenario.

To facilitate this, as well as reduce the notation of Eq.~\eqref{eq:amp_sym_final}, we define a target channel index $\mathfrak{c}_a$ which is either the spin-orbit set $\{L_a,S_a,I_a\}$ or the helicity set $\{\lambda_a,I_a\}$, and a set of \emph{target quantum numbers} $\mathfrak{q} \equiv I(J^P)$. A summed channel index for each pair-spectator channel is denoted $\bar{\mathfrak{c}}_j$, and again can either be helicity or spin-orbit quantum numbers. If additional quantum numbers are required to specify the reaction, \eg, flavor quantum numbers like strangeness, then these can be included trivially by extending the channel index. For a non-zero contribution, each channel $\bar{\mathfrak{c}}_j$ must have a non-zero overlap with $\mathfrak{q}$. Therefore, the symmetrization procedure is
\begin{align}
    \Mc^{\mathfrak{q}}_{\mathfrak{c}'_{a'}\mathfrak{c}_a}(\kappa_{a'}',\kappa_{a}) = \sum_{j,j'}\sum_{\bar{\mathfrak{c}}_j,\bar{\mathfrak{c}}_{j'}'} \left(\Rc^{\mathfrak{q}}_{j'\to a'}\right)_{\bar{\mathfrak{c}}_{j'}'\mathfrak{c}_{a'}}  \, \Mc_{\bar{\mathfrak{c}}_{j'}'\bar{\mathfrak{c}}_j}^{(j',j)\,\mathfrak{q}}(\kappa_{j'}',\kappa_{j}) \, \left(\Rc_{j\to a}^{\mathfrak{q}}\right)_{\bar{\mathfrak{c}}_j \mathfrak{c}_a} \, .
\end{align}
Here $\kappa_j$ (and similarly $\kappa_{j'}'$) represent the set of kinematic variables for the channel involved. That is, if using the helicity basis, then $\kappa_j = \{\sigma\}$, while for the spin-orbit basis, we have $\kappa_j = p_j$.

\section{Dalitz Distributions}
\label{sec:dist}

As highlighted in the introduction, partial wave $\3\to\3$ amplitudes are an essential component in analyzing many physically interesting phenomena such as heavy flavor decays, the excited hadron spectrum, or the structure of nuclear forces. While genuine $\3\to\3$ reactions are difficult to prepare and observe, one could define a reaction rate or cross section which depends, in the usual way, on the modulus-squared scattering amplitude~\cite{weinberg:1995mt}. Computing proper reaction rates or cross sections requires one to ensure that a physically meaningful probabilistic interpretation can be made. However, defining such rates for $\3\to\3$ processes is known to contain subtleties due to physical-energy singularities in both disconnected contributions in T matrix elements and the OPE amplitude, which is an essential feature of three-body processes~\cite{Potapov:1977ux,Potapov:1977sr}. In this work, we present a simple intensity distribution which is meant illustrate dynamically and kinematically enhanced structures of $\3\to\3$ amplitudes. This distribution function closely resembles physical reaction rates and cross-sections.~\footnote{One can view this simple intensity distribution as a piece of the complete $\3\to\3$ reaction rate, see Refs.~\cite{Potapov:1977ux,Potapov:1977sr}} To circumvent the difficulty of working with eight independent kinematic degrees of freedom, as well as charge states specified by isospin projections, we reduce the number of kinematic degrees of freedom by defining the intensity as an average over the angular and isospin degrees of freedom. This intensity function can then be used to construct Dalitz distributions over the remaining kinematic variables.

\begin{figure}[t]
	\centering
	\includegraphics[width=\textwidth]{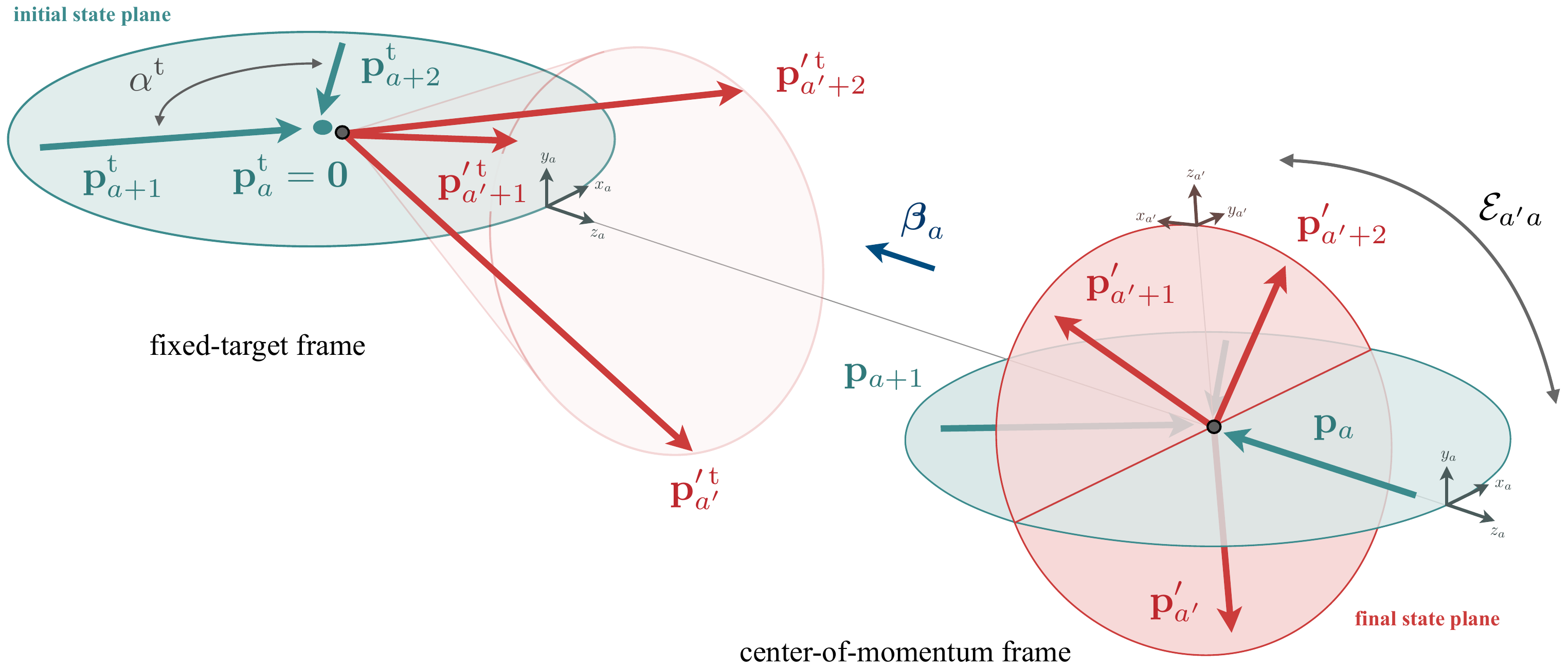}
	\caption{Illustration of the momentum configurations in the fixed-target frame of particle $a$ and the center-of-momentum frame of the three-body reaction.}
	\label{fig:reaction_frames}
\end{figure}

Assume the initial state kinematics are physical and specified through either their invariant masses or CM frame momenta. After the reaction, the final state CM frame kinematics lie in a plane that is oriented according to a set of Euler angles with respect to the initial state plane. We define an intensity $\Ic$ as the angular and isospin average of the modulus-squared of the fully connected $\3\to\3$ scattering amplitude. To compute the angular average, we first choose a set of space-fixed coordinate axes with which we define the Euler angles that orient the initial and final state reaction planes, as discussed in Sec.~\ref{sec:amps}. Our symmetrization procedure is performed by choosing some target standard configurations SC$_a$ and SC$_{a'}$ for initial and final target spectators $a$ and $a'$, respectively. Therefore, we define the Euler angles of the initial and final state planes with respect to these configurations. As discussed in Sec.~\ref{sec:amps}, we combine initial and final state orientations such that the reaction is described by the Euler angles of the final state plane relative to the initial state plane, $\Ec_{a'a}$. This reduces the angular dependence within the $\3 \to \3$ amplitude to the form of Eq.~\eqref{eq:dalitz_exp_one_wigner}. See Fig.~\ref{fig:reaction_frames} for an illustration. Therefore, the angular average is over the Euler angles $\Ec_{a'a}$. It is important to note that, by averaging over $\Ec_{a'a}$, the choice of $a$ and $a'$ is immaterial to the intensity and serves only to define relative coordinates. The intensity $\Ic$ is then given by
\begin{align}
    \label{eq:intensity_def}
    \Ic(\{\sigma'\},\{\sigma\}) & \equiv \Theta(\Phi(\{\sigma'\}))\prod_{j=1}^{3}\frac{1}{2i_j+1} \prod_{j'=1'}^{3'}\frac{1}{2i_{j'} + 1} \nn \\[5pt]
    & \qquad\qquad  \times \sum_{\{\tau_a\},\{\tau'_{a'}\}} \frac{1}{8\pi^2}  \int\!\diff\Ec_{a'a} \, \Big\lvert \Mc_{\{\tau_{a'}'\}\{\tau_a\}}(\{\p_{a'}'\},\{\p_a\})\Big\rvert^2 \, .
\end{align}
The intensity is a scalar function of the five Mandelstam invariants and is independent of the charge state of the particles since we average over all isospin states. Assuming the initial kinematics are physical, the Heaviside step function of the Kibble cubic, Eq.~\eqref{eq:kibble}, ensures that the final state is bounded by the physical region fixed by $\sqrt{s}$.

%

We expand the amplitude in Eq.~\eqref{eq:intensity_def} into Dalitz partial waves using Eq.~\eqref{eq:dalitz_exp_one_wigner}, which in turn allows us to write the angular average as an infinite sum over total angular momenta,
\begin{align}
    \label{eq:intensity_inter1}
    \int\!\frac{\diff\Ec_{a'a}}{8\pi^2} \, \Big\lvert \Mc_{\{\tau'\}\{\tau\}}(\{\p'_{a'}\}, \{\p_a\}) \Big\rvert^2 & = \sum_{J,J'} (2J+1) \sum_{\lambda'_{a'} ,\lambda_a } \sum_{\mu'_{a'}, \mu_a} \Mc^J_{\lambda'_{a'}\{\tau'\},\lambda_a\{\tau\}} \Mc^{J'\,*}_{\mu'_{a'}\{\tau'\},\mu_a\{\tau\}} \nn \\[5pt]
    & \qquad\qquad \times \frac{2J'+1}{8\pi^2}\int\!\diff\Ec_{a'a} \, D^{(J)\,*}_{\lambda_a \lambda'_{a'}}(\Ec_{a'a}) D^{(J')}_{\mu_a \mu_{a'}'}(\Ec_{a'a}) \, , \nn \\[5pt]
    & = \sum_{J} (2J+1) \sum_{\lambda_{a'}',\lambda_a} \Big\lvert \Mc^J_{\lambda'_{a'}\{\tau'\},\lambda_a\{\tau\}} (\{\sigma_{a'}'\},\{\sigma_a\})\Big\rvert^2 \,.
\end{align}
Equation~\eqref{eq:wigner_d_ortho} was used to evaluate the angular integrals in going to the second line. For a particular $J$, the helicity sums run over $-J \le \lambda_a,\lambda_{a'}' \le J$. Furthermore, since parity is conserved in strong interactions, we can break the sum over $J$ into definite $J^P$ sectors. We then expand the amplitude into terms of total isospin amplitudes using Eq.~\eqref{eq:amp_isospin_expand}, and subsequently use the orthogonality of Clebsch-Gordan coefficients as in Eq.~\eqref{eq:R_jj_ortho} to write the isospin average as
\begin{align}
    \label{eq:intensity_inter2}
    \sum_{\{\tau_a\},\{\tau'_{a'}\}}\Big\lvert \Mc^{J^P}_{\lambda'_{a'}\{\tau'\},\lambda_a\{\tau\}} (\{\sigma_{a'}'\},\{\sigma_a\})\Big\rvert^2 = \sum_{I}(2I+1) \sum_{I_a,I_{a'}'}  \Big\lvert \Mc_{I_{a'}'\lambda_{a'}',I_a\lambda_a}^{I(J^P)}(\{\sigma'\},\{\sigma\}) \Big\rvert^2 \, .
\end{align}
The pair isospin sums run over $\lvert I - i_a\rvert\le I_a\le I+i_a$ and $\lvert I - i_{a'}'\rvert \le I_{a'}' \le I+i_{a'}'$. 

Combining Eqs.~\eqref{eq:intensity_inter1} and~\eqref{eq:intensity_inter2}, we write Eq.~\eqref{eq:intensity_def} as a sum over $I(J^P)$ sectors,
\begin{align}
    \Ic(\{\sigma'\},\{\sigma\}) = \sum_{I(J^P)} (2J+1)(2I+1) \prod_{j=1}^{3}\frac{1}{2i_j + 1} \prod_{j'=1'}^{3'}\frac{1}{2i_{j'} + 1} \, \Ic^{I(J^P)}(\{\sigma'\},\{\sigma\}) \, ,
\end{align}
where the definite $I(J^P)$ intensity is defined as
\begin{align}
    \label{eq:intensity}
	\mathcal{I}^{I(J^P)}(\{\sigma'\},\{\sigma\}) \equiv \Theta(\Phi(\{\sigma'\})) \sum_{I_a,I_{a'}'}\sum_{\lambda_a,\lambda_{a'}'}  \Big\lvert \Mc_{I_{a'}'\lambda_{a'}',I_a\lambda_a}^{I(J^P)}(\{\sigma'\},\{\sigma\}) \Big\rvert^2 \, .
\end{align}
In Sec.~\ref{sec:application}, we use Eq.~\eqref{eq:intensity} to study the structures of the symmetric partial wave amplitudes in Eq.~\eqref{eq:amp_sym_final} for a simple model amplitude of $3\pi$ elastic scattering. This serves as a test for future amplitude analyses of lattice QCD $\3\to\3$ amplitudes. The initial state kinematics $\{\sigma\}$ will be fixed using a specific configuration as detailed in the following subsection.

\subsection{Initial kinematic configuration}
\label{sec:kin_cfg}

Consider a configuration where the initial state is prepared such that particle $a$ is at rest, while particles $a+1$ and $a+2$ are directed toward the target with momenta $\p_{a+1}^{\mathsf{t}}$ and $\p_{a+2}^{\mathsf{t}}$, respectively. The $\mathsf{t}$ superscript indicates that kinematic quantities are taken in this fixed-target frame. The angle between the beams is fixed, denoted by $\alpha^{\mathsf{t}}$, and defined via $\cos\alpha^{\mathsf{t}} = \bh{\p}_{a+1}^{\mathsf{t}} \cdot \bh{\p}_{a+2}^{\mathsf{t}}$. The initial state plane is defined by a $z_a$-axis given by $\bh{z}_a \propto \p_{a+1}^{\mathsf{t}} + \p_{a+2}^{\mathsf{t}}$, and a $y_a$-axis given by $\bh{y}_a \propto \p_{a+2}^{\mathsf{t}} \times \p_{a+1}^{\mathsf{t}}$. A boost $\bs{\beta}_a \equiv -\P^{\mathsf{t}}_{\textrm{in}} / E^{\mathsf{t}}_{\textrm{in}}$ takes the system from the fixed-target frame to the overall center-of-momentum frame. Here, $\P_{\mathrm{in}}^{\mathsf{t}} \equiv \p_{a+1}^{\mathsf{t}} + \p_{a+2}^{\mathsf{t}}$ is the total momentum of the initial state while $E_{\mathrm{in}}^{\mathsf{t}}$ is the total energy. In the CM frame, the final state plane with spectator $a'$ is oriented with a set of Euler angles $\Ec_{a'a}$ with respect to the initial state plane.

In the fixed-target frame, the total energy of the system is
\begin{align}
    E_{\mathrm{in}}^{\mathsf{t}} = m_a + \sqrt{m_{a+1}^2 + (p_{a+1}^{\mathsf{t}})^2} + \sqrt{m_{a+2}^2 + (p_{a+2}^{\mathsf{t}})^2} \, ,
\end{align}
while the mangitude of the total momentum is
\begin{align}
    \left(P_{\mathrm{in}}^{\mathsf{t}}\right)^2 = (\p_{a+1}^{\mathsf{t}} + \p_{a+2}^{\mathsf{t}})^2 = (p_{a+1}^{\mathsf{t}})^2 + (p_{a+2}^{\mathsf{t}})^2 + 2(p_{a+1}^{\mathsf{t}})(p_{a+2}^{\mathsf{t}})\,\cos\alpha^{\mathsf{t}} \, .
\end{align}
The three-body invariant mass-squared is then 
\begin{align}
    \label{eq:inv_mass_ini_cfg}
    s &= \left(E_{\mathrm{in}}^{\mathsf{t}}\right)^2 - \left(P_{\mathrm{in}}^{\mathsf{t}}\right)^2 \, \nn \\[5pt]
    & = \left(m_a + \sqrt{m_{a+1}^2 + (p_{a+1} ^{\mathsf{t}})^2} + \sqrt{m_{a+2}^2 + (p_{a+2} ^{\mathsf{t}})^2} \right)^2 \nn \\[5pt]
    & \qquad - (p_{a+1}^{\mathsf{t}})^2 - (p_{a+2}^{\mathsf{t}})^2 - 2(p_{a+1}^{\mathsf{t}})(p_{a+2}^{\mathsf{t}})\,\cos\alpha^{\mathsf{t}} \, ,
\end{align}
while the pair invariant masses can be computed from
\begin{gather}
\begin{aligned}
    \label{eq:sigma_init}
    \sigma_a & = (E_{\mathrm{in}}^{\mathsf{t}} - m_a)^2 - (P_{\mathrm{in}}^{\mathsf{t}})^2 \, , \\[5pt]
    \sigma_{a+1} & = \left( E_{a+2}^{\mathsf{t}} + m_a \right)^2 - (p_{a+2}^{\mathsf{t}})^2 \, ,
\end{aligned}
\end{gather}
with $\sigma_{a+2}$ being fixed from the Mandelstam condition Eq.~\eqref{eq:mandelstam}, \eg $\sigma_{a+2} = \Delta(s) - \sigma_{a} - \sigma_{a+1}$.

In the next section, we fix $\sqrt{s}$, $p_{a+1}^\mathsf{t}$, and $\alpha^{\mathsf{t}}$ to some chosen values, with $p_{a+2}^{\mathsf{t}}$ being determined by solving the nonlinear equation~\eqref{eq:inv_mass_ini_cfg}. Once $p_{a+2}$ is determined, the initial state invariant masses are then computed using Eqs.~\eqref{eq:sigma_init} and~\eqref{eq:mandelstam}. Note that in this scheme, $\sigma_{a+2}$ is constant irrespective of $p_{a+2}^{\mathsf{t}}$ as both $p_a^{\mathsf{t}} = 0$ and $p_{a+1}^{\mathsf{t}}$ are fixed. Therefore, the initial state is completely fixed and physical, giving $\Phi(\{\sigma\}) > 0$ identically, with the final state variables being constrained to $\sqrt{s}$.

A special case of this configuration is worth commenting on. Consider $p_{a+2}^{\mathsf{t}} = p_{a+1}^{\mathsf{t}}$. When boosted to the CM frame, the initial momenta $p_{a+1}^{\mathsf{t}}$ and $p_{a+2}^{\mathsf{t}}$ form bisectors of an iscosceles triangle, and this was one configuration used in Ref.~\cite{Dawid:2025doq} for studying maximal isospin $3\pi^+$, $3K^+$, $2K^+\pi^+$, and $2\pi^+$ amplitudes. If one now chooses $p_{a+1}^{\mathsf{t}}$ such that when boosted to the overall CM frame all momenta are equal, \ie, $p_a = p_{a+1} = p_{a+2}$, then the momenta geometrically form the sides of an equilateral triangle, where the interior angles are all $2\pi/3$. We call these configurations the isosceles configuration and the equilateral configuration, respectively. As noted in Ref.~\cite{Dawid:2025doq}, when working with identical particles, the equilateral configuration forbids certain quantum numbers from having non-zero amplitudes. For non-maximal isospin contributions, we find that the isosceles configuration can also forbid certain quantum numbers from having non-zero amplitudes for identical particle systems, \eg, $I(J^P) = 0(0^-)$. Therefore, in Sec.~\ref{sec:application} we choose more generic values for our initial state configuration to ensure a wide variety of quantum numbers are accessible.

\section{Applications to three pions}
\label{sec:application}

We apply the symmetrization framework discussed in Sec.~\ref{sec:symm} to form elastic $3\pi$ partial wave amplitudes and use the intensity distribution defined in Sec.~\ref{sec:dist} to plot Dalitz distributions and illustrate key features of the corresponding amplitudes. We work in the strong isospin limit, so the pseudoscalar pion is a triplet under SU(2)$_{I}$. The particle masses, parities, and isospins are thus $m_j = m_\pi$, $\eta_j = -1$, and $i_j = 1$ for each $j$, respectively. The product of the intrinsic parities of the pions is then $\eta = \eta' = (-1)^3 = -1$. The physical three-particle CM energy is bounded as $3m_\pi \le \sqrt{s} < \sqrt{s_{\mathrm{inel.}}}$, where $s_{\mathrm{inel.}}$ is the first inelastic threshold.

The 27 isospin combinations can be reduced to seven irreducible representations. The allowed $3\pi$ total isospins are $I = 0,1,2,3$, with mulitplicities dictated by the allowed $2\pi$ subchannels. For $I = 3$, only $2\pi$ states with $I_j = 2$ contribute. For $I = 2$, the pion pair can be in both $I_j = 2$ and $I_j = 1$, while for $I=1$ the pair can be in all three isospins states, $I_j = 2$, $1$, and $0$. Finally, $I=0$ has only $2\pi$ isovector states ($I_j = 1$) populating the system. Generalized Bose symmetry restricts the allowed partial waves for each isospin channel, \cf Table I of Ref.~\cite{Jackura:2023qtp} for a list of low-lying partial wave contributions. The isospin recoupling coefficients are determined from Eq.~\eqref{eq:R_iso_all}. 
As a matrix in the pair isospin space, the cyclic recoupling coefficients are
\begin{gather}
\begin{aligned}
    \label{eq:three_pi_recoup_cyclic}
    \bs{\Rc}^3_{j \to j+1} &= 1 \\
    \bs{\Rc}^{2}_{j \to j+1} & = \begin{pmatrix}
         -\frac{1}{2} & \frac{\sqrt{3}}{2} \\
         -\frac{\sqrt{3}}{2} & -\frac{1}{2} 
    \end{pmatrix} \, , \\[5pt]
    \bs{\Rc}^{1}_{j \to j+1} & = \begin{pmatrix}
         \frac{1}{6} & -\sqrt{\frac{5}{12}} & \frac{\sqrt{5}}{3} \\
         \sqrt{\frac{5}{12}} & -\frac{1}{2} & -\frac{1}{\sqrt{3}} \\
         \frac{\sqrt{5}}{3} & \frac{1}{\sqrt{3}} & \frac{1}{3} \\
    \end{pmatrix} \, , \\[5pt]
    \bs{\Rc}^{0}_{j \to j+1} & = 1 \, ,
\end{aligned}
\end{gather}
while for the anticyclic recoupling coefficients we find
\begin{gather}
\begin{aligned}
    \label{eq:three_pi_recoup_anticyclic}
    \bs{\Rc}^3_{j \to j+2} &= 1 \\
    \bs{\Rc}^{2}_{j \to j+2} & = \begin{pmatrix}
         -\frac{1}{2} & -\frac{\sqrt{3}}{2} \\
         \frac{\sqrt{3}}{2} & -\frac{1}{2} 
    \end{pmatrix} \, , \\[5pt]
    \bs{\Rc}^{1}_{j \to j+2} & = \begin{pmatrix}
         \frac{1}{6} & \sqrt{\frac{5}{12}} & \frac{\sqrt{5}}{3} \\
         -\sqrt{\frac{5}{12}} & -\frac{1}{2} & \frac{1}{\sqrt{3}} \\
         \frac{\sqrt{5}}{3} & -\frac{1}{\sqrt{3}} & \frac{1}{3} \\
    \end{pmatrix} \, , \\[5pt]
    \bs{\Rc}^{0}_{j \to j+2} & = 1 \, .
\end{aligned}
\end{gather}
In these expressions, the first entry of each matrix represents the maximal pair isospin component, and the pair isospins descend by one as you move across the matrix. The expressions are identical to those presented in Ref.~\cite{Hansen:2020zhy}.

Given some dynamical input, one could solve the integral equations for the asymmetric partial wave amplitudes~\cite{Briceno:2024ehy} and use the symmetrization procedure above to form complete partial wave amplitudes. Here we choose a simple amplitude model to illustrate the symmetrization procedure without worrying about the dynamical complexity required to solve the integral equations. We choose to take the partial wave OPE amplitude~\cite{Jackura:2023qtp} as our elastic $3\pi$ scattering amplitude,
\begin{align}
    \label{eq:ope}
    \Mc_{L_{j'}'S_{j'}'I_{j'}';L_jS_jI_j}^{(j',j)\, I(J^P)}(p_{j'}',p_j) = - \Mc_{2,S_{j'}'}^{I_{j'}'}(\sigma_{j'}') \, \Gc_{L_{j'}'S_{j'}'I_{j'}';L_jS_jI_j}^{I(J^P)}(p_{j'}',p_j) \, \Mc_{2,S_j}^{I_j}(\sigma_j) \, ,
\end{align}
where $\Mc_{2,S}^I$ is the two-pion partial wave amplitude and $\Gc^{I(J^P)}$ is the partial wave projected OPE propagator. The two-pion partial wave amplitudes used in this work are discussed in Sec.~\ref{sec:two_body}. 

The partial wave projected OPE propagator is defined in Ref.~\cite{Jackura:2023qtp}, where we refer the reader to Appendix D of that work for details and explicit expressions for the quantum numbers we study here. It is well-known that the OPE, and $\3\to\3$ amplitudes in general, contain kinematic singularities associated with the exchange of on-shell particles~\cite{Jackura:2018xnx,Jackura:2022gib}. For partial wave amplitudes, this singularity manifests as a logarithmic singularity within the physical kinematic domain. For kinematics near this singular region, the amplitude behaves as
\begin{align}
    \label{eq:ope_structure}
    \Gc^{I(J^P)} \sim f^{I(J^P)}(\zeta_{j'j})\,\log\left(\frac{\zeta_{j'j} - 1}{\zeta_{j'j} + 1}\right) \, , \textrm{  as } \zeta_{j'j} \to \pm 1 \, ,
\end{align}
where $f^{I(J^P)}$ is some function which depends on quantum numbers and $\zeta_{j'j}$ is a function of energies and momenta, with $\zeta_{j'j} \to \pm 1$ coinciding with the cosine of the pair-spectator scattering angle when the exchange is physical. For S wave pairs, it can be shown that $f^{I(J^P)}$ approaches a constant as $\zeta_{j'j}\to \pm 1$ for fixed $\sqrt{s}$ and $\sigma$~\cite{Jackura:2023qtp}. Pairs in P wave which recoil against the spectator in P wave produce a functional dependence which vanishes polynomially as $\zeta_{j'j} \to \pm 1$, which softens the singularity. For the reader's convenience, examples of Eq.~\eqref{eq:ope} are shown in App.~\ref{sec:asym_lsi} for the quantum numbers in this study.

\subsection{Two-pion amplitudes}
\label{sec:two_body}

Assuming that the $3\pi$ amplitude is given entirely by the OPE amplitude, Eq.~\eqref{eq:ope}, the only dynamical information needed for this study is the elastic $\pi \pi$ scattering amplitudes. Throughout this section, we drop the spectator notation; thus, the two-pion amplitude is written as $\Mc_{2,S}^{I}(\sigma)$, with the understanding that when this is embedded in the larger $3\pi\to 3\pi$ amplitude, the spectator index will follow. From generalized Bose statistics, $I=0$ and $2$ systems can scatter only in even $S$ partial waves, while $I=1$ systems scatter in odd waves~\cite{Martin:102663}. We represent the elastic $\pi\pi$ partial wave amplitudes through real energy-dependent phase shifts $\delta_{S}^I$, 
\begin{align}
    \label{eq:two_pi_phase_rep}
    \Mc_{2,S}^{I}(\sigma) = \frac{8\pi \sqrt{\sigma}}{\xi} \, \frac{1}{q^\star \cot\delta_{S}^{I} - iq^\star} \, ,
\end{align}
where $\xi = 1/2!$ is the symmetry factor for identical particles and $q^\star$ is the momentum of the first particle in the pair CM frame,
\begin{align}
    q^\star = \frac{1}{2}\sqrt{\sigma - 4m_\pi^2} \, .
\end{align}
The normalization of Eq.~\eqref{eq:two_pi_phase_rep} is fixed from our normalization of single-particle states, Eq.~\eqref{eq:state_norm}, and our definition of amplitudes from S matrix elements, Eq.~\eqref{eq:amp_def}. 

In principle, symmetrizing spin-orbit amplitudes in Eq.~\eqref{eq:amp_sym_final} involves summing over an infinite number of $L$ and $S$ quantum numbers for a target $J^P$.~\footnote{For example, if $J = 0$, then states with $L=S$ will result in a non-zero overlap with the target state.} However, higher angular momentum states are suppressed for energies near threshold since $\Mc_{2,S} \sim (q^\star)^{2S}$ when $q^\star \to 0$~\cite{VonHippel:1972fg}. As our focus involves only elastic $\pi\pi$ scattering, we include only the lowest-lying partial waves, thus truncating the infinite sum. To that end, we include only S and P wave two-pion amplitudes, assuming that all higher waves are negligible.~\footnote{This is supported phenomenologically and by lattice QCD results~\cite{Estabrooks:1975cy,Protopopescu:1973sh,Pelaez:2004vs,Dudek:2012gj,Dudek:2012xn}.} Thus we set $\Mc_{2,S}^{I} = 0$ for $S \ge 2$ throughout this work.

%
\begin{figure}[t]
	\centering
	\includegraphics[width=0.5\textwidth]{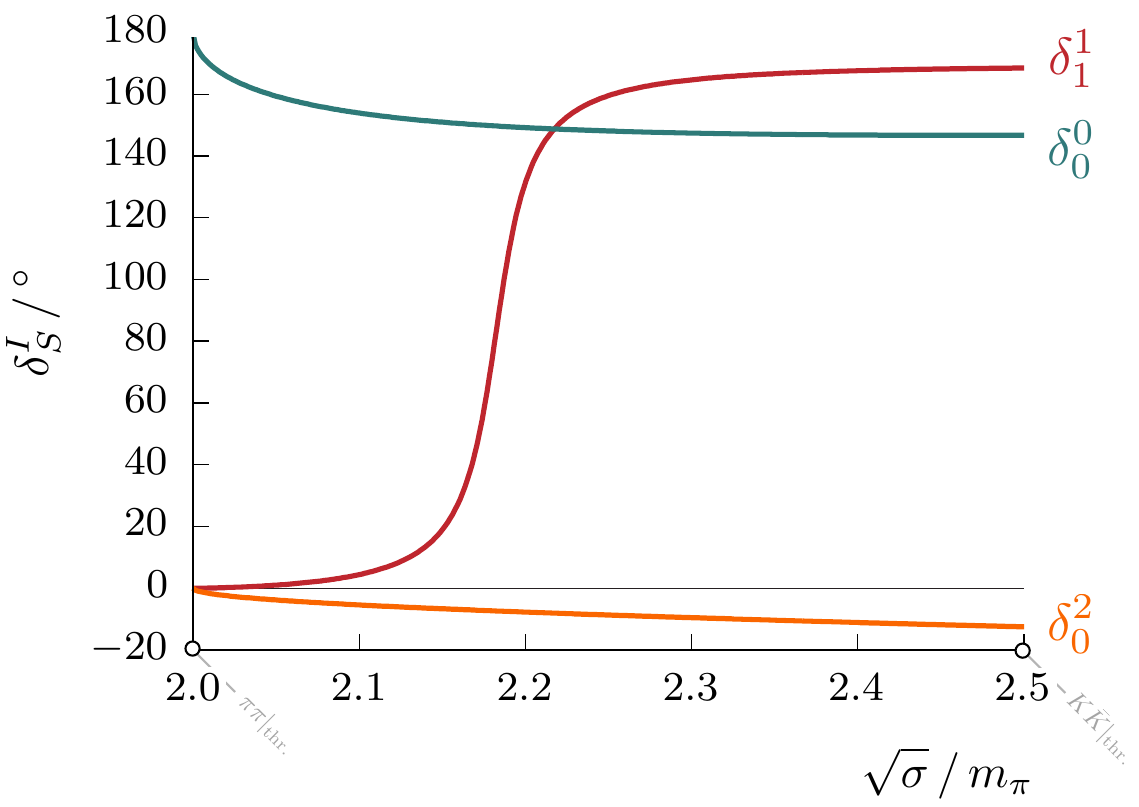}
	\caption{Input phase shifts for the elastic $\pi\pi$ scattering amplitudes as a function of the $\pi\pi$ invariant mass. Parameters are taken from lattice QCD analyses at $m_\pi \approx 400$ MeV, see Refs.~\cite{Dudek:2012gj,Dudek:2012xn,Briceno:2016mjc}.}
	\label{fig:two_body_phases}
\end{figure}
In our applications, we take the parameters for elastic $\pi\pi$ scattering to be approximately consistent with ones for quark masses, where $m_\pi \approx 400\,\mathrm{MeV}$ as computed by the Hadron Spectrum Collaboration. At this pion mass, the first inelastic threshold is $K\bar{K}$ for all two-pion quantum numbers, thus $2m_\pi \le \sqrt{\sigma} < 2m_K$ with $m_K\approx 500\,\mathrm{MeV}$.

For S-wave scattering, we parameterize the phase shift by a second-order effective range expansion,
\begin{align}
	q^\star\cot\delta_0^I = -\frac{1}{a_0^I} + \frac{1}{2} r_0^I \, q^{\star\,2} \, ,
\end{align}
where for the isotensor channel we use for the scattering length and effective range $m_\pi\,a_0^2 = 0.296$ and $r_0^2 = 0$, respectively~\cite{Dudek:2012gj}. For the isoscalar channel, we use $m_\pi \, a_0^0 = 1.84$ and $r_0^0  = -2.13\, m_\pi$, which results in a bound state $\sigma$ meson with mass $m_\sigma \approx 760\,\mathrm{MeV}\approx 0.95\,m_\pi$~\cite{Briceno:2016mjc}. The isovector P wave amplitude is parameterized by a Breit-Wigner phase shift,
\begin{align}
    q^\star \cot\delta_1^1 
    & = \frac{m_{\mathrm{bw}}^2-\sigma}{\sqrt{\sigma}\,\Gamma_1^{\mathrm{bw}}(\sigma)},  \qquad  \Gamma_1^{\mathrm{bw}}(\sigma)= \frac{g^{2}_{\mathrm{bw}}}{6\pi \sigma} q^{\star 2} \, ,
\end{align}
where the Breit-Wigner mass and coupling are fixed to be $m_{\mathrm{bw}} = 2.18\,m_\pi$ and $g_{\mathrm{bw}} = 5.80$~\cite{Dudek:2012xn}. At $m_\pi\approx 400\,\mathrm{MeV}$, the P wave amplitude has a narrow $\rho$ meson resonance with a pole mass and width $m_\rho\approx 860 \,\mathrm{MeV} \approx 2.15\,m_\pi$ and $\gamma_{\rho} \approx 10\,\mathrm{MeV} \approx 2.5\times 10^{-2}\,m_{\pi}$, respectively~\cite{Dudek:2012xn}. All $\pi\pi$ phase shifts used in this work are shown in Fig.~\ref{fig:two_body_phases}.

\subsection{Numerical results}
\label{sec:num_res}

%
\begin{figure}[t]
	\centering
	\includegraphics[width=0.7\textwidth]{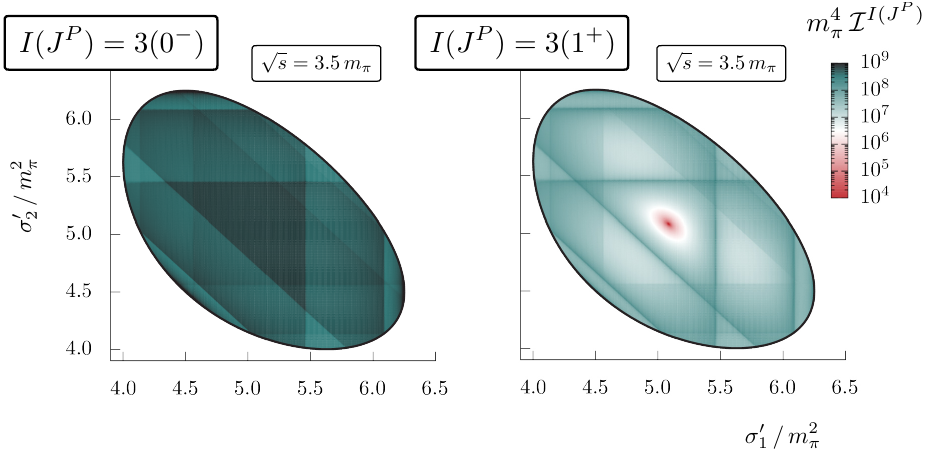}
	\caption{Dalitz distributions for the OPE amplitude in the $I=3$ sector at $\sqrt{s} \, / m_\pi = 3.5$ for (a) $J^P = 0^-$ and (b) $J^P = 1^+$. Distributions are shown with initial state kinematics as described in the text. 
    }
	\label{fig:dalitz_3_0m_and_3_1p}
\end{figure}
Here we showcase various $J^P$ to illustrate features of the Dalitz distribution for elastic $3\pi$ scattering, verify results known from the literature, and highlight various non-trivial cases for the interested reader to reproduce. Since $S=0$ and $1$ are the only allowed two-pion amplitudes, the number of contributing partial waves is limited for a fixed $J$. In this work, we consider $J = 0$ and $1$; therefore, $3\pi$ orbital angular momenta are limited to $L \leq 2$. The reader can consult Table 1 of Ref.~\cite{Jackura:2023qtp} for the allowed partial waves within these limits. The partial waves which enter the symmetrized amplitudes are labeled by $([\pi\pi]_{S_j}^{I_j}\pi)_{L_j}$, where both $S_j$ and $L_j$ will be labeled by their spectroscopic label, \ie $\mathrm{S}$ for $S_j,L_j = 0$, $\mathrm{P}$ for $S_j,L_j = 1$, and $\mathrm{D}$ for $L_j = 2$. In each of the cases shown, we plot the intensity distribution, Eq.~\eqref{eq:intensity}, for the OPE amplitude Eq.~\eqref{eq:ope}. For all cases, we choose target spectators $a = a' = 1$.  We fix the initial state as described in the next paragraph and create Dalitz distributions in the final state variables $\sigma_1'$ and $\sigma_2'$.

%
\begin{figure}[t]
	\centering
	\includegraphics[width=\textwidth]{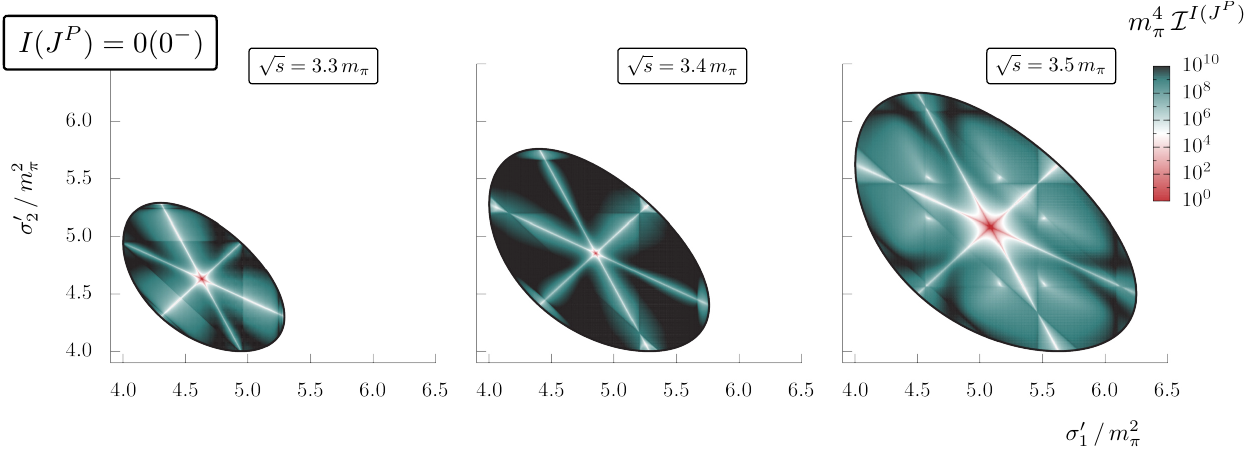}
	\caption{Dalitz distributions for the $I(J^P) = 0(0^-)$ OPE amplitude at $\sqrt{s}\,/m_\pi = 3.3$, $3.4$, and $3.5$. The initial state is fixed as in Fig.~\ref{fig:dalitz_3_0m_and_3_1p}. 
    }
	\label{fig:dalitz_0_0m}
\end{figure}

Given $m_\pi \approx 400\,\mathrm{MeV}$, the first inelastic threshold (taken as $\pi K\bar{K}$) is $\sqrt{s_{\mathrm{inel.}}} = m_\pi + 2m_K \approx 1,400 \,\mathrm{MeV} \approx 3.5 \,m_\pi$. Therefore, we take kinematics bounded within $3\,m_\pi \le \sqrt{s} \lesssim 3.5\,m_\pi$.~\footnote{Some channels in total $I=0$ and $1$ can couple to $K\bar{K}$, \eg, the $\phi(1020)$ meson in $I(J^P) = 0(1^-)$ can decay to both $K\bar{K}$ and $3\pi$ modes~\cite{ParticleDataGroup:2024cfk}. We ignore such $\2\longleftrightarrow\3$ couplings here for simplicity.} The initial state is fixed according to the discussion in Sec.~\ref{sec:kin_cfg}. In the fixed-target frame, the momentum of particle $a+1$ is $p_{a+1}^{\mathsf{t}} = 0.7\,m_\pi$, and we choose $\alpha^{\mathsf{t}} = \pi / 4$ for the separation angle between particles $a+1$ and $a+2$. We fix the kinematics for CM frame energies $\sqrt{s} \, / \, m_\pi = 3.3$, $3.4$, and $3.5$. For these three CM energies, the momentum of $a+2$ in the fixed-target frame is computed to be $p_{a+2}^{\mathsf{t}} \, / \, m_\pi = 1.07$, $1.35$, and $1.60$. The pair invariant masses are then $\{\sigma\} = \{4.52,4.93,4.44\}\,m_\pi^2$ for $\sqrt{s} = 3.3\,m_\pi$, $\{\sigma\} = \{4.76,5.36,4.44\}\,m_\pi^2$ for $\sqrt{s} = 3.4\,m_\pi$, and $\{\sigma\} = \{5.03,5.78,4.44\}\,m_\pi^2$ for $\sqrt{s} = 3.5\,m_\pi$. The boundary of the physical region for the final state given by Eq.~\eqref{eq:kibble} simplifies to $\Phi(\{\sigma'\}) = \sigma_1'\sigma_2'(s + 3m_\pi^2 - \sigma_1' - \sigma_2') - m_\pi^2 (s - m_\pi^2)^2 \ge 0$.

First, we show the Dalitz distributions for the $I=3$ sector in both $J^P = 0^-$ and $J^P = 1^+$ at $\sqrt{s} = 3.5\,m_\pi$. The asymmetric spin-orbit amplitudes contain physical singularities associated with the exchange of a physical pion, see Fig.~\ref{fig:lsi_amp_3_0m_and_3_1p} in App.~\ref{sec:asym_lsi}. Upon symmetrization with Eq.~\eqref{eq:amp_sym_final}, the singularities form triangular structures throughout the Dalitz region. The location of the OPE singularities in $\sigma_{1,2,3}'$ depends on both $s$ and $\{\sigma\}$; however, they are independent of the two-body sub-channel dynamics. See App.~\ref{sec:asym_lsi} for further discussion on their locations. 

As shown in panel (a) of Fig.~\ref{fig:dalitz_3_0m_and_3_1p}, besides the OPE singularities, the $J^P = 0^-$ distribution is relatively flat in the center of the plot. This is expected since the only contributing partial wave is $([\pi\pi]_{\mathrm{S}}^2\pi)_{\mathrm{S}}$, and the isospin and angular momentum recoupling coefficients are simply the identity, \cf Eqs.~\eqref{eq:R_LS_to_hel},~\eqref{eq:three_pi_recoup_cyclic}, and~\eqref{eq:three_pi_recoup_anticyclic}. Near the Dalitz boundary, the amplitude increases in magnitude since the $\mathrm{S}$ wave OPE diverges at threshold. For $J^P = 1^+$, although still only $[\pi\pi]_{\mathrm{S}}^2$ pairs, they couple in a relative $\mathrm{P}$ wave with the spectator pion. Thus, non-trivial structures occur in the Dalitz distribution as seen in Fig.~\ref{fig:dalitz_3_0m_and_3_1p} (b). Notably, there exist regions where the distribution vanishes (shown in red). It should be noted that the locations of vanishing intensity correspond directly with Zemach's work on three pion Dalitz distributions in $3\pi$ decays, see Fig.~2 of Ref.~\cite{Zemach:1963bc}.~\footnote{A small caveat in comparing our results to Ref.~\cite{Zemach:1963bc} is the fact that they present separated charge modes, not isospin averaged decay amplitudes.}

%
\begin{figure}[t]
	\centering
	\includegraphics[width=\textwidth]{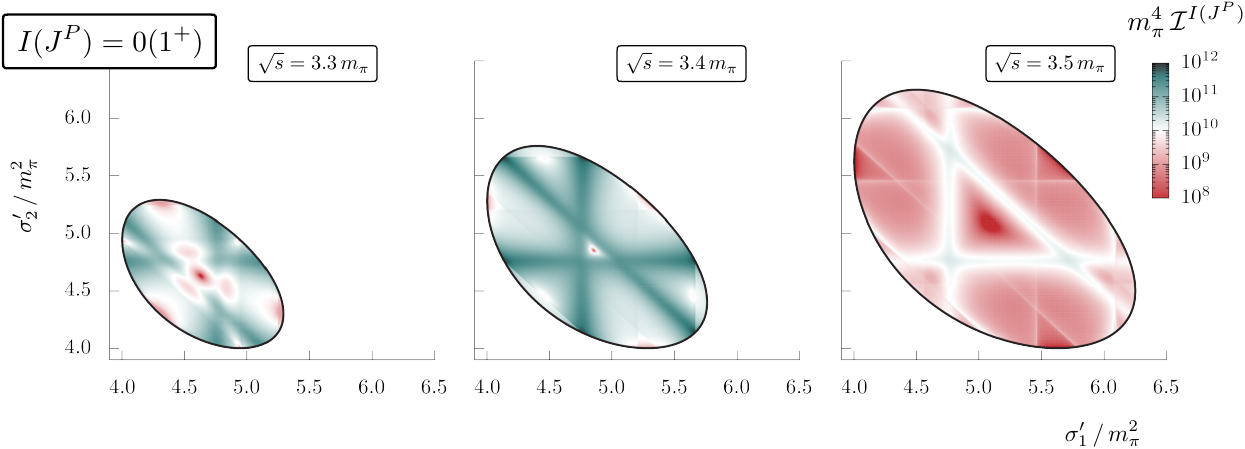}
	\caption{Same as Fig.~\ref{fig:dalitz_0_0m} except for $I(J^P) = 0(1^+)$.
    }
	\label{fig:dalitz_0_1p}
\end{figure}
Next, we examine $I(J^P) = 0(0^-)$. Restricting the allowed partial waves to be $S \le 1$ means that there is only a single contributing amplitude, $([\pi\pi]_{\mathrm{P}}^1\pi)_{\mathrm{P}} \to ([\pi\pi]_{\mathrm{P}}^1\pi)_{\mathrm{P}}$. We compute the Dalitz distributions for each $\sqrt{s} = 3.3\,m_\pi $, $3.4\,m_\pi$, and $3.5\,m_\pi$. 
Notable features in the distribution are bands corresponding to the $\rho$ resonance at $\sigma_{1,2,3}'\approx 4.75\,m_\pi^2$, which is the highest concentration of events save for the OPE singularities. This is most easily seen for $\sqrt{s} = 3.5\,m_\pi$ due to the large phase space. Other notable features in Fig.~\ref{fig:dalitz_0_0m} are lines of vanishing intensity. These are requirements from symmetry of three identical pions assuming exact $\mathrm{SU}(2)$ isospin symmetry, \cf~\cite{Zemach:1963bc}. Another $I=0$ case we examine is $J^P = 1^+$, shown in Fig.~\ref{fig:dalitz_0_1p}. Here, there are two waves which contribute, $([\pi\pi]_\mathrm{P}^1\pi)_\mathrm{S}$ and $([\pi\pi]_\mathrm{P}^1\pi)_\mathrm{D}$. There are 4 regions where the intensity is expected to vanish, at the center and near each of the three thresholds, as expected from the analysis of Ref.~\cite{Zemach:1963bc}. Prominently featured are the three bands of the $\rho$ meson.

\begin{figure}[t]
	\centering
	\includegraphics[width=\textwidth]{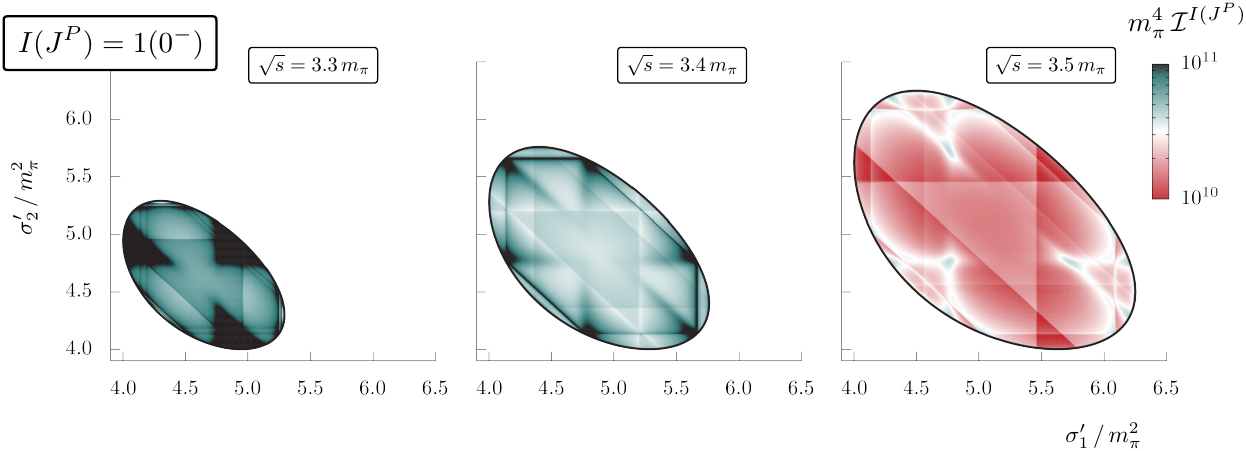}
	\caption{Same as Fig.~\ref{fig:dalitz_0_0m} except for $I(J^P) = 1(0^-)$. 
    }
	\label{fig:dalitz_1_0m}
\end{figure}

For $I=1$, we consider $J^P = 0^-$.  Here there are three contributing waves, $([\pi\pi]_\mathrm{S}^0\pi)_\mathrm{S}$, $([\pi\pi]_\mathrm{S}^2\pi)_\mathrm{S}$,  and $([\pi\pi]_\mathrm{P}^1\pi)_\mathrm{P}$. For some fixed initial and final kinematics, this means there are $(3\times 3)^2 = 81$ terms in the symmetrization sum. The Dalitz distributions shown in Fig.~\ref{fig:dalitz_1_0m} have heavy interference effects at the $\rho$ bands due the isospin symmetry suppressing the events at the center of the plot.

%

\begin{figure}[t]
	\centering
	\includegraphics[width=\textwidth]{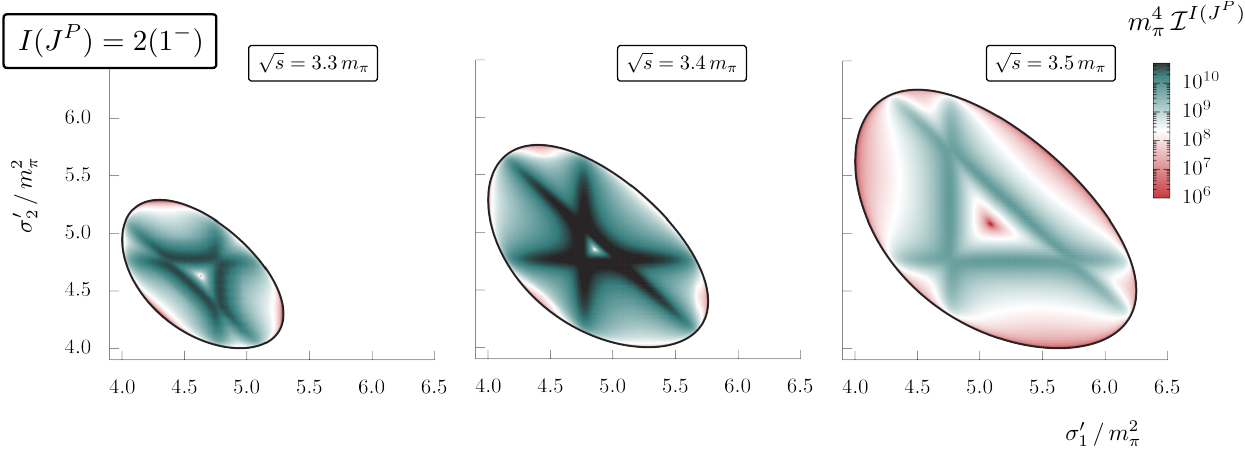}
	\caption{Same as Fig.~\ref{fig:dalitz_0_0m} except for $I(J^P) = 2(1^-)$. 
    }
	\label{fig:dalitz_2_1m}
\end{figure}

For $I=2$, we consider $J^P = 1^-$, where again only a single partial wave enters, $([\pi\pi]_P^1\pi)_P$. Here, we find that the Dalitz distribution vanishes at the physical boundary, as indicated in red in Fig.~\ref{fig:dalitz_2_1m}. Additionally, the distribution also vanishes at the center, which is required by isospin symmetry~\cite{Zemach:1963bc}. Bands corresponding to the $\rho$ meson are clearly visible. Notably, no OPE singularities are featured. This is due to the fact that for pure $\mathrm{P}$ wave systems, that is $S_j = L_j = 1$, in $J^P = 1^-$, the factor multiplying the logarithm in the OPE amplitude, as discussed with Eq.~\eqref{eq:ope_structure} and shown in Ref.~\cite{Jackura:2023qtp}, vanishes as the function approaches the singularities. Thus, the OPE singularities are integrable for this case.

%

\section{Summary}
\label{sec:summary}

We have presented an analysis framework to construct $\3\to\3$ partial wave amplitudes which include all possible pair-spectator amplitudes. The result is valid for arbitrary $J^P$, as well as definite isospin $I$ if the particles have flavor symmetry, so long as the reacting particles are spinless. We show that these amplitudes are useful for constructing three-body Dalitz plots where one can examine structures of the amplitude in invariant mass or energy variables. As our motivation stems from lattice QCD, this work gives a methodology to compute $\3\to\3$ hadronic observables from QCD, which have some resemblance to distributions studied in many experiments.

We give numerical examples for elastic $3\pi$ scattering in various quantum numbers, showing resonance structures like the $\rho$. These examples also exhibit symmetry features in the distributions, which are consistent with what is known from the literature.

As emphasized in the introduction, this work completes a workflow for accessing scattering observables for three-body systems starting from key inputs into their defining integral equations. However, there remains some outstanding challenges in constructing a general framework for studying \emph{any} three-body scattering amplitude in low to medium energies. Examples include extending the formalism to include reacting particles of non-zero intrinsic spin and allowing two- and three-body systems to couple.~\footnote{See Refs.~\cite{Briceno:2017tce} and \cite{Draper:2023xvu} for first steps in this direction.}

\section*{Acknowledgements}

The authors would like to thank Sebastian~Dawid, Robert~Edwards, Maxwell~Hansen, and Christopher~Thomas for useful discussions. AWJ acknowledges the support of the USDOE ExoHad Topical Collaboration, contract DE-SC0023598. RAB was supported in part by the U.S. Department of Energy, Office of Science, Office of Nuclear Physics under Awards No. DE-SC0025665 and No. DE-AC02-05CH11231.

\appendix

\section{Spin-orbit recoupling coefficients}
\label{sec:ls_recoupling}

In addition to symmetrized amplitudes in the Dalitz basis, one can also form symmetrized in the spin-orbit basis. Here, the amplitudes are a function of only three kinematic variables, $\sqrt{s}$, one initial state variable (either $p_j$ or $\sigma_j$), and one final state variable (either $p_{j'}'$ or $\sigma_{j'}'$). These amplitudes would be useful in looking at certain classes of integrated distributions, \eg similar to those studied by the COMPASS experiment~\cite{COMPASS:2015gxz,COMPASS:2018uzl}, as well as investigating the spectral properties of the amplitudes. To construct these class of symmetrized amplitudes, we must form \emph{spin-orbit angular momentum recoupling coefficients}.~\footnote{These coefficients were also presented in Ref.~\cite{Jackura:2018xnx}.} Starting from Eq.~\eqref{eq:dalitz_JP_to_LS}, repeated here for convenience,
\begin{align}
    \ket{\{\sigma\},Jm_J\lambda_a} = \sum_{L_a,S_a} \ket{p_a,Jm_J,L_aS_a}\, \sqrt{2S_a + 1} \, d_{\lambda_a 0 }^{(S_a)}(\vartheta_a^\star)\, \Pc_{\lambda_a}({}^{2S_a+1}({L_a})_J) \,, \nn
\end{align}
we find the inverse relationship
\begin{align}
    \label{eq:dalitz_to_ls_appA}
    \ket{p_a,Jm_J,L_aS_a} &=  \sqrt{2S_a + 1} \sum_{\lambda_a} \Pc_{\lambda_a}({}^{2S_a+1}({L_a})_J) \, \nn \\
    &\qquad\qquad\qquad\qquad\times \frac{1}{2}\int_{-1}^{1}\!\diff\cos\vartheta_a^\star \, \ket{\{\sigma\},Jm_J\lambda_a}  d_{\lambda_a 0 }^{(S_a)}(\vartheta_a^\star) \, . 
\end{align}

As in Sec.~\ref{sec:symm_ang_mom}, we connect the Dalitz state on the right-hand-side of Eq.~\eqref{eq:dalitz_to_ls_appA} to the corresponding state in SC$_j$ by applying Eq.~\eqref{eq:dalitz_symm_state},
\begin{align}
    \ket{p_a,Jm_J,L_aS_a} & =  \sqrt{2S_a + 1}\sum_{\lambda_a} \Pc_{\lambda_a}({}^{2S_a+1}({L_a})_J) \, \nn\\[5pt]
    & \qquad \times \sum_{\mu_j} \frac{1}{2}\int_{-1}^{1}\!\diff\cos\vartheta_a^\star \, \ket{\{\sigma\},Jm_J\mu_j} \, d_{\lambda_a \mu_j}^{(J)}(\omega_{j \to a})\,  d_{\lambda_a 0 }^{(S_a)}(\vartheta_a^\star) \, , \nn \\[5pt]
    & = \sum_{L_j',S_j'}  \sqrt{(2S_a + 1)(2S_j'+1)} \sum_{\lambda_a,\mu_j}  \Pc_{\lambda_a}({}^{2S_a+1}({L_a})_J) \Pc_{\mu_j}({}^{2S_j'+1}({L_j'})_J) \nn \\[5pt]
    & \qquad  \times \frac{1}{2} \int_{-1}^{1}\!\diff\cos\vartheta_a^\star  \, \ket{p_j,Jm_J,L_j'S_j'} \, d_{\mu_j 0}^{(S_j')}(\vartheta_j^\star)\, d_{\lambda_a \mu_j}^{(J)}(\omega_{j \to a})\,  d_{\lambda_a 0 }^{(S_a)}(\vartheta_a^\star) \, ,
\end{align}
where in going to the second line we used Eq.~\eqref{eq:dalitz_JP_to_LS} to convert the Dalitz state back to an $LS$ state. Note that $p_j$, $\vartheta_j^\star$, and $\omega_{j \to a}$ depend nontrivially on $\vartheta_a^\star$ when $a \neq j$, which can be most easily seen by considering the Lorentz transformation between the three-body and pair CM frames. For example, when $j = a+1$ (see Fig.~\ref{fig:pair_cm_frame}), the Lorentz transformation gives
\begin{align}
    \label{eq:mom_boost}
    p_{a+1}\cos (\pi - \alpha_{a,a+1}) = \gamma_j(p_{a+1}^\star \cos \vartheta_a^\star + \beta_a E_{a+1}^\star) \, ,
\end{align}
with $\gamma_a = (\sqrt{s} - E_a)/\sqrt{\sigma_a}$ and $\beta_a = p_a / (\sqrt{s} - E_a)$. To capture this non-trivial dependence on $\vartheta_a^\star$, we define the recoupling from one $LS$ basis to another to be
\begin{align}
    \label{eq:recoupling_app_a}
    \ket{p_a,Jm_J,L_aS_a} = \sum_{L_j',S_j'}\int_{-1}^{1}\diff\cos\vartheta_a^\star \, (\Rc_{j \to a}^{J^P})_{L_j'S_j',L_aS_a}(p_a,\vartheta_a^\star) \, \ket{p_j(p_a,\vartheta_a^\star),Jm_J,L_j'S_j'} \, ,
\end{align}
where the $LS$ basis recoupling coefficient is given by
\begin{align}
    \label{eq:so_to_so_recoup_app_a}
    (\Rc_{j\to a}^{J^P})_{L_j'S_j',L_aS_s}(p_a,\vartheta_a^\star) & \equiv \frac{1}{2} \delta_{PP_j}\delta_{PP_a} \sqrt{(2S_a + 1)(2S_j'+1)} \nn \\[5pt]
    & \qquad \times \sum_{\lambda_a,\mu_j}  \Pc_{\lambda_a}({}^{2S_a+1}({L_a})_J) \Pc_{\mu_j}({}^{2S_j'+1}({L_j'})_J) \nn \\[5pt]
    & \qquad \qquad \times  \, d_{\lambda_a 0 }^{(S_a)}(\vartheta_a^\star) \, d_{\lambda_a \mu_j }^{(J)}(\omega_{j\to a})\,  d_{\mu_j 0}^{(S_j')}(\vartheta_j^\star) \, .
\end{align}
The symmetrization of the $LS$ amplitude is then given by
\begin{align}
    & \Mc_{L_{a'}'S_{a'}';L_aS_a}^{J^P}(p'_{a'},p_{a}) = \sum_{j,j'} \sum_{L_{j'}',S_{j'}'}\sum_{L_j,S_j} \int_{-1}^{1}\!\diff\cos\vartheta_{a'}'^\star \, \int_{-1}^{1}\!\diff\cos\vartheta_{a}^\star \, \nn \\[5pt]
    & \qquad \times (\Rc_{j'\to a'}^{J^P})_{L_{j'}'S_{j'}';L_{a'}' S_{a'}'}(p_{a'}',\vartheta_{a'}'^\star) \, \Mc_{L_{j'}'S_{j'}';L_j S_j}^{(j',j)\,J} (p_{j'}';p_j) \, (\Rc_{j\to a}^{J^P})_{L_jS_j;L_a S_a}(p_{a},\vartheta_{a}^\star) \, ,
\end{align}
where $p_j = p_j(p_{a},\vartheta_{a}^\star)$ and $p_{j'}' = p_{j'}'(p_{a'}',\vartheta_{a'}'^\star)$ according to Eq.~\eqref{eq:mom_boost}.

When $j = a$, Eqs.~\eqref{eq:so_to_so_recoup_app_a} and~\eqref{eq:recoupling_app_a} simplify since $p_a$ is independent of $\vartheta_a^\star$, thus Clebsch-Gordan sum rules of and Wigner $d$ matrix orthogonality can be used to reduce the relationship to the identity. For $j\neq a$, this is not the case due to the complicated dependence between the $p_j$ and $\vartheta_{a}^\star$. This differs from the Dalitz states discussed in Sec.~\ref{sec:symm_ang_mom} as those state completely separate the invariant masses $\{\sigma\}$ (which do not depend on a spectator choice) from the angular kinematics which dictate the global orientation of the three-particle configuration in a single frame of reference. This convenience in the recoupling scheme illustrates the advantage of using the Dalitz basis in studying the $\3 \to \3$ reaction amplitude. In contrast, Eq.~\eqref{eq:so_to_so_recoup_app_a} is formed from the spin-orbit states which involve two references frames which result in the non-trivial angular-kinematic relationship Eq.~\eqref{eq:mom_boost}, see Sec.~\ref{sec:ls_recoupling} and Ref.~\cite{Jackura:2023qtp} for more details.

\section{Asymmetric spin-orbit amplitudes}
\label{sec:asym_lsi}

%
\begin{figure}[t]
	\centering
	\includegraphics[width=0.95\textwidth]{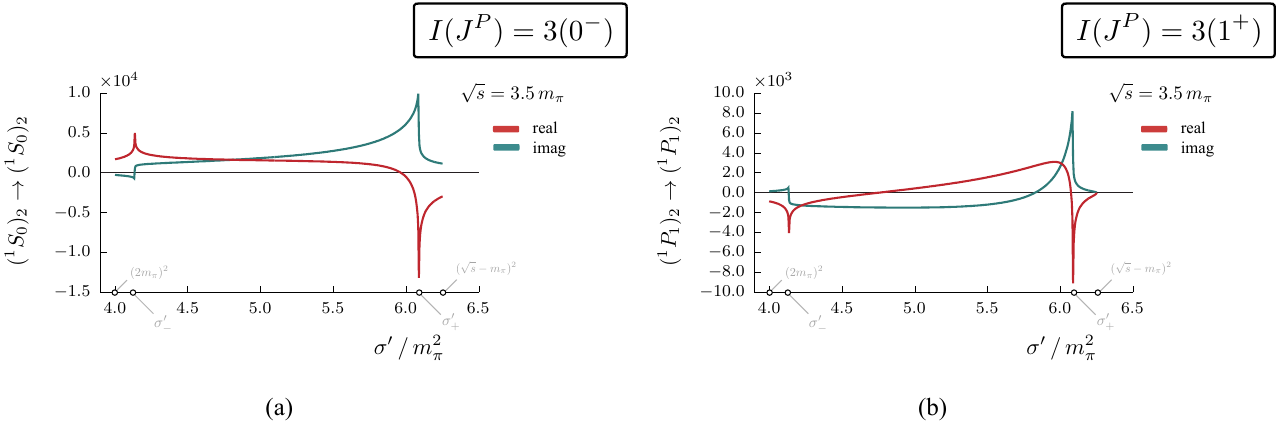}
	\caption{Real and imaginary parts for the asymmetric spin-orbit amplitudes as a function of $\sigma'$ for (a) $I(J^P) = 3(0^-)$ and (b) $I(J^P) = 3(1^+)$ both at $\sqrt{s} = 3.5\, m_\pi$. Important kinematic points are indicated as described in the text. In our restricted waveset, only isotensor $\pi\pi$ S wave pairs are allowed, which couple to the spectator pion in both (a) S wave and (b) P wave.}
	\label{fig:lsi_amp_3_0m_and_3_1p}
\end{figure}
The purpose of this appendix is to provide the reader the amplitudes used in intermediate calculations to produce the symmetrized amplitudes and, subsequently, the intensity distributions of Sec.~\ref{sec:application}. We focus on elastic $3\pi$ scattering and the kinematics discussed in Sec.~\ref{sec:application}, showing the amplitudes for the target $I(J^P)$ quantum numbers shown there. For some target $J^P$, we truncate the spin and orbital angular momentum sums in Eq.~\eqref{eq:amp_symm_JP} such that only $L_{j'}',L_j \le 2$ and $S_{j'}',S_j \le 1$ terms are included. This leaves only a few asymmetric amplitudes, $\Mc_{L_{j'}'S_{j'}'I_{j'}'; L_jS_jI_j}^{(j',j)\,I(J^P)}(\sigma_{j'}', \sigma_j)$, that contribute to the symmetrized amplitude in Eq.~\eqref{eq:amp_sym_final}. For elastic $3\pi$ scattering, the choice of $j$ and $j'$ in the asymmetric amplitudes is irrelevant as all three spectators and associated quantum numbers are identical.

%
\begin{figure}[t]
	\centering
	\includegraphics[width=0.85\textwidth]{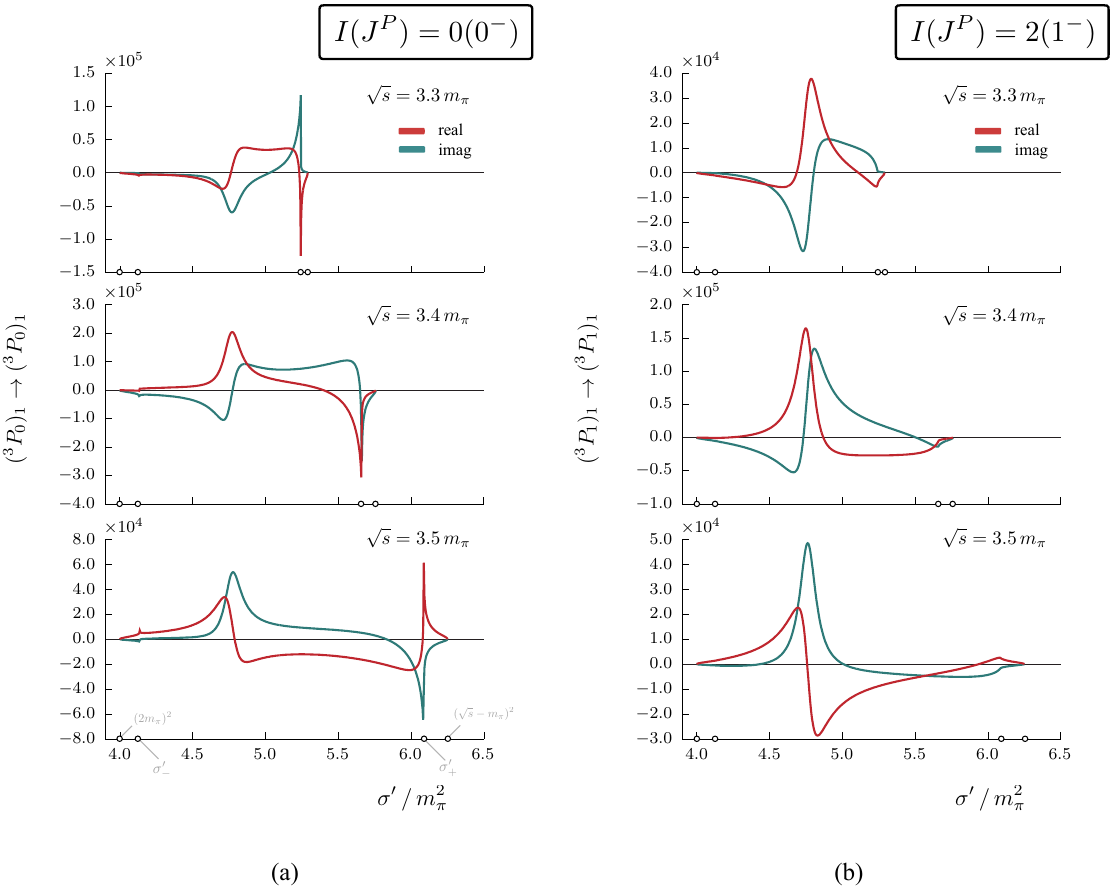}
	\caption{Real and imaginary parts for the asymmetric spin-orbit amplitudes as a function of $\sigma'$ for (a) $I(J^P) = 0(0^-)$ and (b) $T(J^P) = 2(1^-)$. With the restricted waveset, both amplitudes have only a single contribution from isovector $\pi\pi$ in P wave, which couples to the spectator pion in an orbital P wave. Each amplitude is shown for $\sqrt{s} \, / m_\pi = 3.3$, $3.4$, and $3.5$. Notable kinematic points are indicated for each $\sqrt{s}$ as described in the text.}
	\label{fig:lsi_amp_0_0m_and_2_1m}
\end{figure}
In this section, we denote the partial wave amplitudes contributing to the truncated sum by $({}^{2S_j+1}{L_j}_J)_{I_j} \to ({}^{2S_{j'}'+1}{L_{j'}'}_J)_{I_{j'}'}$ (see Sec.~\ref{sec:so_to_dalitz} for a reminder on the definition of these quantum numbers). Refer to Table~I of Ref.~\cite{Jackura:2023qtp} as well as Sec.~\ref{sec:application} for a list of partial waves that contribute to each set of target quantum numbers $I(J^P)$ of interest. For a given $\sqrt{s}$ and associated $\sigma_j$ (or $p_j$), we plot each asymmetric partial wave amplitude as a function of physical $\sigma_{j'}' \equiv \sigma'$, that is $2m_\pi \le \sqrt{\sigma'} < \sqrt{s} - m_\pi$.
The initial state kinematics are fixed as discussed in Sec.~\ref{sec:application}, where $\sqrt{s} \, /\, m_\pi = \{3.3, 3.4, 3.5\}$, with $\alpha^{\mathsf{t}} = \pi/4$, $p_{j+1}^{\mathsf{t}} \, / \, m_\pi = 0.7$, and $p_j^{\mathsf{t}} = 0$ for each of these values. This configuration gives $\sigma_j \, / \, m_\pi^2 = \{4.52, 4.76, 5.02\}$ for each of our chosen values of $\sqrt{s}$. The initial spectator in the three-pion CM frame is then computed by Eq.~\eqref{eq:momentum}, which gives $p_j \, / m_\pi \, = \{0.50, 0.56, 0.62\}$ for each $\sqrt{s}$.

%
\begin{figure}[t]
	\centering
	\includegraphics[width=\textwidth]{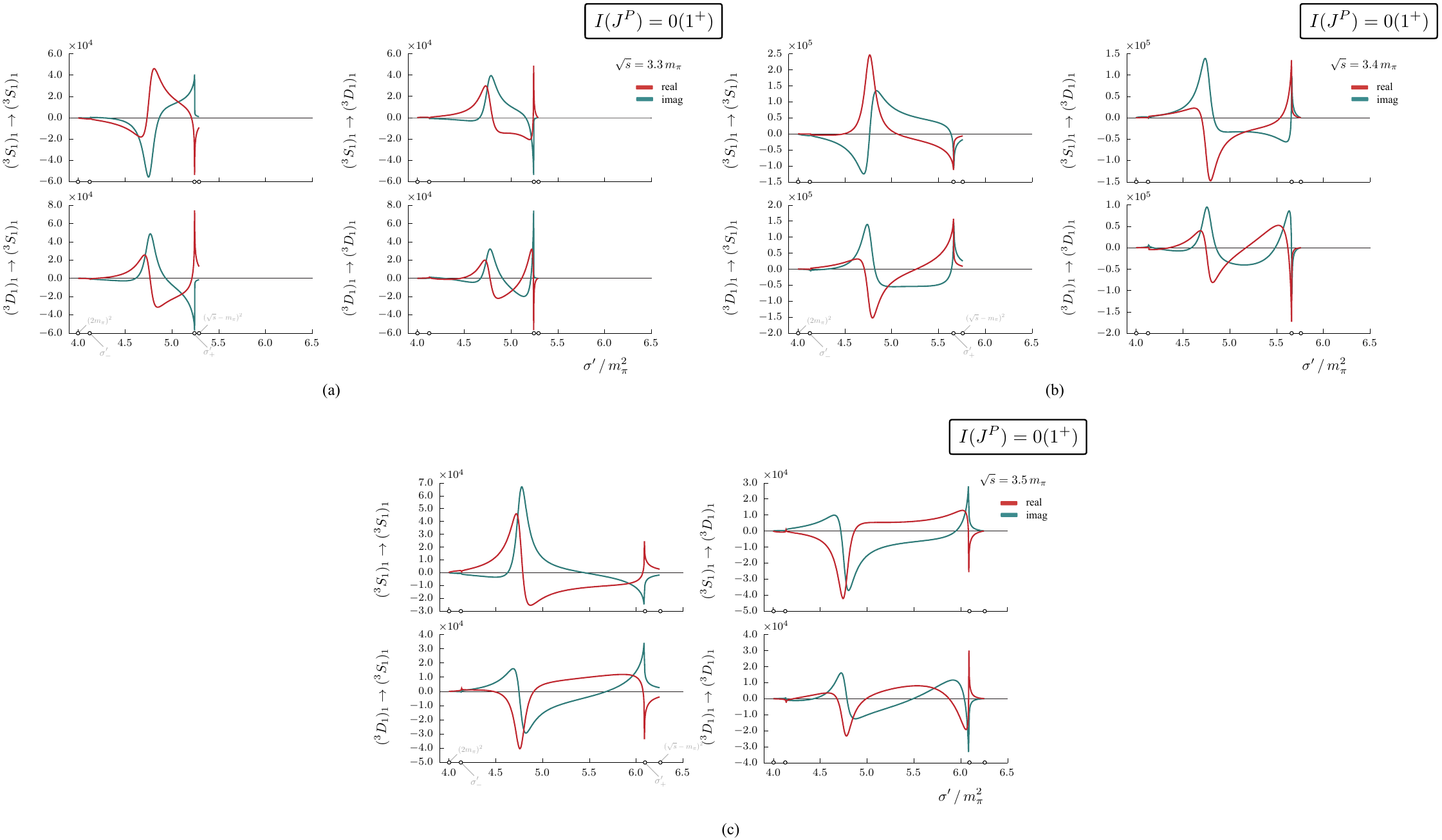}
	\caption{Real and imaginary parts for the asymmetric spin-orbit amplitudes as a function of $\sigma'$ for $I(J^P) = 0(1^+)$ at (a) $\sqrt{s} = 3.3\,m_\pi$, (b) $\sqrt{s} = 3.4\,m_\pi$, and (c) $\sqrt{s} = 3.5\,m_\pi$. Only two channels are allowed in our restricted waveset, $(^{3}S_0)_1$ and $(^{3}D_0)_1$. Important kinematic points are indicated.}
	\label{fig:lsi_amp_0_1p}
\end{figure}
\begin{figure}[t]
	\centering
	\includegraphics[width=\textwidth]{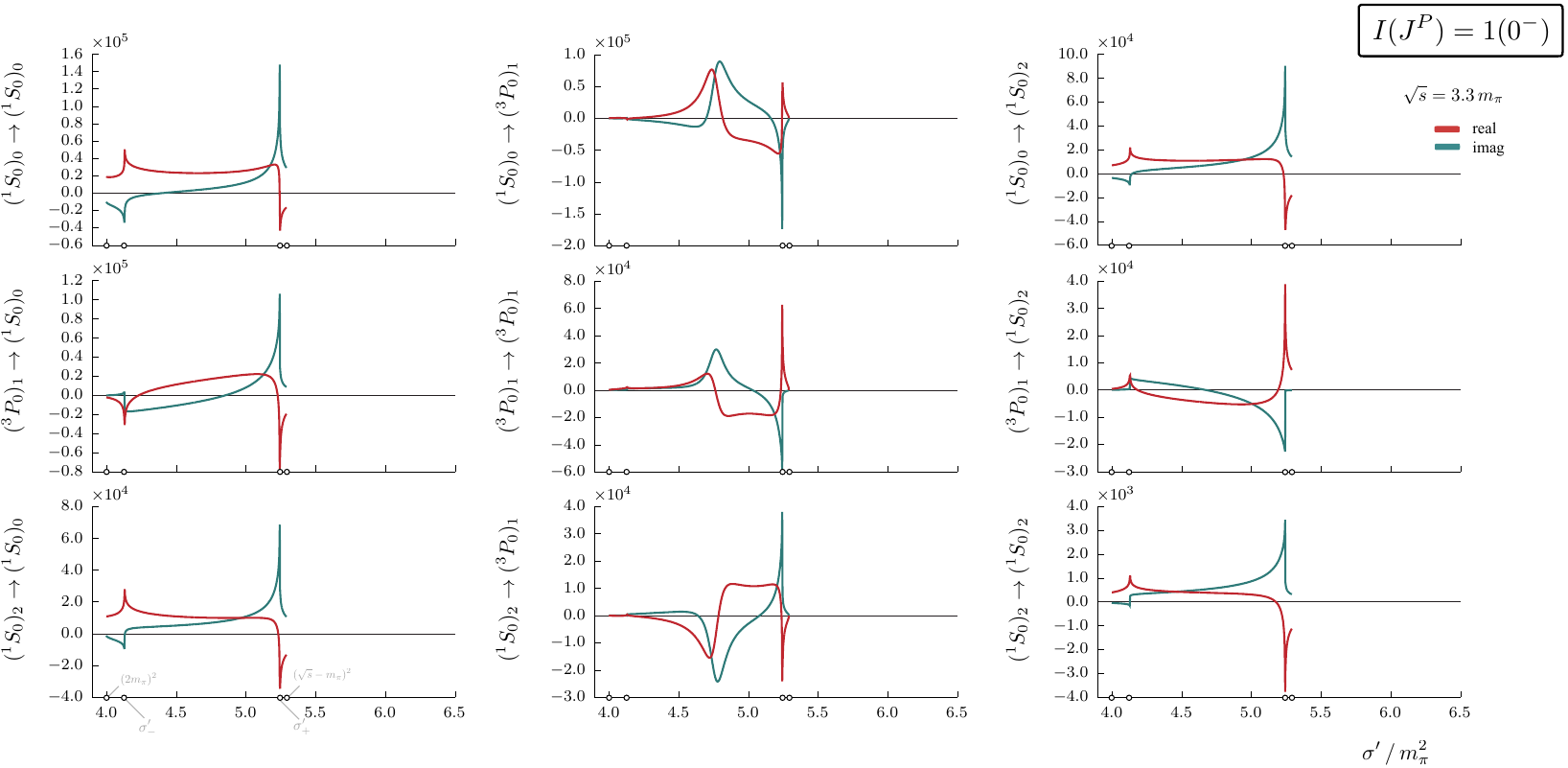}
	\caption{Real and imaginary parts for the asymmetric spin-orbit amplitudes as a function of $\sigma'$ for $I(J^P) = 1(0^-)$ at $\sqrt{s} = 3.3\,m_\pi$. Three channels are allowed in our restricted waveset, $(^{1}S_0)_0$, $(^{3}P_0)_1$, and $(^1S_0)_2$. Important kinematic points are indicated as discussed in the text.}
	\label{fig:lsi_amp_1_0m_pt1}
\end{figure}
\begin{figure}[t]
	\centering
	\includegraphics[width=\textwidth]{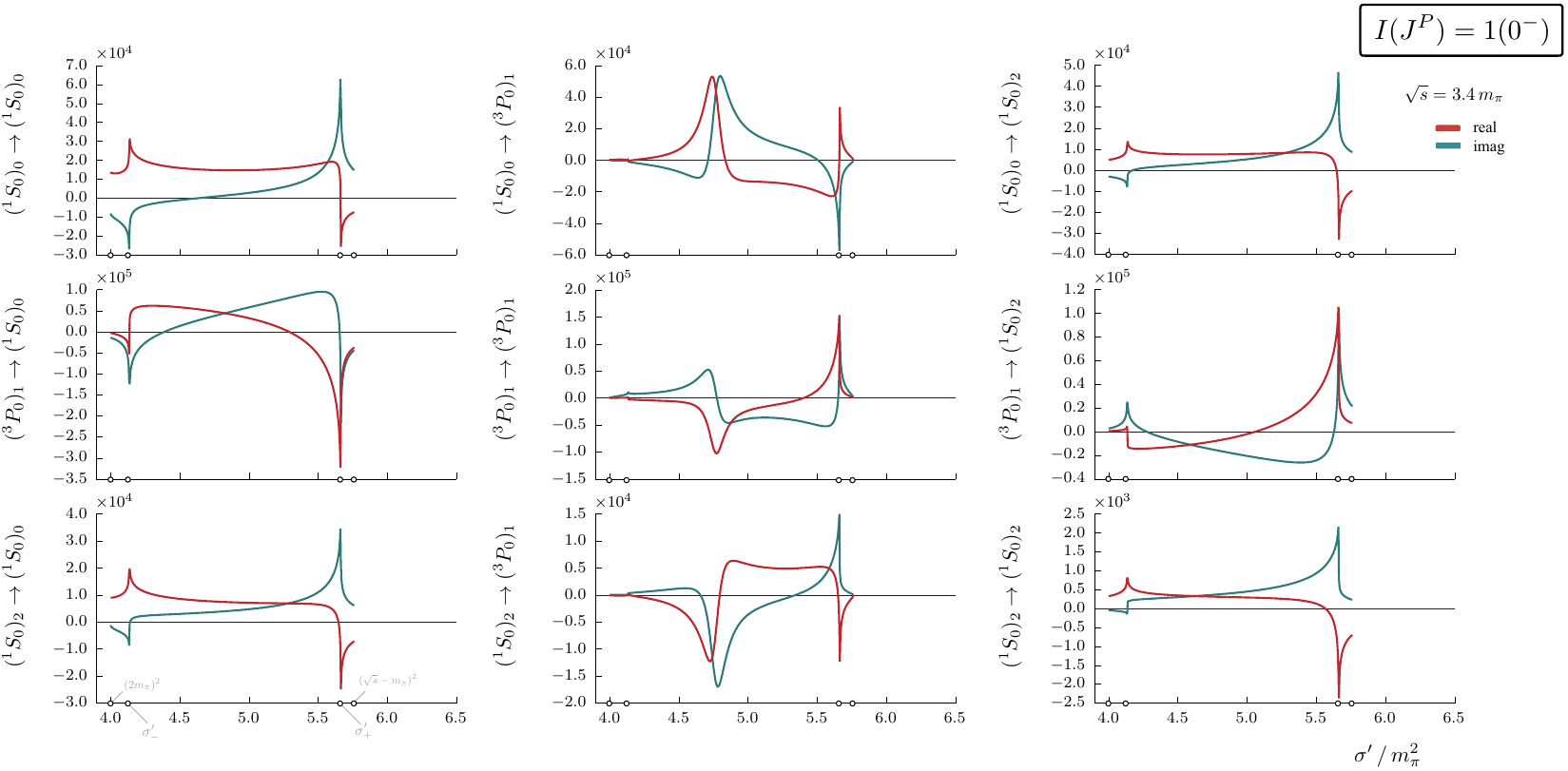}
	\caption{Same as Fig.~\ref{fig:lsi_amp_1_0m_pt1} except for $\sqrt{s} = 3.4\,  m_\pi$.}
	\label{fig:lsi_amp_1_0m_pt2}
\end{figure}
\begin{figure}[t]
	\centering
	\includegraphics[width=\textwidth]{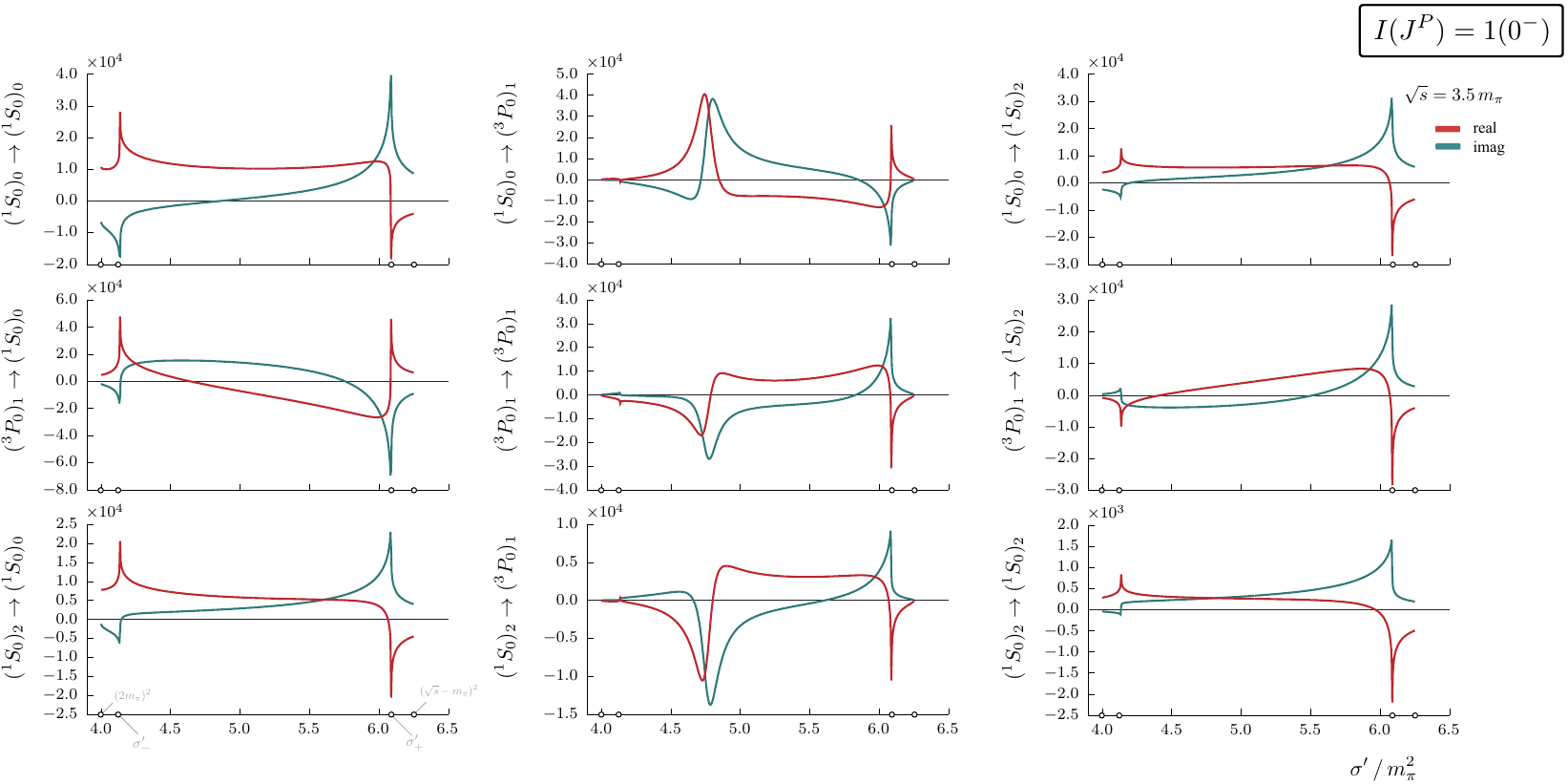}
	\caption{Same as Fig.~\ref{fig:lsi_amp_1_0m_pt1} except for $\sqrt{s} = 3.5\,  m_\pi$.}
	\label{fig:lsi_amp_1_0m_pt3}
\end{figure}

The asymmetric OPE amplitude is shown in Figs.~\ref{fig:lsi_amp_3_0m_and_3_1p} through~\ref{fig:lsi_amp_1_0m_pt3}, with the definition of the OPE given by Eq.~\eqref{eq:ope} and repeated here for convenience,
\begin{align}
    \Mc_{L_{j'}'S_{j'}'I_{j'}';L_jS_jI_j}^{(j',j)\, I(J^P)}(p_{j'}',p_j) = - \Mc_{2,S_{j'}'}^{I_{j'}'}(\sigma_{j'}') \, \Gc_{L_{j'}'S_{j'}'I_{j'}';L_jS_jI_j}^{I(J^P)}(p_{j'}',p_j) \, \Mc_{2,S_j}^{I_j}(\sigma_j)\, , \nn 
\end{align}
where $\Mc_{2,S}^I$ is the two-body amplitude given in Sec.~\ref{sec:two_body} and $\Gc$ is the partial wave exchange propagator as defined in Ref.~\cite{Jackura:2023qtp}. The specific form of $\Gc$ for each $I(J^P)$ considered is given in Ref.~\cite{Jackura:2023qtp} (Appendix D of Ref.~\cite{Jackura:2023qtp} summarizes the results.~\footnote{Note that $T$ is used as total isospin throughout Ref.~\cite{Jackura:2023qtp}.}) and will not be restated here. However, the general structure of the amplitude is given by
\begin{align}
    \label{eq:ope_app}
    \Gc^{I(J^P)}(p',p) = \Kc_\Gc^{I(J^P)}(p',p) + \Tc^{I(J^P)}(p',p)\, Q_0(\zeta(p',p)) \, ,
\end{align}
where $\Kc_\Gc$ and $\Tc$ are analytic functions of energies in the physical region, $Q_0(z)$ is the Legendre function of the second kind,
\begin{align}
    Q_0(z) = \frac{1}{2}\log\left( \frac{z+1}{z-1} \right) \, ,
\end{align}
and $\zeta(p',p)$ is a function of kinematics (see Ref.~\cite{Jackura:2023qtp}). An important feature of the exchange function is that it has singularities associated with the exchange of an on-shell particle, which occur at $\zeta(p',p) = \pm 1$ in Eq.~\eqref{eq:ope_app}. For fixed, physical values of $s$ and $\sigma \equiv \sigma_j$ as well as equal particle masses $m_\pi$, the singularities $\sigma_{\pm}' = \sigma_{\pm}'(s,\sigma)$ are located at
\begin{align}
    \sigma'_{\pm} = \frac{1}{2} \Bigg[\, 3m_\pi^2 + s - \sigma \pm  \lambda^{1/2}(s,m_\pi^2,\sigma) \, \sqrt{1 - \frac{4m_\pi^2}{\sigma}} \, \Bigg] \, .
\end{align}
For the $s$ values chosen, we find numerically $\sigma'_{\pm} \, / \, m_\pi^2 \approx \{4.13, 5.24\}$ for $\sqrt{s} \, / \, m_\pi = 3.3$, $\sigma'_{\pm} \, / \, m_\pi^2 \approx \{4.14, 5.67\}$ for $\sqrt{s} \, / \, m_\pi = 3.4$, and $\sigma'_{\pm} \, / \, m_\pi^2 \approx \{4.14, 6.09\}$ for $\sqrt{s} \, / \, m_\pi = 3.5$.

%

%

%

\bibliography{bibi.bib}

\begin{thebibliography}{111}%
\makeatletter
\providecommand \@ifxundefined [1]{%
 \@ifx{#1\undefined}
}%
\providecommand \@ifnum [1]{%
 \ifnum #1\expandafter \@firstoftwo
 \else \expandafter \@secondoftwo
 \fi
}%
\providecommand \@ifx [1]{%
 \ifx #1\expandafter \@firstoftwo
 \else \expandafter \@secondoftwo
 \fi
}%
\providecommand \natexlab [1]{#1}%
\providecommand \enquote  [1]{``#1''}%
\providecommand \bibnamefont  [1]{#1}%
\providecommand \bibfnamefont [1]{#1}%
\providecommand \citenamefont [1]{#1}%
\providecommand \href@noop [0]{\@secondoftwo}%
\providecommand \href [0]{\begingroup \@sanitize@url \@href}%
\providecommand \@href[1]{\@@startlink{#1}\@@href}%
\providecommand \@@href[1]{\endgroup#1\@@endlink}%
\providecommand \@sanitize@url [0]{\catcode `\\12\catcode `\$12\catcode `\&12\catcode `\#12\catcode `\^12\catcode `\_12\catcode `\%12\relax}%
\providecommand \@@startlink[1]{}%
\providecommand \@@endlink[0]{}%
\providecommand \url  [0]{\begingroup\@sanitize@url \@url }%
\providecommand \@url [1]{\endgroup\@href {#1}{\urlprefix }}%
\providecommand \urlprefix  [0]{URL }%
\providecommand \Eprint [0]{\href }%
\providecommand \doibase [0]{http://dx.doi.org/}%
\providecommand \selectlanguage [0]{\@gobble}%
\providecommand \bibinfo  [0]{\@secondoftwo}%
\providecommand \bibfield  [0]{\@secondoftwo}%
\providecommand \translation [1]{[#1]}%
\providecommand \BibitemOpen [0]{}%
\providecommand \bibitemStop [0]{}%
\providecommand \bibitemNoStop [0]{.\EOS\space}%
\providecommand \EOS [0]{\spacefactor3000\relax}%
\providecommand \BibitemShut  [1]{\csname bibitem#1\endcsname}%
\let\auto@bib@innerbib\@empty
\bibitem [{\citenamefont {Navas}\ \emph {et~al.}(2024)\citenamefont {Navas} \emph {et~al.}}]{ParticleDataGroup:2024cfk}%
  \BibitemOpen
  \bibfield  {author} {\bibinfo {author} {\bibfnamefont {S.}~\bibnamefont {Navas}} \emph {et~al.} (\bibinfo {collaboration} {Particle Data Group}),\ }\href {\doibase 10.1103/PhysRevD.110.030001} {\bibfield  {journal} {\bibinfo  {journal} {Phys. Rev. D}\ }\textbf {\bibinfo {volume} {110}},\ \bibinfo {pages} {030001} (\bibinfo {year} {2024})}\BibitemShut {NoStop}%
\bibitem [{\citenamefont {Acciarri}\ \emph {et~al.}(2016)\citenamefont {Acciarri} \emph {et~al.}}]{DUNE:2016hlj}%
  \BibitemOpen
  \bibfield  {author} {\bibinfo {author} {\bibfnamefont {R.}~\bibnamefont {Acciarri}} \emph {et~al.} (\bibinfo {collaboration} {DUNE}),\ }\href@noop {} {\  (\bibinfo {year} {2016})},\ \Eprint {http://arxiv.org/abs/1601.05471} {arXiv:1601.05471 [physics.ins-det]} \BibitemShut {NoStop}%
\bibitem [{\citenamefont {Aaij}\ \emph {et~al.}(2019)\citenamefont {Aaij} \emph {et~al.}}]{LHCb:2019xmb}%
  \BibitemOpen
  \bibfield  {author} {\bibinfo {author} {\bibfnamefont {R.}~\bibnamefont {Aaij}} \emph {et~al.} (\bibinfo {collaboration} {LHCb}),\ }\href {\doibase 10.1103/PhysRevLett.123.231802} {\bibfield  {journal} {\bibinfo  {journal} {Phys. Rev. Lett.}\ }\textbf {\bibinfo {volume} {123}},\ \bibinfo {pages} {231802} (\bibinfo {year} {2019})},\ \Eprint {http://arxiv.org/abs/1905.09244} {arXiv:1905.09244 [hep-ex]} \BibitemShut {NoStop}%
\bibitem [{\citenamefont {Aaij}\ \emph {et~al.}(2023)\citenamefont {Aaij} \emph {et~al.}}]{LHCb:2022fpg}%
  \BibitemOpen
  \bibfield  {author} {\bibinfo {author} {\bibfnamefont {R.}~\bibnamefont {Aaij}} \emph {et~al.} (\bibinfo {collaboration} {LHCb}),\ }\href {\doibase 10.1103/PhysRevD.108.012008} {\bibfield  {journal} {\bibinfo  {journal} {Phys. Rev. D}\ }\textbf {\bibinfo {volume} {108}},\ \bibinfo {pages} {012008} (\bibinfo {year} {2023})},\ \Eprint {http://arxiv.org/abs/2206.07622} {arXiv:2206.07622 [hep-ex]} \BibitemShut {NoStop}%
\bibitem [{\citenamefont {Aaij}\ \emph {et~al.}(2014{\natexlab{a}})\citenamefont {Aaij} \emph {et~al.}}]{LHCb:2014mir}%
  \BibitemOpen
  \bibfield  {author} {\bibinfo {author} {\bibfnamefont {R.}~\bibnamefont {Aaij}} \emph {et~al.} (\bibinfo {collaboration} {LHCb}),\ }\href {\doibase 10.1103/PhysRevD.90.112004} {\bibfield  {journal} {\bibinfo  {journal} {Phys. Rev. D}\ }\textbf {\bibinfo {volume} {90}},\ \bibinfo {pages} {112004} (\bibinfo {year} {2014}{\natexlab{a}})},\ \Eprint {http://arxiv.org/abs/1408.5373} {arXiv:1408.5373 [hep-ex]} \BibitemShut {NoStop}%
\bibitem [{\citenamefont {Aaij}\ \emph {et~al.}(2020)\citenamefont {Aaij} \emph {et~al.}}]{LHCb:2019jta}%
  \BibitemOpen
  \bibfield  {author} {\bibinfo {author} {\bibfnamefont {R.}~\bibnamefont {Aaij}} \emph {et~al.} (\bibinfo {collaboration} {LHCb}),\ }\href {\doibase 10.1103/PhysRevLett.124.031801} {\bibfield  {journal} {\bibinfo  {journal} {Phys. Rev. Lett.}\ }\textbf {\bibinfo {volume} {124}},\ \bibinfo {pages} {031801} (\bibinfo {year} {2020})},\ \Eprint {http://arxiv.org/abs/1909.05211} {arXiv:1909.05211 [hep-ex]} \BibitemShut {NoStop}%
\bibitem [{\citenamefont {Aaij}\ \emph {et~al.}(2014{\natexlab{b}})\citenamefont {Aaij} \emph {et~al.}}]{LHCb:2013lcl}%
  \BibitemOpen
  \bibfield  {author} {\bibinfo {author} {\bibfnamefont {R.}~\bibnamefont {Aaij}} \emph {et~al.} (\bibinfo {collaboration} {LHCb}),\ }\href {\doibase 10.1103/PhysRevLett.112.011801} {\bibfield  {journal} {\bibinfo  {journal} {Phys. Rev. Lett.}\ }\textbf {\bibinfo {volume} {112}},\ \bibinfo {pages} {011801} (\bibinfo {year} {2014}{\natexlab{b}})},\ \Eprint {http://arxiv.org/abs/1310.4740} {arXiv:1310.4740 [hep-ex]} \BibitemShut {NoStop}%
\bibitem [{\citenamefont {Suzuki}\ and\ \citenamefont {Wolfenstein}(1999)}]{Suzuki:1999uc}%
  \BibitemOpen
  \bibfield  {author} {\bibinfo {author} {\bibfnamefont {M.}~\bibnamefont {Suzuki}}\ and\ \bibinfo {author} {\bibfnamefont {L.}~\bibnamefont {Wolfenstein}},\ }\href {\doibase 10.1103/PhysRevD.60.074019} {\bibfield  {journal} {\bibinfo  {journal} {Phys. Rev. D}\ }\textbf {\bibinfo {volume} {60}},\ \bibinfo {pages} {074019} (\bibinfo {year} {1999})},\ \Eprint {http://arxiv.org/abs/hep-ph/9903477} {arXiv:hep-ph/9903477} \BibitemShut {NoStop}%
\bibitem [{\citenamefont {Wolfenstein}(1991)}]{Wolfenstein:1990ks}%
  \BibitemOpen
  \bibfield  {author} {\bibinfo {author} {\bibfnamefont {L.}~\bibnamefont {Wolfenstein}},\ }\href {\doibase 10.1103/PhysRevD.43.151} {\bibfield  {journal} {\bibinfo  {journal} {Phys. Rev. D}\ }\textbf {\bibinfo {volume} {43}},\ \bibinfo {pages} {151} (\bibinfo {year} {1991})}\BibitemShut {NoStop}%
\bibitem [{\citenamefont {Suzuki}(2008)}]{Suzuki:2007je}%
  \BibitemOpen
  \bibfield  {author} {\bibinfo {author} {\bibfnamefont {M.}~\bibnamefont {Suzuki}},\ }\href {\doibase 10.1103/PhysRevD.77.054021} {\bibfield  {journal} {\bibinfo  {journal} {Phys. Rev. D}\ }\textbf {\bibinfo {volume} {77}},\ \bibinfo {pages} {054021} (\bibinfo {year} {2008})},\ \Eprint {http://arxiv.org/abs/0710.5534} {arXiv:0710.5534 [hep-ph]} \BibitemShut {NoStop}%
\bibitem [{\citenamefont {Alvarenga~Nogueira}\ \emph {et~al.}(2015)\citenamefont {Alvarenga~Nogueira}, \citenamefont {Bediaga}, \citenamefont {Cavalcante}, \citenamefont {Frederico},\ and\ \citenamefont {Louren\c{c}o}}]{AlvarengaNogueira:2015wpj}%
  \BibitemOpen
  \bibfield  {author} {\bibinfo {author} {\bibfnamefont {J.~H.}\ \bibnamefont {Alvarenga~Nogueira}}, \bibinfo {author} {\bibfnamefont {I.}~\bibnamefont {Bediaga}}, \bibinfo {author} {\bibfnamefont {A.~B.~R.}\ \bibnamefont {Cavalcante}}, \bibinfo {author} {\bibfnamefont {T.}~\bibnamefont {Frederico}}, \ and\ \bibinfo {author} {\bibfnamefont {O.}~\bibnamefont {Louren\c{c}o}},\ }\href {\doibase 10.1103/PhysRevD.92.054010} {\bibfield  {journal} {\bibinfo  {journal} {Phys. Rev. D}\ }\textbf {\bibinfo {volume} {92}},\ \bibinfo {pages} {054010} (\bibinfo {year} {2015})},\ \Eprint {http://arxiv.org/abs/1506.08332} {arXiv:1506.08332 [hep-ph]} \BibitemShut {NoStop}%
\bibitem [{\citenamefont {Bediaga}\ \emph {et~al.}(2014)\citenamefont {Bediaga}, \citenamefont {Frederico},\ and\ \citenamefont {Louren\c{c}o}}]{Bediaga:2013ela}%
  \BibitemOpen
  \bibfield  {author} {\bibinfo {author} {\bibfnamefont {I.}~\bibnamefont {Bediaga}}, \bibinfo {author} {\bibfnamefont {T.}~\bibnamefont {Frederico}}, \ and\ \bibinfo {author} {\bibfnamefont {O.}~\bibnamefont {Louren\c{c}o}},\ }\href {\doibase 10.1103/PhysRevD.89.094013} {\bibfield  {journal} {\bibinfo  {journal} {Phys. Rev. D}\ }\textbf {\bibinfo {volume} {89}},\ \bibinfo {pages} {094013} (\bibinfo {year} {2014})},\ \Eprint {http://arxiv.org/abs/1307.8164} {arXiv:1307.8164 [hep-ph]} \BibitemShut {NoStop}%
\bibitem [{\citenamefont {Garrote}\ \emph {et~al.}(2023)\citenamefont {Garrote}, \citenamefont {Cuervo}, \citenamefont {Magalh\~aes},\ and\ \citenamefont {Pel\'aez}}]{Garrote:2022uub}%
  \BibitemOpen
  \bibfield  {author} {\bibinfo {author} {\bibfnamefont {R.~A.}\ \bibnamefont {Garrote}}, \bibinfo {author} {\bibfnamefont {J.}~\bibnamefont {Cuervo}}, \bibinfo {author} {\bibfnamefont {P.~C.}\ \bibnamefont {Magalh\~aes}}, \ and\ \bibinfo {author} {\bibfnamefont {J.~R.}\ \bibnamefont {Pel\'aez}},\ }\href {\doibase 10.1103/PhysRevLett.130.201901} {\bibfield  {journal} {\bibinfo  {journal} {Phys. Rev. Lett.}\ }\textbf {\bibinfo {volume} {130}},\ \bibinfo {pages} {201901} (\bibinfo {year} {2023})},\ \Eprint {http://arxiv.org/abs/2210.08354} {arXiv:2210.08354 [hep-ph]} \BibitemShut {NoStop}%
\bibitem [{\citenamefont {Briceno}\ \emph {et~al.}(2018{\natexlab{a}})\citenamefont {Briceno}, \citenamefont {Dudek},\ and\ \citenamefont {Young}}]{Briceno:2017max}%
  \BibitemOpen
  \bibfield  {author} {\bibinfo {author} {\bibfnamefont {R.~A.}\ \bibnamefont {Briceno}}, \bibinfo {author} {\bibfnamefont {J.~J.}\ \bibnamefont {Dudek}}, \ and\ \bibinfo {author} {\bibfnamefont {R.~D.}\ \bibnamefont {Young}},\ }\href {\doibase 10.1103/RevModPhys.90.025001} {\bibfield  {journal} {\bibinfo  {journal} {Rev. Mod. Phys.}\ }\textbf {\bibinfo {volume} {90}},\ \bibinfo {pages} {025001} (\bibinfo {year} {2018}{\natexlab{a}})},\ \Eprint {http://arxiv.org/abs/1706.06223} {arXiv:1706.06223 [hep-lat]} \BibitemShut {NoStop}%
\bibitem [{\citenamefont {Hansen}\ and\ \citenamefont {Sharpe}(2019)}]{Hansen:2019nir}%
  \BibitemOpen
  \bibfield  {author} {\bibinfo {author} {\bibfnamefont {M.~T.}\ \bibnamefont {Hansen}}\ and\ \bibinfo {author} {\bibfnamefont {S.~R.}\ \bibnamefont {Sharpe}},\ }\href {\doibase 10.1146/annurev-nucl-101918-023723} {\bibfield  {journal} {\bibinfo  {journal} {Annual Review of Nuclear and Particle Science}\ }\textbf {\bibinfo {volume} {69}},\ \bibinfo {pages} {null} (\bibinfo {year} {2019})},\ \Eprint {http://arxiv.org/abs/1901.00483} {arXiv:1901.00483 [hep-lat]} \BibitemShut {NoStop}%
\bibitem [{\citenamefont {Mai}\ \emph {et~al.}(2021)\citenamefont {Mai}, \citenamefont {D\"oring},\ and\ \citenamefont {Rusetsky}}]{Mai:2021lwb}%
  \BibitemOpen
  \bibfield  {author} {\bibinfo {author} {\bibfnamefont {M.}~\bibnamefont {Mai}}, \bibinfo {author} {\bibfnamefont {M.}~\bibnamefont {D\"oring}}, \ and\ \bibinfo {author} {\bibfnamefont {A.}~\bibnamefont {Rusetsky}},\ }\href {\doibase 10.1140/epjs/s11734-021-00146-5} {\bibfield  {journal} {\bibinfo  {journal} {Eur. Phys. J. ST}\ }\textbf {\bibinfo {volume} {230}},\ \bibinfo {pages} {1623} (\bibinfo {year} {2021})},\ \Eprint {http://arxiv.org/abs/2103.00577} {arXiv:2103.00577 [hep-lat]} \BibitemShut {NoStop}%
\bibitem [{\citenamefont {Jackura}\ \emph {et~al.}(2019)\citenamefont {Jackura}, \citenamefont {Fern\'andez-Ram\'\i{}rez}, \citenamefont {Mathieu}, \citenamefont {Mikhasenko}, \citenamefont {Nys}, \citenamefont {Pilloni}, \citenamefont {Salda\~na}, \citenamefont {Sherrill},\ and\ \citenamefont {Szczepaniak}}]{Jackura:2018xnx}%
  \BibitemOpen
  \bibfield  {author} {\bibinfo {author} {\bibfnamefont {A.}~\bibnamefont {Jackura}}, \bibinfo {author} {\bibfnamefont {C.}~\bibnamefont {Fern\'andez-Ram\'\i{}rez}}, \bibinfo {author} {\bibfnamefont {V.}~\bibnamefont {Mathieu}}, \bibinfo {author} {\bibfnamefont {M.}~\bibnamefont {Mikhasenko}}, \bibinfo {author} {\bibfnamefont {J.}~\bibnamefont {Nys}}, \bibinfo {author} {\bibfnamefont {A.}~\bibnamefont {Pilloni}}, \bibinfo {author} {\bibfnamefont {K.}~\bibnamefont {Salda\~na}}, \bibinfo {author} {\bibfnamefont {N.}~\bibnamefont {Sherrill}}, \ and\ \bibinfo {author} {\bibfnamefont {A.}~\bibnamefont {Szczepaniak}} (\bibinfo {collaboration} {JPAC}),\ }\href {\doibase 10.1140/epjc/s10052-019-6566-1} {\bibfield  {journal} {\bibinfo  {journal} {Eur. Phys. J. C}\ }\textbf {\bibinfo {volume} {79}},\ \bibinfo {pages} {56} (\bibinfo {year} {2019})},\ \Eprint {http://arxiv.org/abs/1809.10523} {arXiv:1809.10523 [hep-ph]} \BibitemShut {NoStop}%
\bibitem [{\citenamefont {Hansen}\ and\ \citenamefont {Sharpe}(2015)}]{Hansen:2015zga}%
  \BibitemOpen
  \bibfield  {author} {\bibinfo {author} {\bibfnamefont {M.~T.}\ \bibnamefont {Hansen}}\ and\ \bibinfo {author} {\bibfnamefont {S.~R.}\ \bibnamefont {Sharpe}},\ }\href {\doibase 10.1103/PhysRevD.92.114509} {\bibfield  {journal} {\bibinfo  {journal} {Phys. Rev.}\ }\textbf {\bibinfo {volume} {D92}},\ \bibinfo {pages} {114509} (\bibinfo {year} {2015})},\ \Eprint {http://arxiv.org/abs/1504.04248} {arXiv:1504.04248 [hep-lat]} \BibitemShut {NoStop}%
\bibitem [{\citenamefont {Dawid}\ \emph {et~al.}(2023{\natexlab{a}})\citenamefont {Dawid}, \citenamefont {Islam},\ and\ \citenamefont {Brice\~no}}]{Dawid:2023jrj}%
  \BibitemOpen
  \bibfield  {author} {\bibinfo {author} {\bibfnamefont {S.~M.}\ \bibnamefont {Dawid}}, \bibinfo {author} {\bibfnamefont {M.~H.~E.}\ \bibnamefont {Islam}}, \ and\ \bibinfo {author} {\bibfnamefont {R.~A.}\ \bibnamefont {Brice\~no}},\ }\href {\doibase 10.1103/PhysRevD.108.034016} {\bibfield  {journal} {\bibinfo  {journal} {Phys. Rev. D}\ }\textbf {\bibinfo {volume} {108}},\ \bibinfo {pages} {034016} (\bibinfo {year} {2023}{\natexlab{a}})},\ \Eprint {http://arxiv.org/abs/2303.04394} {arXiv:2303.04394 [nucl-th]} \BibitemShut {NoStop}%
\bibitem [{\citenamefont {Dawid}\ \emph {et~al.}(2023{\natexlab{b}})\citenamefont {Dawid}, \citenamefont {Islam}, \citenamefont {Brice\~no},\ and\ \citenamefont {Jackura}}]{Dawid:2023kxu}%
  \BibitemOpen
  \bibfield  {author} {\bibinfo {author} {\bibfnamefont {S.~M.}\ \bibnamefont {Dawid}}, \bibinfo {author} {\bibfnamefont {M.~H.~E.}\ \bibnamefont {Islam}}, \bibinfo {author} {\bibfnamefont {R.~A.}\ \bibnamefont {Brice\~no}}, \ and\ \bibinfo {author} {\bibfnamefont {A.~W.}\ \bibnamefont {Jackura}},\ }\href@noop {} {\  (\bibinfo {year} {2023}{\natexlab{b}})},\ \Eprint {http://arxiv.org/abs/2309.01732} {arXiv:2309.01732 [nucl-th]} \BibitemShut {NoStop}%
\bibitem [{\citenamefont {Jackura}\ \emph {et~al.}(2021)\citenamefont {Jackura}, \citenamefont {Brice{\~n}o}, \citenamefont {Dawid}, \citenamefont {Islam},\ and\ \citenamefont {McCarty}}]{Jackura:2020bsk}%
  \BibitemOpen
  \bibfield  {author} {\bibinfo {author} {\bibfnamefont {A.~W.}\ \bibnamefont {Jackura}}, \bibinfo {author} {\bibfnamefont {R.~A.}\ \bibnamefont {Brice{\~n}o}}, \bibinfo {author} {\bibfnamefont {S.~M.}\ \bibnamefont {Dawid}}, \bibinfo {author} {\bibfnamefont {M.~H.~E.}\ \bibnamefont {Islam}}, \ and\ \bibinfo {author} {\bibfnamefont {C.}~\bibnamefont {McCarty}},\ }\href {\doibase 10.1103/PhysRevD.104.014507} {\bibfield  {journal} {\bibinfo  {journal} {Phys. Rev. D}\ }\textbf {\bibinfo {volume} {104}},\ \bibinfo {pages} {014507} (\bibinfo {year} {2021})},\ \Eprint {http://arxiv.org/abs/2010.09820} {arXiv:2010.09820 [hep-lat]} \BibitemShut {NoStop}%
\bibitem [{\citenamefont {Jackura}\ and\ \citenamefont {Brice\~no}(2024)}]{Jackura:2023qtp}%
  \BibitemOpen
  \bibfield  {author} {\bibinfo {author} {\bibfnamefont {A.~W.}\ \bibnamefont {Jackura}}\ and\ \bibinfo {author} {\bibfnamefont {R.~A.}\ \bibnamefont {Brice\~no}},\ }\href {\doibase 10.1103/PhysRevD.109.096030} {\bibfield  {journal} {\bibinfo  {journal} {Phys. Rev. D}\ }\textbf {\bibinfo {volume} {109}},\ \bibinfo {pages} {096030} (\bibinfo {year} {2024})},\ \Eprint {http://arxiv.org/abs/2312.00625} {arXiv:2312.00625 [hep-ph]} \BibitemShut {NoStop}%
\bibitem [{\citenamefont {Jackura}(2023)}]{Jackura:2022gib}%
  \BibitemOpen
  \bibfield  {author} {\bibinfo {author} {\bibfnamefont {A.~W.}\ \bibnamefont {Jackura}},\ }\href {\doibase 10.1103/PhysRevD.108.034505} {\bibfield  {journal} {\bibinfo  {journal} {Phys. Rev. D}\ }\textbf {\bibinfo {volume} {108}},\ \bibinfo {pages} {034505} (\bibinfo {year} {2023})},\ \Eprint {http://arxiv.org/abs/2208.10587} {arXiv:2208.10587 [hep-lat]} \BibitemShut {NoStop}%
\bibitem [{\citenamefont {Mai}\ \emph {et~al.}(2017)\citenamefont {Mai}, \citenamefont {Hu}, \citenamefont {Doring}, \citenamefont {Pilloni},\ and\ \citenamefont {Szczepaniak}}]{Mai:2017vot}%
  \BibitemOpen
  \bibfield  {author} {\bibinfo {author} {\bibfnamefont {M.}~\bibnamefont {Mai}}, \bibinfo {author} {\bibfnamefont {B.}~\bibnamefont {Hu}}, \bibinfo {author} {\bibfnamefont {M.}~\bibnamefont {Doring}}, \bibinfo {author} {\bibfnamefont {A.}~\bibnamefont {Pilloni}}, \ and\ \bibinfo {author} {\bibfnamefont {A.}~\bibnamefont {Szczepaniak}},\ }\href {\doibase 10.1140/epja/i2017-12368-4} {\bibfield  {journal} {\bibinfo  {journal} {Eur. Phys. J. A}\ }\textbf {\bibinfo {volume} {53}},\ \bibinfo {pages} {177} (\bibinfo {year} {2017})},\ \Eprint {http://arxiv.org/abs/1706.06118} {arXiv:1706.06118 [nucl-th]} \BibitemShut {NoStop}%
\bibitem [{\citenamefont {Mikhasenko}\ \emph {et~al.}(2019)\citenamefont {Mikhasenko}, \citenamefont {Wunderlich}, \citenamefont {Jackura}, \citenamefont {Mathieu}, \citenamefont {Pilloni}, \citenamefont {Ketzer},\ and\ \citenamefont {Szczepaniak}}]{Mikhasenko:2019vhk}%
  \BibitemOpen
  \bibfield  {author} {\bibinfo {author} {\bibfnamefont {M.}~\bibnamefont {Mikhasenko}}, \bibinfo {author} {\bibfnamefont {Y.}~\bibnamefont {Wunderlich}}, \bibinfo {author} {\bibfnamefont {A.}~\bibnamefont {Jackura}}, \bibinfo {author} {\bibfnamefont {V.}~\bibnamefont {Mathieu}}, \bibinfo {author} {\bibfnamefont {A.}~\bibnamefont {Pilloni}}, \bibinfo {author} {\bibfnamefont {B.}~\bibnamefont {Ketzer}}, \ and\ \bibinfo {author} {\bibfnamefont {A.}~\bibnamefont {Szczepaniak}},\ }\href {\doibase 10.1007/JHEP08(2019)080} {\bibfield  {journal} {\bibinfo  {journal} {JHEP}\ }\textbf {\bibinfo {volume} {08}},\ \bibinfo {pages} {080} (\bibinfo {year} {2019})},\ \Eprint {http://arxiv.org/abs/1904.11894} {arXiv:1904.11894 [hep-ph]} \BibitemShut {NoStop}%
\bibitem [{\citenamefont {Dawid}\ and\ \citenamefont {Szczepaniak}(2021)}]{Dawid:2020uhn}%
  \BibitemOpen
  \bibfield  {author} {\bibinfo {author} {\bibfnamefont {S.~M.}\ \bibnamefont {Dawid}}\ and\ \bibinfo {author} {\bibfnamefont {A.~P.}\ \bibnamefont {Szczepaniak}},\ }\href {\doibase 10.1103/PhysRevD.103.014009} {\bibfield  {journal} {\bibinfo  {journal} {Phys. Rev. D}\ }\textbf {\bibinfo {volume} {103}},\ \bibinfo {pages} {014009} (\bibinfo {year} {2021})},\ \Eprint {http://arxiv.org/abs/2010.08084} {arXiv:2010.08084 [nucl-th]} \BibitemShut {NoStop}%
\bibitem [{\citenamefont {Feng}\ \emph {et~al.}(2024)\citenamefont {Feng}, \citenamefont {Gil}, \citenamefont {D{\"o}ring}, \citenamefont {Molina}, \citenamefont {Mai}, \citenamefont {Shastry},\ and\ \citenamefont {Szczepaniak}}]{Feng:2024wyg}%
  \BibitemOpen
  \bibfield  {author} {\bibinfo {author} {\bibfnamefont {Y.}~\bibnamefont {Feng}}, \bibinfo {author} {\bibfnamefont {F.}~\bibnamefont {Gil}}, \bibinfo {author} {\bibfnamefont {M.}~\bibnamefont {D{\"o}ring}}, \bibinfo {author} {\bibfnamefont {R.}~\bibnamefont {Molina}}, \bibinfo {author} {\bibfnamefont {M.}~\bibnamefont {Mai}}, \bibinfo {author} {\bibfnamefont {V.}~\bibnamefont {Shastry}}, \ and\ \bibinfo {author} {\bibfnamefont {A.}~\bibnamefont {Szczepaniak}},\ }\href {\doibase 10.1103/PhysRevD.110.094002} {\bibfield  {journal} {\bibinfo  {journal} {Phys. Rev. D}\ }\textbf {\bibinfo {volume} {110}},\ \bibinfo {pages} {094002} (\bibinfo {year} {2024})},\ \Eprint {http://arxiv.org/abs/2407.08721} {arXiv:2407.08721 [nucl-th]} \BibitemShut {NoStop}%
\bibitem [{\citenamefont {Polejaeva}\ and\ \citenamefont {Rusetsky}(2012)}]{Polejaeva:2012ut}%
  \BibitemOpen
  \bibfield  {author} {\bibinfo {author} {\bibfnamefont {K.}~\bibnamefont {Polejaeva}}\ and\ \bibinfo {author} {\bibfnamefont {A.}~\bibnamefont {Rusetsky}},\ }\href {\doibase 10.1140/epja/i2012-12067-8} {\bibfield  {journal} {\bibinfo  {journal} {Eur. Phys. J.}\ }\textbf {\bibinfo {volume} {A48}},\ \bibinfo {pages} {67} (\bibinfo {year} {2012})},\ \Eprint {http://arxiv.org/abs/1203.1241} {arXiv:1203.1241 [hep-lat]} \BibitemShut {NoStop}%
\bibitem [{\citenamefont {Brice\~no}\ and\ \citenamefont {Davoudi}(2013{\natexlab{a}})}]{Briceno:2012rv}%
  \BibitemOpen
  \bibfield  {author} {\bibinfo {author} {\bibfnamefont {R.~A.}\ \bibnamefont {Brice\~no}}\ and\ \bibinfo {author} {\bibfnamefont {Z.}~\bibnamefont {Davoudi}},\ }\href {\doibase 10.1103/PhysRevD.87.094507} {\bibfield  {journal} {\bibinfo  {journal} {Phys. Rev.}\ }\textbf {\bibinfo {volume} {D87}},\ \bibinfo {pages} {094507} (\bibinfo {year} {2013}{\natexlab{a}})},\ \Eprint {http://arxiv.org/abs/1212.3398} {arXiv:1212.3398 [hep-lat]} \BibitemShut {NoStop}%
\bibitem [{\citenamefont {Luscher}(1986{\natexlab{a}})}]{Luscher:1985dn}%
  \BibitemOpen
  \bibfield  {author} {\bibinfo {author} {\bibfnamefont {M.}~\bibnamefont {Luscher}},\ }\href {\doibase 10.1007/BF01211589} {\bibfield  {journal} {\bibinfo  {journal} {Commun. Math. Phys.}\ }\textbf {\bibinfo {volume} {104}},\ \bibinfo {pages} {177} (\bibinfo {year} {1986}{\natexlab{a}})}\BibitemShut {NoStop}%
\bibitem [{\citenamefont {Luscher}(1986{\natexlab{b}})}]{Luscher:1986n2}%
  \BibitemOpen
  \bibfield  {author} {\bibinfo {author} {\bibfnamefont {M.}~\bibnamefont {Luscher}},\ }\href {\doibase 10.1007/BF01211097} {\bibfield  {journal} {\bibinfo  {journal} {Commun.Math.Phys.}\ }\textbf {\bibinfo {volume} {105}},\ \bibinfo {pages} {153} (\bibinfo {year} {1986}{\natexlab{b}})}\BibitemShut {NoStop}%
\bibitem [{\citenamefont {Luscher}(1991)}]{Luscher:1990ux}%
  \BibitemOpen
  \bibfield  {author} {\bibinfo {author} {\bibfnamefont {M.}~\bibnamefont {Luscher}},\ }\href {\doibase 10.1016/0550-3213(91)90366-6} {\bibfield  {journal} {\bibinfo  {journal} {Nucl. Phys.}\ }\textbf {\bibinfo {volume} {B354}},\ \bibinfo {pages} {531} (\bibinfo {year} {1991})}\BibitemShut {NoStop}%
\bibitem [{\citenamefont {Rummukainen}\ and\ \citenamefont {Gottlieb}(1995)}]{Rummukainen:1995vs}%
  \BibitemOpen
  \bibfield  {author} {\bibinfo {author} {\bibfnamefont {K.}~\bibnamefont {Rummukainen}}\ and\ \bibinfo {author} {\bibfnamefont {S.~A.}\ \bibnamefont {Gottlieb}},\ }\href {\doibase 10.1016/0550-3213(95)00313-H} {\bibfield  {journal} {\bibinfo  {journal} {Nucl. Phys.}\ }\textbf {\bibinfo {volume} {B450}},\ \bibinfo {pages} {397} (\bibinfo {year} {1995})},\ \Eprint {http://arxiv.org/abs/hep-lat/9503028} {arXiv:hep-lat/9503028 [hep-lat]} \BibitemShut {NoStop}%
\bibitem [{\citenamefont {Kim}\ \emph {et~al.}(2005)\citenamefont {Kim}, \citenamefont {Sachrajda},\ and\ \citenamefont {Sharpe}}]{Kim:2005gf}%
  \BibitemOpen
  \bibfield  {author} {\bibinfo {author} {\bibfnamefont {C.~h.}\ \bibnamefont {Kim}}, \bibinfo {author} {\bibfnamefont {C.~T.}\ \bibnamefont {Sachrajda}}, \ and\ \bibinfo {author} {\bibfnamefont {S.~R.}\ \bibnamefont {Sharpe}},\ }\href {\doibase 10.1016/j.nuclphysb.2005.08.029} {\bibfield  {journal} {\bibinfo  {journal} {Nucl. Phys.}\ }\textbf {\bibinfo {volume} {B727}},\ \bibinfo {pages} {218} (\bibinfo {year} {2005})},\ \Eprint {http://arxiv.org/abs/hep-lat/0507006} {arXiv:hep-lat/0507006 [hep-lat]} \BibitemShut {NoStop}%
\bibitem [{\citenamefont {He}\ \emph {et~al.}(2005)\citenamefont {He}, \citenamefont {Feng},\ and\ \citenamefont {Liu}}]{He:2005ey}%
  \BibitemOpen
  \bibfield  {author} {\bibinfo {author} {\bibfnamefont {S.}~\bibnamefont {He}}, \bibinfo {author} {\bibfnamefont {X.}~\bibnamefont {Feng}}, \ and\ \bibinfo {author} {\bibfnamefont {C.}~\bibnamefont {Liu}},\ }\href {\doibase 10.1088/1126-6708/2005/07/011} {\bibfield  {journal} {\bibinfo  {journal} {JHEP}\ }\textbf {\bibinfo {volume} {07}},\ \bibinfo {pages} {011} (\bibinfo {year} {2005})},\ \Eprint {http://arxiv.org/abs/hep-lat/0504019} {arXiv:hep-lat/0504019 [hep-lat]} \BibitemShut {NoStop}%
\bibitem [{\citenamefont {Hansen}\ and\ \citenamefont {Sharpe}(2012)}]{Hansen:2012tf}%
  \BibitemOpen
  \bibfield  {author} {\bibinfo {author} {\bibfnamefont {M.~T.}\ \bibnamefont {Hansen}}\ and\ \bibinfo {author} {\bibfnamefont {S.~R.}\ \bibnamefont {Sharpe}},\ }\href {\doibase 10.1103/PhysRevD.86.016007} {\bibfield  {journal} {\bibinfo  {journal} {Phys. Rev.}\ }\textbf {\bibinfo {volume} {D86}},\ \bibinfo {pages} {016007} (\bibinfo {year} {2012})},\ \Eprint {http://arxiv.org/abs/1204.0826} {arXiv:1204.0826 [hep-lat]} \BibitemShut {NoStop}%
\bibitem [{\citenamefont {Brice\~no}\ and\ \citenamefont {Davoudi}(2013{\natexlab{b}})}]{Briceno:2012yi}%
  \BibitemOpen
  \bibfield  {author} {\bibinfo {author} {\bibfnamefont {R.~A.}\ \bibnamefont {Brice\~no}}\ and\ \bibinfo {author} {\bibfnamefont {Z.}~\bibnamefont {Davoudi}},\ }\href {\doibase 10.1103/PhysRevD.88.094507} {\bibfield  {journal} {\bibinfo  {journal} {Phys. Rev.}\ }\textbf {\bibinfo {volume} {D88}},\ \bibinfo {pages} {094507} (\bibinfo {year} {2013}{\natexlab{b}})},\ \Eprint {http://arxiv.org/abs/1204.1110} {arXiv:1204.1110 [hep-lat]} \BibitemShut {NoStop}%
\bibitem [{\citenamefont {Brice\~no}\ \emph {et~al.}(2013)\citenamefont {Brice\~no}, \citenamefont {Davoudi},\ and\ \citenamefont {Luu}}]{Briceno:2013lba}%
  \BibitemOpen
  \bibfield  {author} {\bibinfo {author} {\bibfnamefont {R.~A.}\ \bibnamefont {Brice\~no}}, \bibinfo {author} {\bibfnamefont {Z.}~\bibnamefont {Davoudi}}, \ and\ \bibinfo {author} {\bibfnamefont {T.~C.}\ \bibnamefont {Luu}},\ }\href {\doibase 10.1103/PhysRevD.88.034502} {\bibfield  {journal} {\bibinfo  {journal} {Phys. Rev.}\ }\textbf {\bibinfo {volume} {D88}},\ \bibinfo {pages} {034502} (\bibinfo {year} {2013})},\ \Eprint {http://arxiv.org/abs/1305.4903} {arXiv:1305.4903 [hep-lat]} \BibitemShut {NoStop}%
\bibitem [{\citenamefont {Brice\~no}(2014)}]{Briceno:2014oea}%
  \BibitemOpen
  \bibfield  {author} {\bibinfo {author} {\bibfnamefont {R.~A.}\ \bibnamefont {Brice\~no}},\ }\href {\doibase 10.1103/PhysRevD.89.074507} {\bibfield  {journal} {\bibinfo  {journal} {Phys. Rev.}\ }\textbf {\bibinfo {volume} {D89}},\ \bibinfo {pages} {074507} (\bibinfo {year} {2014})},\ \Eprint {http://arxiv.org/abs/1401.3312} {arXiv:1401.3312 [hep-lat]} \BibitemShut {NoStop}%
\bibitem [{\citenamefont {Hansen}\ and\ \citenamefont {Sharpe}(2014)}]{Hansen:2014eka}%
  \BibitemOpen
  \bibfield  {author} {\bibinfo {author} {\bibfnamefont {M.~T.}\ \bibnamefont {Hansen}}\ and\ \bibinfo {author} {\bibfnamefont {S.~R.}\ \bibnamefont {Sharpe}},\ }\href {\doibase 10.1103/PhysRevD.90.116003} {\bibfield  {journal} {\bibinfo  {journal} {Phys. Rev.}\ }\textbf {\bibinfo {volume} {D90}},\ \bibinfo {pages} {116003} (\bibinfo {year} {2014})},\ \Eprint {http://arxiv.org/abs/1408.5933} {arXiv:1408.5933 [hep-lat]} \BibitemShut {NoStop}%
\bibitem [{\citenamefont {Briceno}\ \emph {et~al.}(2017)\citenamefont {Briceno}, \citenamefont {Hansen},\ and\ \citenamefont {Sharpe}}]{Briceno:2017tce}%
  \BibitemOpen
  \bibfield  {author} {\bibinfo {author} {\bibfnamefont {R.~A.}\ \bibnamefont {Briceno}}, \bibinfo {author} {\bibfnamefont {M.~T.}\ \bibnamefont {Hansen}}, \ and\ \bibinfo {author} {\bibfnamefont {S.~R.}\ \bibnamefont {Sharpe}},\ }\href {\doibase 10.1103/PhysRevD.95.074510} {\bibfield  {journal} {\bibinfo  {journal} {Phys. Rev.}\ }\textbf {\bibinfo {volume} {D95}},\ \bibinfo {pages} {074510} (\bibinfo {year} {2017})},\ \Eprint {http://arxiv.org/abs/1701.07465} {arXiv:1701.07465 [hep-lat]} \BibitemShut {NoStop}%
\bibitem [{\citenamefont {Brice\~no}\ \emph {et~al.}(2018)\citenamefont {Brice\~no}, \citenamefont {Hansen},\ and\ \citenamefont {Sharpe}}]{Briceno:2018mlh}%
  \BibitemOpen
  \bibfield  {author} {\bibinfo {author} {\bibfnamefont {R.~A.}\ \bibnamefont {Brice\~no}}, \bibinfo {author} {\bibfnamefont {M.~T.}\ \bibnamefont {Hansen}}, \ and\ \bibinfo {author} {\bibfnamefont {S.~R.}\ \bibnamefont {Sharpe}},\ }\href {\doibase 10.1103/PhysRevD.98.014506} {\bibfield  {journal} {\bibinfo  {journal} {Phys. Rev.}\ }\textbf {\bibinfo {volume} {D98}},\ \bibinfo {pages} {014506} (\bibinfo {year} {2018})},\ \Eprint {http://arxiv.org/abs/1803.04169} {arXiv:1803.04169 [hep-lat]} \BibitemShut {NoStop}%
\bibitem [{\citenamefont {Briceno}\ \emph {et~al.}(2019)\citenamefont {Briceno}, \citenamefont {Hansen},\ and\ \citenamefont {Sharpe}}]{Briceno:2018aml}%
  \BibitemOpen
  \bibfield  {author} {\bibinfo {author} {\bibfnamefont {R.~A.}\ \bibnamefont {Briceno}}, \bibinfo {author} {\bibfnamefont {M.~T.}\ \bibnamefont {Hansen}}, \ and\ \bibinfo {author} {\bibfnamefont {S.~R.}\ \bibnamefont {Sharpe}},\ }\href {\doibase 10.1103/PhysRevD.99.014516} {\bibfield  {journal} {\bibinfo  {journal} {Phys. Rev.}\ }\textbf {\bibinfo {volume} {D99}},\ \bibinfo {pages} {014516} (\bibinfo {year} {2019})},\ \Eprint {http://arxiv.org/abs/1810.01429} {arXiv:1810.01429 [hep-lat]} \BibitemShut {NoStop}%
\bibitem [{\citenamefont {Briceño}\ \emph {et~al.}(2019)\citenamefont {Briceño}, \citenamefont {Hansen}, \citenamefont {Sharpe},\ and\ \citenamefont {Szczepaniak}}]{Briceno:2019muc}%
  \BibitemOpen
  \bibfield  {author} {\bibinfo {author} {\bibfnamefont {R.~A.}\ \bibnamefont {Briceño}}, \bibinfo {author} {\bibfnamefont {M.~T.}\ \bibnamefont {Hansen}}, \bibinfo {author} {\bibfnamefont {S.~R.}\ \bibnamefont {Sharpe}}, \ and\ \bibinfo {author} {\bibfnamefont {A.~P.}\ \bibnamefont {Szczepaniak}},\ }\href {\doibase 10.1103/PhysRevD.100.054508} {\bibfield  {journal} {\bibinfo  {journal} {Phys. Rev. D}\ }\textbf {\bibinfo {volume} {100}},\ \bibinfo {pages} {054508} (\bibinfo {year} {2019})},\ \Eprint {http://arxiv.org/abs/1905.11188} {arXiv:1905.11188 [hep-lat]} \BibitemShut {NoStop}%
\bibitem [{\citenamefont {Blanton}\ \emph {et~al.}(2019)\citenamefont {Blanton}, \citenamefont {Romero-L{\'o}pez},\ and\ \citenamefont {Sharpe}}]{Blanton:2019igq}%
  \BibitemOpen
  \bibfield  {author} {\bibinfo {author} {\bibfnamefont {T.~D.}\ \bibnamefont {Blanton}}, \bibinfo {author} {\bibfnamefont {F.}~\bibnamefont {Romero-L{\'o}pez}}, \ and\ \bibinfo {author} {\bibfnamefont {S.~R.}\ \bibnamefont {Sharpe}},\ }\href {\doibase 10.1007/JHEP03(2019)106} {\bibfield  {journal} {\bibinfo  {journal} {JHEP}\ }\textbf {\bibinfo {volume} {03}},\ \bibinfo {pages} {106} (\bibinfo {year} {2019})},\ \Eprint {http://arxiv.org/abs/1901.07095} {arXiv:1901.07095 [hep-lat]} \BibitemShut {NoStop}%
\bibitem [{\citenamefont {Hansen}\ \emph {et~al.}(2020)\citenamefont {Hansen}, \citenamefont {Romero-L\'opez},\ and\ \citenamefont {Sharpe}}]{Hansen:2020zhy}%
  \BibitemOpen
  \bibfield  {author} {\bibinfo {author} {\bibfnamefont {M.~T.}\ \bibnamefont {Hansen}}, \bibinfo {author} {\bibfnamefont {F.}~\bibnamefont {Romero-L\'opez}}, \ and\ \bibinfo {author} {\bibfnamefont {S.~R.}\ \bibnamefont {Sharpe}},\ }\href {\doibase 10.1007/JHEP07(2020)047} {\bibfield  {journal} {\bibinfo  {journal} {JHEP}\ }\textbf {\bibinfo {volume} {07}},\ \bibinfo {pages} {047} (\bibinfo {year} {2020})},\ \bibinfo {note} {[Erratum: JHEP 02, 014 (2021)]},\ \Eprint {http://arxiv.org/abs/2003.10974} {arXiv:2003.10974 [hep-lat]} \BibitemShut {NoStop}%
\bibitem [{\citenamefont {Blanton}\ and\ \citenamefont {Sharpe}(2020)}]{Blanton:2020gha}%
  \BibitemOpen
  \bibfield  {author} {\bibinfo {author} {\bibfnamefont {T.~D.}\ \bibnamefont {Blanton}}\ and\ \bibinfo {author} {\bibfnamefont {S.~R.}\ \bibnamefont {Sharpe}},\ }\href {\doibase 10.1103/PhysRevD.102.054520} {\bibfield  {journal} {\bibinfo  {journal} {Phys. Rev. D}\ }\textbf {\bibinfo {volume} {102}},\ \bibinfo {pages} {054520} (\bibinfo {year} {2020})},\ \Eprint {http://arxiv.org/abs/2007.16188} {arXiv:2007.16188 [hep-lat]} \BibitemShut {NoStop}%
\bibitem [{\citenamefont {Hammer}\ \emph {et~al.}(2017{\natexlab{a}})\citenamefont {Hammer}, \citenamefont {Pang},\ and\ \citenamefont {Rusetsky}}]{Hammer:2017uqm}%
  \BibitemOpen
  \bibfield  {author} {\bibinfo {author} {\bibfnamefont {H.-W.}\ \bibnamefont {Hammer}}, \bibinfo {author} {\bibfnamefont {J.-Y.}\ \bibnamefont {Pang}}, \ and\ \bibinfo {author} {\bibfnamefont {A.}~\bibnamefont {Rusetsky}},\ }\href {\doibase 10.1007/JHEP09(2017)109} {\bibfield  {journal} {\bibinfo  {journal} {JHEP}\ }\textbf {\bibinfo {volume} {09}},\ \bibinfo {pages} {109} (\bibinfo {year} {2017}{\natexlab{a}})},\ \Eprint {http://arxiv.org/abs/1706.07700} {arXiv:1706.07700 [hep-lat]} \BibitemShut {NoStop}%
\bibitem [{\citenamefont {Hammer}\ \emph {et~al.}(2017{\natexlab{b}})\citenamefont {Hammer}, \citenamefont {Pang},\ and\ \citenamefont {Rusetsky}}]{Hammer:2017kms}%
  \BibitemOpen
  \bibfield  {author} {\bibinfo {author} {\bibfnamefont {H.~W.}\ \bibnamefont {Hammer}}, \bibinfo {author} {\bibfnamefont {J.~Y.}\ \bibnamefont {Pang}}, \ and\ \bibinfo {author} {\bibfnamefont {A.}~\bibnamefont {Rusetsky}},\ }\href {\doibase 10.1007/JHEP10(2017)115} {\bibfield  {journal} {\bibinfo  {journal} {JHEP}\ }\textbf {\bibinfo {volume} {10}},\ \bibinfo {pages} {115} (\bibinfo {year} {2017}{\natexlab{b}})},\ \Eprint {http://arxiv.org/abs/1707.02176} {arXiv:1707.02176 [hep-lat]} \BibitemShut {NoStop}%
\bibitem [{\citenamefont {Meng}\ \emph {et~al.}(2018)\citenamefont {Meng}, \citenamefont {Liu}, \citenamefont {Mei\ss{}ner},\ and\ \citenamefont {Rusetsky}}]{Meng:2017jgx}%
  \BibitemOpen
  \bibfield  {author} {\bibinfo {author} {\bibfnamefont {Y.}~\bibnamefont {Meng}}, \bibinfo {author} {\bibfnamefont {C.}~\bibnamefont {Liu}}, \bibinfo {author} {\bibfnamefont {U.-G.}\ \bibnamefont {Mei\ss{}ner}}, \ and\ \bibinfo {author} {\bibfnamefont {A.}~\bibnamefont {Rusetsky}},\ }\href {\doibase 10.1103/PhysRevD.98.014508} {\bibfield  {journal} {\bibinfo  {journal} {Phys. Rev. D}\ }\textbf {\bibinfo {volume} {98}},\ \bibinfo {pages} {014508} (\bibinfo {year} {2018})},\ \Eprint {http://arxiv.org/abs/1712.08464} {arXiv:1712.08464 [hep-lat]} \BibitemShut {NoStop}%
\bibitem [{\citenamefont {Pang}\ \emph {et~al.}(2019)\citenamefont {Pang}, \citenamefont {Wu}, \citenamefont {Hammer}, \citenamefont {Mei\ss{}ner},\ and\ \citenamefont {Rusetsky}}]{Pang:2019dfe}%
  \BibitemOpen
  \bibfield  {author} {\bibinfo {author} {\bibfnamefont {J.-Y.}\ \bibnamefont {Pang}}, \bibinfo {author} {\bibfnamefont {J.-J.}\ \bibnamefont {Wu}}, \bibinfo {author} {\bibfnamefont {H.~W.}\ \bibnamefont {Hammer}}, \bibinfo {author} {\bibfnamefont {U.-G.}\ \bibnamefont {Mei\ss{}ner}}, \ and\ \bibinfo {author} {\bibfnamefont {A.}~\bibnamefont {Rusetsky}},\ }\href {\doibase 10.1103/PhysRevD.99.074513} {\bibfield  {journal} {\bibinfo  {journal} {Phys. Rev. D}\ }\textbf {\bibinfo {volume} {99}},\ \bibinfo {pages} {074513} (\bibinfo {year} {2019})},\ \Eprint {http://arxiv.org/abs/1902.01111} {arXiv:1902.01111 [hep-lat]} \BibitemShut {NoStop}%
\bibitem [{\citenamefont {M\"uller}\ \emph {et~al.}(2022)\citenamefont {M\"uller}, \citenamefont {Pang}, \citenamefont {Rusetsky},\ and\ \citenamefont {Wu}}]{Muller:2021uur}%
  \BibitemOpen
  \bibfield  {author} {\bibinfo {author} {\bibfnamefont {F.}~\bibnamefont {M\"uller}}, \bibinfo {author} {\bibfnamefont {J.-Y.}\ \bibnamefont {Pang}}, \bibinfo {author} {\bibfnamefont {A.}~\bibnamefont {Rusetsky}}, \ and\ \bibinfo {author} {\bibfnamefont {J.-J.}\ \bibnamefont {Wu}},\ }\href {\doibase 10.1007/JHEP02(2022)158} {\bibfield  {journal} {\bibinfo  {journal} {JHEP}\ }\textbf {\bibinfo {volume} {02}},\ \bibinfo {pages} {158} (\bibinfo {year} {2022})},\ \Eprint {http://arxiv.org/abs/2110.09351} {arXiv:2110.09351 [hep-lat]} \BibitemShut {NoStop}%
\bibitem [{\citenamefont {Blanton}\ and\ \citenamefont {Sharpe}(2021{\natexlab{a}})}]{Blanton:2020gmf}%
  \BibitemOpen
  \bibfield  {author} {\bibinfo {author} {\bibfnamefont {T.~D.}\ \bibnamefont {Blanton}}\ and\ \bibinfo {author} {\bibfnamefont {S.~R.}\ \bibnamefont {Sharpe}},\ }\href {\doibase 10.1103/PhysRevD.103.054503} {\bibfield  {journal} {\bibinfo  {journal} {Phys. Rev. D}\ }\textbf {\bibinfo {volume} {103}},\ \bibinfo {pages} {054503} (\bibinfo {year} {2021}{\natexlab{a}})},\ \Eprint {http://arxiv.org/abs/2011.05520} {arXiv:2011.05520 [hep-lat]} \BibitemShut {NoStop}%
\bibitem [{\citenamefont {Blanton}\ and\ \citenamefont {Sharpe}(2021{\natexlab{b}})}]{Blanton:2021mih}%
  \BibitemOpen
  \bibfield  {author} {\bibinfo {author} {\bibfnamefont {T.~D.}\ \bibnamefont {Blanton}}\ and\ \bibinfo {author} {\bibfnamefont {S.~R.}\ \bibnamefont {Sharpe}},\ }\href {\doibase 10.1103/PhysRevD.104.034509} {\bibfield  {journal} {\bibinfo  {journal} {Phys. Rev. D}\ }\textbf {\bibinfo {volume} {104}},\ \bibinfo {pages} {034509} (\bibinfo {year} {2021}{\natexlab{b}})},\ \Eprint {http://arxiv.org/abs/2105.12094} {arXiv:2105.12094 [hep-lat]} \BibitemShut {NoStop}%
\bibitem [{\citenamefont {Hansen}\ \emph {et~al.}(2024)\citenamefont {Hansen}, \citenamefont {Romero-L\'opez},\ and\ \citenamefont {Sharpe}}]{Hansen:2024ffk}%
  \BibitemOpen
  \bibfield  {author} {\bibinfo {author} {\bibfnamefont {M.~T.}\ \bibnamefont {Hansen}}, \bibinfo {author} {\bibfnamefont {F.}~\bibnamefont {Romero-L\'opez}}, \ and\ \bibinfo {author} {\bibfnamefont {S.~R.}\ \bibnamefont {Sharpe}},\ }\href {\doibase 10.1007/JHEP06(2024)051} {\bibfield  {journal} {\bibinfo  {journal} {JHEP}\ }\textbf {\bibinfo {volume} {06}},\ \bibinfo {pages} {051} (\bibinfo {year} {2024})},\ \Eprint {http://arxiv.org/abs/2401.06609} {arXiv:2401.06609 [hep-lat]} \BibitemShut {NoStop}%
\bibitem [{\citenamefont {Raposo}\ and\ \citenamefont {Hansen}(2024)}]{Raposo:2023oru}%
  \BibitemOpen
  \bibfield  {author} {\bibinfo {author} {\bibfnamefont {A.~B.}\ \bibnamefont {Raposo}}\ and\ \bibinfo {author} {\bibfnamefont {M.~T.}\ \bibnamefont {Hansen}},\ }\href {\doibase 10.1007/JHEP08(2024)075} {\bibfield  {journal} {\bibinfo  {journal} {JHEP}\ }\textbf {\bibinfo {volume} {08}},\ \bibinfo {pages} {075} (\bibinfo {year} {2024})},\ \Eprint {http://arxiv.org/abs/2311.18793} {arXiv:2311.18793 [hep-lat]} \BibitemShut {NoStop}%
\bibitem [{\citenamefont {Raposo}\ \emph {et~al.}(2025)\citenamefont {Raposo}, \citenamefont {Brice{\~n}o}, \citenamefont {Hansen},\ and\ \citenamefont {Jackura}}]{Raposo:2025dkb}%
  \BibitemOpen
  \bibfield  {author} {\bibinfo {author} {\bibfnamefont {A.~B.}\ \bibnamefont {Raposo}}, \bibinfo {author} {\bibfnamefont {R.~A.}\ \bibnamefont {Brice{\~n}o}}, \bibinfo {author} {\bibfnamefont {M.~T.}\ \bibnamefont {Hansen}}, \ and\ \bibinfo {author} {\bibfnamefont {A.~W.}\ \bibnamefont {Jackura}},\ }\href {\doibase 10.1007/JHEP06(2025)186} {\bibfield  {journal} {\bibinfo  {journal} {JHEP}\ }\textbf {\bibinfo {volume} {06}},\ \bibinfo {pages} {186} (\bibinfo {year} {2025})},\ \Eprint {http://arxiv.org/abs/2502.19375} {arXiv:2502.19375 [hep-lat]} \BibitemShut {NoStop}%
\bibitem [{\citenamefont {Detmold}\ \emph {et~al.}(2008)\citenamefont {Detmold}, \citenamefont {Savage}, \citenamefont {Torok}, \citenamefont {Beane}, \citenamefont {Luu}, \citenamefont {Orginos},\ and\ \citenamefont {Parreno}}]{Detmold:2008fn}%
  \BibitemOpen
  \bibfield  {author} {\bibinfo {author} {\bibfnamefont {W.}~\bibnamefont {Detmold}}, \bibinfo {author} {\bibfnamefont {M.~J.}\ \bibnamefont {Savage}}, \bibinfo {author} {\bibfnamefont {A.}~\bibnamefont {Torok}}, \bibinfo {author} {\bibfnamefont {S.~R.}\ \bibnamefont {Beane}}, \bibinfo {author} {\bibfnamefont {T.~C.}\ \bibnamefont {Luu}}, \bibinfo {author} {\bibfnamefont {K.}~\bibnamefont {Orginos}}, \ and\ \bibinfo {author} {\bibfnamefont {A.}~\bibnamefont {Parreno}},\ }\href {\doibase 10.1103/PhysRevD.78.014507} {\bibfield  {journal} {\bibinfo  {journal} {Phys. Rev. D}\ }\textbf {\bibinfo {volume} {78}},\ \bibinfo {pages} {014507} (\bibinfo {year} {2008})},\ \Eprint {http://arxiv.org/abs/0803.2728} {arXiv:0803.2728 [hep-lat]} \BibitemShut {NoStop}%
\bibitem [{\citenamefont {Culver}\ \emph {et~al.}(2020)\citenamefont {Culver}, \citenamefont {Mai}, \citenamefont {Brett}, \citenamefont {Alexandru},\ and\ \citenamefont {D\"oring}}]{Culver:2019vvu}%
  \BibitemOpen
  \bibfield  {author} {\bibinfo {author} {\bibfnamefont {C.}~\bibnamefont {Culver}}, \bibinfo {author} {\bibfnamefont {M.}~\bibnamefont {Mai}}, \bibinfo {author} {\bibfnamefont {R.}~\bibnamefont {Brett}}, \bibinfo {author} {\bibfnamefont {A.}~\bibnamefont {Alexandru}}, \ and\ \bibinfo {author} {\bibfnamefont {M.}~\bibnamefont {D\"oring}},\ }\href {\doibase 10.1103/PhysRevD.101.114507} {\bibfield  {journal} {\bibinfo  {journal} {Phys. Rev. D}\ }\textbf {\bibinfo {volume} {101}},\ \bibinfo {pages} {114507} (\bibinfo {year} {2020})},\ \Eprint {http://arxiv.org/abs/1911.09047} {arXiv:1911.09047 [hep-lat]} \BibitemShut {NoStop}%
\bibitem [{\citenamefont {Alexandru}\ \emph {et~al.}(2020)\citenamefont {Alexandru}, \citenamefont {Brett}, \citenamefont {Culver}, \citenamefont {D\"oring}, \citenamefont {Guo}, \citenamefont {Lee},\ and\ \citenamefont {Mai}}]{Alexandru:2020xqf}%
  \BibitemOpen
  \bibfield  {author} {\bibinfo {author} {\bibfnamefont {A.}~\bibnamefont {Alexandru}}, \bibinfo {author} {\bibfnamefont {R.}~\bibnamefont {Brett}}, \bibinfo {author} {\bibfnamefont {C.}~\bibnamefont {Culver}}, \bibinfo {author} {\bibfnamefont {M.}~\bibnamefont {D\"oring}}, \bibinfo {author} {\bibfnamefont {D.}~\bibnamefont {Guo}}, \bibinfo {author} {\bibfnamefont {F.~X.}\ \bibnamefont {Lee}}, \ and\ \bibinfo {author} {\bibfnamefont {M.}~\bibnamefont {Mai}},\ }\href {\doibase 10.1103/PhysRevD.102.114523} {\bibfield  {journal} {\bibinfo  {journal} {Phys. Rev. D}\ }\textbf {\bibinfo {volume} {102}},\ \bibinfo {pages} {114523} (\bibinfo {year} {2020})},\ \Eprint {http://arxiv.org/abs/2009.12358} {arXiv:2009.12358 [hep-lat]} \BibitemShut {NoStop}%
\bibitem [{\citenamefont {Hansen}\ \emph {et~al.}(2021)\citenamefont {Hansen}, \citenamefont {Brice\~no}, \citenamefont {Edwards}, \citenamefont {Thomas},\ and\ \citenamefont {Wilson}}]{Hansen:2020otl}%
  \BibitemOpen
  \bibfield  {author} {\bibinfo {author} {\bibfnamefont {M.~T.}\ \bibnamefont {Hansen}}, \bibinfo {author} {\bibfnamefont {R.~A.}\ \bibnamefont {Brice\~no}}, \bibinfo {author} {\bibfnamefont {R.~G.}\ \bibnamefont {Edwards}}, \bibinfo {author} {\bibfnamefont {C.~E.}\ \bibnamefont {Thomas}}, \ and\ \bibinfo {author} {\bibfnamefont {D.~J.}\ \bibnamefont {Wilson}} (\bibinfo {collaboration} {Hadron Spectrum}),\ }\href {\doibase 10.1103/PhysRevLett.126.012001} {\bibfield  {journal} {\bibinfo  {journal} {Phys. Rev. Lett.}\ }\textbf {\bibinfo {volume} {126}},\ \bibinfo {pages} {012001} (\bibinfo {year} {2021})},\ \Eprint {http://arxiv.org/abs/2009.04931} {arXiv:2009.04931 [hep-lat]} \BibitemShut {NoStop}%
\bibitem [{\citenamefont {Draper}\ \emph {et~al.}(2023{\natexlab{a}})\citenamefont {Draper}, \citenamefont {Hanlon}, \citenamefont {H\"orz}, \citenamefont {Morningstar}, \citenamefont {Romero-L\'opez},\ and\ \citenamefont {Sharpe}}]{Draper:2023boj}%
  \BibitemOpen
  \bibfield  {author} {\bibinfo {author} {\bibfnamefont {Z.~T.}\ \bibnamefont {Draper}}, \bibinfo {author} {\bibfnamefont {A.~D.}\ \bibnamefont {Hanlon}}, \bibinfo {author} {\bibfnamefont {B.}~\bibnamefont {H\"orz}}, \bibinfo {author} {\bibfnamefont {C.}~\bibnamefont {Morningstar}}, \bibinfo {author} {\bibfnamefont {F.}~\bibnamefont {Romero-L\'opez}}, \ and\ \bibinfo {author} {\bibfnamefont {S.~R.}\ \bibnamefont {Sharpe}},\ }\href {\doibase 10.1007/JHEP05(2023)137} {\bibfield  {journal} {\bibinfo  {journal} {JHEP}\ }\textbf {\bibinfo {volume} {05}},\ \bibinfo {pages} {137} (\bibinfo {year} {2023}{\natexlab{a}})},\ \Eprint {http://arxiv.org/abs/2302.13587} {arXiv:2302.13587 [hep-lat]} \BibitemShut {NoStop}%
\bibitem [{\citenamefont {Dudek}\ \emph {et~al.}(2011)\citenamefont {Dudek}, \citenamefont {Edwards}, \citenamefont {Peardon}, \citenamefont {Richards},\ and\ \citenamefont {Thomas}}]{Dudek:2010ew}%
  \BibitemOpen
  \bibfield  {author} {\bibinfo {author} {\bibfnamefont {J.~J.}\ \bibnamefont {Dudek}}, \bibinfo {author} {\bibfnamefont {R.~G.}\ \bibnamefont {Edwards}}, \bibinfo {author} {\bibfnamefont {M.~J.}\ \bibnamefont {Peardon}}, \bibinfo {author} {\bibfnamefont {D.~G.}\ \bibnamefont {Richards}}, \ and\ \bibinfo {author} {\bibfnamefont {C.~E.}\ \bibnamefont {Thomas}},\ }\href {\doibase 10.1103/PhysRevD.83.071504} {\bibfield  {journal} {\bibinfo  {journal} {Phys. Rev.}\ }\textbf {\bibinfo {volume} {D83}},\ \bibinfo {pages} {071504} (\bibinfo {year} {2011})},\ \Eprint {http://arxiv.org/abs/1011.6352} {arXiv:1011.6352 [hep-ph]} \BibitemShut {NoStop}%
\bibitem [{\citenamefont {Pelissier}\ and\ \citenamefont {Alexandru}(2013)}]{Pelissier:2012pi}%
  \BibitemOpen
  \bibfield  {author} {\bibinfo {author} {\bibfnamefont {C.}~\bibnamefont {Pelissier}}\ and\ \bibinfo {author} {\bibfnamefont {A.}~\bibnamefont {Alexandru}},\ }\href {\doibase 10.1103/PhysRevD.87.014503} {\bibfield  {journal} {\bibinfo  {journal} {Phys. Rev.}\ }\textbf {\bibinfo {volume} {D87}},\ \bibinfo {pages} {014503} (\bibinfo {year} {2013})},\ \Eprint {http://arxiv.org/abs/1211.0092} {arXiv:1211.0092 [hep-lat]} \BibitemShut {NoStop}%
\bibitem [{\citenamefont {Dudek}\ \emph {et~al.}(2013)\citenamefont {Dudek}, \citenamefont {Edwards},\ and\ \citenamefont {Thomas}}]{Dudek:2012xn}%
  \BibitemOpen
  \bibfield  {author} {\bibinfo {author} {\bibfnamefont {J.~J.}\ \bibnamefont {Dudek}}, \bibinfo {author} {\bibfnamefont {R.~G.}\ \bibnamefont {Edwards}}, \ and\ \bibinfo {author} {\bibfnamefont {C.~E.}\ \bibnamefont {Thomas}} (\bibinfo {collaboration} {Hadron Spectrum}),\ }\href {\doibase 10.1103/PhysRevD.87.034505, 10.1103/PhysRevD.90.099902} {\bibfield  {journal} {\bibinfo  {journal} {Phys. Rev.}\ }\textbf {\bibinfo {volume} {D87}},\ \bibinfo {pages} {034505} (\bibinfo {year} {2013})},\ \bibinfo {note} {[Erratum: Phys. Rev.D90,no.9,099902(2014)]},\ \Eprint {http://arxiv.org/abs/1212.0830} {arXiv:1212.0830 [hep-ph]} \BibitemShut {NoStop}%
\bibitem [{\citenamefont {Liu}\ \emph {et~al.}(2013)\citenamefont {Liu}, \citenamefont {Orginos}, \citenamefont {Guo}, \citenamefont {Hanhart},\ and\ \citenamefont {Meissner}}]{Liu:2012zya}%
  \BibitemOpen
  \bibfield  {author} {\bibinfo {author} {\bibfnamefont {L.}~\bibnamefont {Liu}}, \bibinfo {author} {\bibfnamefont {K.}~\bibnamefont {Orginos}}, \bibinfo {author} {\bibfnamefont {F.-K.}\ \bibnamefont {Guo}}, \bibinfo {author} {\bibfnamefont {C.}~\bibnamefont {Hanhart}}, \ and\ \bibinfo {author} {\bibfnamefont {U.-G.}\ \bibnamefont {Meissner}},\ }\href {\doibase 10.1103/PhysRevD.87.014508} {\bibfield  {journal} {\bibinfo  {journal} {Phys. Rev.}\ }\textbf {\bibinfo {volume} {D87}},\ \bibinfo {pages} {014508} (\bibinfo {year} {2013})},\ \Eprint {http://arxiv.org/abs/1208.4535} {arXiv:1208.4535 [hep-lat]} \BibitemShut {NoStop}%
\bibitem [{\citenamefont {Wilson}\ \emph {et~al.}(2015{\natexlab{a}})\citenamefont {Wilson}, \citenamefont {Dudek}, \citenamefont {Edwards},\ and\ \citenamefont {Thomas}}]{Wilson:2014cna}%
  \BibitemOpen
  \bibfield  {author} {\bibinfo {author} {\bibfnamefont {D.~J.}\ \bibnamefont {Wilson}}, \bibinfo {author} {\bibfnamefont {J.~J.}\ \bibnamefont {Dudek}}, \bibinfo {author} {\bibfnamefont {R.~G.}\ \bibnamefont {Edwards}}, \ and\ \bibinfo {author} {\bibfnamefont {C.~E.}\ \bibnamefont {Thomas}},\ }\href {\doibase 10.1103/PhysRevD.91.054008} {\bibfield  {journal} {\bibinfo  {journal} {Phys. Rev.}\ }\textbf {\bibinfo {volume} {D91}},\ \bibinfo {pages} {054008} (\bibinfo {year} {2015}{\natexlab{a}})},\ \Eprint {http://arxiv.org/abs/1411.2004} {arXiv:1411.2004 [hep-ph]} \BibitemShut {NoStop}%
\bibitem [{\citenamefont {Dudek}\ \emph {et~al.}(2014)\citenamefont {Dudek}, \citenamefont {Edwards}, \citenamefont {Thomas},\ and\ \citenamefont {Wilson}}]{Dudek:2014qha}%
  \BibitemOpen
  \bibfield  {author} {\bibinfo {author} {\bibfnamefont {J.~J.}\ \bibnamefont {Dudek}}, \bibinfo {author} {\bibfnamefont {R.~G.}\ \bibnamefont {Edwards}}, \bibinfo {author} {\bibfnamefont {C.~E.}\ \bibnamefont {Thomas}}, \ and\ \bibinfo {author} {\bibfnamefont {D.~J.}\ \bibnamefont {Wilson}} (\bibinfo {collaboration} {Hadron Spectrum}),\ }\href {\doibase 10.1103/PhysRevLett.113.182001} {\bibfield  {journal} {\bibinfo  {journal} {Phys. Rev. Lett.}\ }\textbf {\bibinfo {volume} {113}},\ \bibinfo {pages} {182001} (\bibinfo {year} {2014})},\ \Eprint {http://arxiv.org/abs/1406.4158} {arXiv:1406.4158 [hep-ph]} \BibitemShut {NoStop}%
\bibitem [{\citenamefont {Lang}\ \emph {et~al.}(2015)\citenamefont {Lang}, \citenamefont {Mohler}, \citenamefont {Prelovsek},\ and\ \citenamefont {Woloshyn}}]{Lang:2015hza}%
  \BibitemOpen
  \bibfield  {author} {\bibinfo {author} {\bibfnamefont {C.~B.}\ \bibnamefont {Lang}}, \bibinfo {author} {\bibfnamefont {D.}~\bibnamefont {Mohler}}, \bibinfo {author} {\bibfnamefont {S.}~\bibnamefont {Prelovsek}}, \ and\ \bibinfo {author} {\bibfnamefont {R.~M.}\ \bibnamefont {Woloshyn}},\ }\href {\doibase 10.1016/j.physletb.2015.08.038} {\bibfield  {journal} {\bibinfo  {journal} {Phys. Lett.}\ }\textbf {\bibinfo {volume} {B750}},\ \bibinfo {pages} {17} (\bibinfo {year} {2015})},\ \Eprint {http://arxiv.org/abs/1501.01646} {arXiv:1501.01646 [hep-lat]} \BibitemShut {NoStop}%
\bibitem [{\citenamefont {Wilson}\ \emph {et~al.}(2015{\natexlab{b}})\citenamefont {Wilson}, \citenamefont {Brice\~no}, \citenamefont {Dudek}, \citenamefont {Edwards},\ and\ \citenamefont {Thomas}}]{Wilson:2015dqa}%
  \BibitemOpen
  \bibfield  {author} {\bibinfo {author} {\bibfnamefont {D.~J.}\ \bibnamefont {Wilson}}, \bibinfo {author} {\bibfnamefont {R.~A.}\ \bibnamefont {Brice\~no}}, \bibinfo {author} {\bibfnamefont {J.~J.}\ \bibnamefont {Dudek}}, \bibinfo {author} {\bibfnamefont {R.~G.}\ \bibnamefont {Edwards}}, \ and\ \bibinfo {author} {\bibfnamefont {C.~E.}\ \bibnamefont {Thomas}},\ }\href {\doibase 10.1103/PhysRevD.92.094502} {\bibfield  {journal} {\bibinfo  {journal} {Phys. Rev.}\ }\textbf {\bibinfo {volume} {D92}},\ \bibinfo {pages} {094502} (\bibinfo {year} {2015}{\natexlab{b}})},\ \Eprint {http://arxiv.org/abs/1507.02599} {arXiv:1507.02599 [hep-ph]} \BibitemShut {NoStop}%
\bibitem [{\citenamefont {Dudek}\ \emph {et~al.}(2016)\citenamefont {Dudek}, \citenamefont {Edwards},\ and\ \citenamefont {Wilson}}]{Dudek:2016cru}%
  \BibitemOpen
  \bibfield  {author} {\bibinfo {author} {\bibfnamefont {J.~J.}\ \bibnamefont {Dudek}}, \bibinfo {author} {\bibfnamefont {R.~G.}\ \bibnamefont {Edwards}}, \ and\ \bibinfo {author} {\bibfnamefont {D.~J.}\ \bibnamefont {Wilson}} (\bibinfo {collaboration} {Hadron Spectrum}),\ }\href {\doibase 10.1103/PhysRevD.93.094506} {\bibfield  {journal} {\bibinfo  {journal} {Phys. Rev.}\ }\textbf {\bibinfo {volume} {D93}},\ \bibinfo {pages} {094506} (\bibinfo {year} {2016})},\ \Eprint {http://arxiv.org/abs/1602.05122} {arXiv:1602.05122 [hep-ph]} \BibitemShut {NoStop}%
\bibitem [{\citenamefont {Brice\~no}\ \emph {et~al.}(2017)\citenamefont {Brice\~no}, \citenamefont {Dudek}, \citenamefont {Edwards},\ and\ \citenamefont {Wilson}}]{Briceno:2016mjc}%
  \BibitemOpen
  \bibfield  {author} {\bibinfo {author} {\bibfnamefont {R.~A.}\ \bibnamefont {Brice\~no}}, \bibinfo {author} {\bibfnamefont {J.~J.}\ \bibnamefont {Dudek}}, \bibinfo {author} {\bibfnamefont {R.~G.}\ \bibnamefont {Edwards}}, \ and\ \bibinfo {author} {\bibfnamefont {D.~J.}\ \bibnamefont {Wilson}},\ }\href {\doibase 10.1103/PhysRevLett.118.022002} {\bibfield  {journal} {\bibinfo  {journal} {Phys. Rev. Lett.}\ }\textbf {\bibinfo {volume} {118}},\ \bibinfo {pages} {022002} (\bibinfo {year} {2017})},\ \Eprint {http://arxiv.org/abs/1607.05900} {arXiv:1607.05900 [hep-ph]} \BibitemShut {NoStop}%
\bibitem [{\citenamefont {Moir}\ \emph {et~al.}(2016)\citenamefont {Moir}, \citenamefont {Peardon}, \citenamefont {Ryan}, \citenamefont {Thomas},\ and\ \citenamefont {Wilson}}]{Moir:2016srx}%
  \BibitemOpen
  \bibfield  {author} {\bibinfo {author} {\bibfnamefont {G.}~\bibnamefont {Moir}}, \bibinfo {author} {\bibfnamefont {M.}~\bibnamefont {Peardon}}, \bibinfo {author} {\bibfnamefont {S.~M.}\ \bibnamefont {Ryan}}, \bibinfo {author} {\bibfnamefont {C.~E.}\ \bibnamefont {Thomas}}, \ and\ \bibinfo {author} {\bibfnamefont {D.~J.}\ \bibnamefont {Wilson}},\ }\href {\doibase 10.1007/JHEP10(2016)011} {\bibfield  {journal} {\bibinfo  {journal} {JHEP}\ }\textbf {\bibinfo {volume} {10}},\ \bibinfo {pages} {011} (\bibinfo {year} {2016})},\ \Eprint {http://arxiv.org/abs/1607.07093} {arXiv:1607.07093 [hep-lat]} \BibitemShut {NoStop}%
\bibitem [{\citenamefont {Bulava}\ \emph {et~al.}(2016)\citenamefont {Bulava}, \citenamefont {Fahy}, \citenamefont {Horz}, \citenamefont {Juge}, \citenamefont {Morningstar},\ and\ \citenamefont {Wong}}]{Bulava:2016mks}%
  \BibitemOpen
  \bibfield  {author} {\bibinfo {author} {\bibfnamefont {J.}~\bibnamefont {Bulava}}, \bibinfo {author} {\bibfnamefont {B.}~\bibnamefont {Fahy}}, \bibinfo {author} {\bibfnamefont {B.}~\bibnamefont {Horz}}, \bibinfo {author} {\bibfnamefont {K.~J.}\ \bibnamefont {Juge}}, \bibinfo {author} {\bibfnamefont {C.}~\bibnamefont {Morningstar}}, \ and\ \bibinfo {author} {\bibfnamefont {C.~H.}\ \bibnamefont {Wong}},\ }\href {\doibase 10.1016/j.nuclphysb.2016.07.024} {\bibfield  {journal} {\bibinfo  {journal} {Nucl. Phys.}\ }\textbf {\bibinfo {volume} {B910}},\ \bibinfo {pages} {842} (\bibinfo {year} {2016})},\ \Eprint {http://arxiv.org/abs/1604.05593} {arXiv:1604.05593 [hep-lat]} \BibitemShut {NoStop}%
\bibitem [{\citenamefont {Hu}\ \emph {et~al.}(2016)\citenamefont {Hu}, \citenamefont {Molina}, \citenamefont {Doring},\ and\ \citenamefont {Alexandru}}]{Hu:2016shf}%
  \BibitemOpen
  \bibfield  {author} {\bibinfo {author} {\bibfnamefont {B.}~\bibnamefont {Hu}}, \bibinfo {author} {\bibfnamefont {R.}~\bibnamefont {Molina}}, \bibinfo {author} {\bibfnamefont {M.}~\bibnamefont {Doring}}, \ and\ \bibinfo {author} {\bibfnamefont {A.}~\bibnamefont {Alexandru}},\ }\href {\doibase 10.1103/PhysRevLett.117.122001} {\bibfield  {journal} {\bibinfo  {journal} {Phys. Rev. Lett.}\ }\textbf {\bibinfo {volume} {117}},\ \bibinfo {pages} {122001} (\bibinfo {year} {2016})},\ \Eprint {http://arxiv.org/abs/1605.04823} {arXiv:1605.04823 [hep-lat]} \BibitemShut {NoStop}%
\bibitem [{\citenamefont {Alexandrou}\ \emph {et~al.}(2017)\citenamefont {Alexandrou}, \citenamefont {Leskovec}, \citenamefont {Meinel}, \citenamefont {Negele}, \citenamefont {Paul}, \citenamefont {Petschlies}, \citenamefont {Pochinsky}, \citenamefont {Rendon},\ and\ \citenamefont {Syritsyn}}]{Alexandrou:2017mpi}%
  \BibitemOpen
  \bibfield  {author} {\bibinfo {author} {\bibfnamefont {C.}~\bibnamefont {Alexandrou}}, \bibinfo {author} {\bibfnamefont {L.}~\bibnamefont {Leskovec}}, \bibinfo {author} {\bibfnamefont {S.}~\bibnamefont {Meinel}}, \bibinfo {author} {\bibfnamefont {J.}~\bibnamefont {Negele}}, \bibinfo {author} {\bibfnamefont {S.}~\bibnamefont {Paul}}, \bibinfo {author} {\bibfnamefont {M.}~\bibnamefont {Petschlies}}, \bibinfo {author} {\bibfnamefont {A.}~\bibnamefont {Pochinsky}}, \bibinfo {author} {\bibfnamefont {G.}~\bibnamefont {Rendon}}, \ and\ \bibinfo {author} {\bibfnamefont {S.}~\bibnamefont {Syritsyn}},\ }\href {\doibase 10.1103/PhysRevD.96.034525} {\bibfield  {journal} {\bibinfo  {journal} {Phys. Rev.}\ }\textbf {\bibinfo {volume} {D96}},\ \bibinfo {pages} {034525} (\bibinfo {year} {2017})},\ \Eprint {http://arxiv.org/abs/1704.05439} {arXiv:1704.05439 [hep-lat]} \BibitemShut {NoStop}%
\bibitem [{\citenamefont {Bali}\ \emph {et~al.}(2017)\citenamefont {Bali}, \citenamefont {Collins}, \citenamefont {Cox},\ and\ \citenamefont {Schäfer}}]{Bali:2017pdv}%
  \BibitemOpen
  \bibfield  {author} {\bibinfo {author} {\bibfnamefont {G.~S.}\ \bibnamefont {Bali}}, \bibinfo {author} {\bibfnamefont {S.}~\bibnamefont {Collins}}, \bibinfo {author} {\bibfnamefont {A.}~\bibnamefont {Cox}}, \ and\ \bibinfo {author} {\bibfnamefont {A.}~\bibnamefont {Schäfer}},\ }\href {\doibase 10.1103/PhysRevD.96.074501} {\bibfield  {journal} {\bibinfo  {journal} {Phys. Rev.}\ }\textbf {\bibinfo {volume} {D96}},\ \bibinfo {pages} {074501} (\bibinfo {year} {2017})},\ \Eprint {http://arxiv.org/abs/1706.01247} {arXiv:1706.01247 [hep-lat]} \BibitemShut {NoStop}%
\bibitem [{\citenamefont {Wagman}\ \emph {et~al.}(2017)\citenamefont {Wagman}, \citenamefont {Winter}, \citenamefont {Chang}, \citenamefont {Davoudi}, \citenamefont {Detmold}, \citenamefont {Orginos}, \citenamefont {Savage},\ and\ \citenamefont {Shanahan}}]{Wagman:2017tmp}%
  \BibitemOpen
  \bibfield  {author} {\bibinfo {author} {\bibfnamefont {M.~L.}\ \bibnamefont {Wagman}}, \bibinfo {author} {\bibfnamefont {F.}~\bibnamefont {Winter}}, \bibinfo {author} {\bibfnamefont {E.}~\bibnamefont {Chang}}, \bibinfo {author} {\bibfnamefont {Z.}~\bibnamefont {Davoudi}}, \bibinfo {author} {\bibfnamefont {W.}~\bibnamefont {Detmold}}, \bibinfo {author} {\bibfnamefont {K.}~\bibnamefont {Orginos}}, \bibinfo {author} {\bibfnamefont {M.~J.}\ \bibnamefont {Savage}}, \ and\ \bibinfo {author} {\bibfnamefont {P.~E.}\ \bibnamefont {Shanahan}},\ }\href {\doibase 10.1103/PhysRevD.96.114510} {\bibfield  {journal} {\bibinfo  {journal} {Phys. Rev.}\ }\textbf {\bibinfo {volume} {D96}},\ \bibinfo {pages} {114510} (\bibinfo {year} {2017})},\ \Eprint {http://arxiv.org/abs/1706.06550} {arXiv:1706.06550 [hep-lat]} \BibitemShut {NoStop}%
\bibitem [{\citenamefont {Andersen}\ \emph {et~al.}(2018)\citenamefont {Andersen}, \citenamefont {Bulava}, \citenamefont {Horz},\ and\ \citenamefont {Morningstar}}]{Andersen:2017una}%
  \BibitemOpen
  \bibfield  {author} {\bibinfo {author} {\bibfnamefont {C.~W.}\ \bibnamefont {Andersen}}, \bibinfo {author} {\bibfnamefont {J.}~\bibnamefont {Bulava}}, \bibinfo {author} {\bibfnamefont {B.}~\bibnamefont {Horz}}, \ and\ \bibinfo {author} {\bibfnamefont {C.}~\bibnamefont {Morningstar}},\ }\href {\doibase 10.1103/PhysRevD.97.014506} {\bibfield  {journal} {\bibinfo  {journal} {Phys. Rev.}\ }\textbf {\bibinfo {volume} {D97}},\ \bibinfo {pages} {014506} (\bibinfo {year} {2018})},\ \Eprint {http://arxiv.org/abs/1710.01557} {arXiv:1710.01557 [hep-lat]} \BibitemShut {NoStop}%
\bibitem [{\citenamefont {Briceno}\ \emph {et~al.}(2018{\natexlab{b}})\citenamefont {Briceno}, \citenamefont {Dudek}, \citenamefont {Edwards},\ and\ \citenamefont {Wilson}}]{Briceno:2017qmb}%
  \BibitemOpen
  \bibfield  {author} {\bibinfo {author} {\bibfnamefont {R.~A.}\ \bibnamefont {Briceno}}, \bibinfo {author} {\bibfnamefont {J.~J.}\ \bibnamefont {Dudek}}, \bibinfo {author} {\bibfnamefont {R.~G.}\ \bibnamefont {Edwards}}, \ and\ \bibinfo {author} {\bibfnamefont {D.~J.}\ \bibnamefont {Wilson}},\ }\href {\doibase 10.1103/PhysRevD.97.054513} {\bibfield  {journal} {\bibinfo  {journal} {Phys. Rev.}\ }\textbf {\bibinfo {volume} {D97}},\ \bibinfo {pages} {054513} (\bibinfo {year} {2018}{\natexlab{b}})},\ \Eprint {http://arxiv.org/abs/1708.06667} {arXiv:1708.06667 [hep-lat]} \BibitemShut {NoStop}%
\bibitem [{\citenamefont {Woss}\ \emph {et~al.}(2018)\citenamefont {Woss}, \citenamefont {Thomas}, \citenamefont {Dudek}, \citenamefont {Edwards},\ and\ \citenamefont {Wilson}}]{Woss:2018irj}%
  \BibitemOpen
  \bibfield  {author} {\bibinfo {author} {\bibfnamefont {A.}~\bibnamefont {Woss}}, \bibinfo {author} {\bibfnamefont {C.~E.}\ \bibnamefont {Thomas}}, \bibinfo {author} {\bibfnamefont {J.~J.}\ \bibnamefont {Dudek}}, \bibinfo {author} {\bibfnamefont {R.~G.}\ \bibnamefont {Edwards}}, \ and\ \bibinfo {author} {\bibfnamefont {D.~J.}\ \bibnamefont {Wilson}},\ }\href {\doibase 10.1007/JHEP07(2018)043} {\bibfield  {journal} {\bibinfo  {journal} {JHEP}\ }\textbf {\bibinfo {volume} {07}},\ \bibinfo {pages} {043} (\bibinfo {year} {2018})},\ \Eprint {http://arxiv.org/abs/1802.05580} {arXiv:1802.05580 [hep-lat]} \BibitemShut {NoStop}%
\bibitem [{\citenamefont {Brett}\ \emph {et~al.}(2018)\citenamefont {Brett}, \citenamefont {Bulava}, \citenamefont {Fallica}, \citenamefont {Hanlon}, \citenamefont {Horz},\ and\ \citenamefont {Morningstar}}]{Brett:2018jqw}%
  \BibitemOpen
  \bibfield  {author} {\bibinfo {author} {\bibfnamefont {R.}~\bibnamefont {Brett}}, \bibinfo {author} {\bibfnamefont {J.}~\bibnamefont {Bulava}}, \bibinfo {author} {\bibfnamefont {J.}~\bibnamefont {Fallica}}, \bibinfo {author} {\bibfnamefont {A.}~\bibnamefont {Hanlon}}, \bibinfo {author} {\bibfnamefont {B.}~\bibnamefont {Horz}}, \ and\ \bibinfo {author} {\bibfnamefont {C.}~\bibnamefont {Morningstar}},\ }\href {\doibase 10.1016/j.nuclphysb.2018.05.008} {\bibfield  {journal} {\bibinfo  {journal} {Nucl. Phys.}\ }\textbf {\bibinfo {volume} {B932}},\ \bibinfo {pages} {29} (\bibinfo {year} {2018})},\ \Eprint {http://arxiv.org/abs/1802.03100} {arXiv:1802.03100 [hep-lat]} \BibitemShut {NoStop}%
\bibitem [{\citenamefont {Mai}\ \emph {et~al.}(2019)\citenamefont {Mai}, \citenamefont {Culver}, \citenamefont {Alexandru}, \citenamefont {D\"oring},\ and\ \citenamefont {Lee}}]{Mai:2019pqr}%
  \BibitemOpen
  \bibfield  {author} {\bibinfo {author} {\bibfnamefont {M.}~\bibnamefont {Mai}}, \bibinfo {author} {\bibfnamefont {C.}~\bibnamefont {Culver}}, \bibinfo {author} {\bibfnamefont {A.}~\bibnamefont {Alexandru}}, \bibinfo {author} {\bibfnamefont {M.}~\bibnamefont {D\"oring}}, \ and\ \bibinfo {author} {\bibfnamefont {F.~X.}\ \bibnamefont {Lee}},\ }\href {\doibase 10.1103/PhysRevD.100.114514} {\bibfield  {journal} {\bibinfo  {journal} {Phys. Rev. D}\ }\textbf {\bibinfo {volume} {100}},\ \bibinfo {pages} {114514} (\bibinfo {year} {2019})},\ \Eprint {http://arxiv.org/abs/1908.01847} {arXiv:1908.01847 [hep-lat]} \BibitemShut {NoStop}%
\bibitem [{\citenamefont {Woss}\ \emph {et~al.}(2019)\citenamefont {Woss}, \citenamefont {Thomas}, \citenamefont {Dudek}, \citenamefont {Edwards},\ and\ \citenamefont {Wilson}}]{Woss:2019hse}%
  \BibitemOpen
  \bibfield  {author} {\bibinfo {author} {\bibfnamefont {A.~J.}\ \bibnamefont {Woss}}, \bibinfo {author} {\bibfnamefont {C.~E.}\ \bibnamefont {Thomas}}, \bibinfo {author} {\bibfnamefont {J.~J.}\ \bibnamefont {Dudek}}, \bibinfo {author} {\bibfnamefont {R.~G.}\ \bibnamefont {Edwards}}, \ and\ \bibinfo {author} {\bibfnamefont {D.~J.}\ \bibnamefont {Wilson}},\ }\href {\doibase 10.1103/PhysRevD.100.054506} {\bibfield  {journal} {\bibinfo  {journal} {Phys. Rev. D}\ }\textbf {\bibinfo {volume} {100}},\ \bibinfo {pages} {054506} (\bibinfo {year} {2019})},\ \Eprint {http://arxiv.org/abs/1904.04136} {arXiv:1904.04136 [hep-lat]} \BibitemShut {NoStop}%
\bibitem [{\citenamefont {Wilson}\ \emph {et~al.}(2019)\citenamefont {Wilson}, \citenamefont {Briceno}, \citenamefont {Dudek}, \citenamefont {Edwards},\ and\ \citenamefont {Thomas}}]{Wilson:2019wfr}%
  \BibitemOpen
  \bibfield  {author} {\bibinfo {author} {\bibfnamefont {D.~J.}\ \bibnamefont {Wilson}}, \bibinfo {author} {\bibfnamefont {R.~A.}\ \bibnamefont {Briceno}}, \bibinfo {author} {\bibfnamefont {J.~J.}\ \bibnamefont {Dudek}}, \bibinfo {author} {\bibfnamefont {R.~G.}\ \bibnamefont {Edwards}}, \ and\ \bibinfo {author} {\bibfnamefont {C.~E.}\ \bibnamefont {Thomas}},\ }\href {\doibase 10.1103/PhysRevLett.123.042002} {\bibfield  {journal} {\bibinfo  {journal} {Phys. Rev. Lett.}\ }\textbf {\bibinfo {volume} {123}},\ \bibinfo {pages} {042002} (\bibinfo {year} {2019})},\ \Eprint {http://arxiv.org/abs/1904.03188} {arXiv:1904.03188 [hep-lat]} \BibitemShut {NoStop}%
\bibitem [{\citenamefont {Cheung}\ \emph {et~al.}(2021)\citenamefont {Cheung}, \citenamefont {Thomas}, \citenamefont {Wilson}, \citenamefont {Moir}, \citenamefont {Peardon},\ and\ \citenamefont {Ryan}}]{Cheung:2020mql}%
  \BibitemOpen
  \bibfield  {author} {\bibinfo {author} {\bibfnamefont {G.~K.~C.}\ \bibnamefont {Cheung}}, \bibinfo {author} {\bibfnamefont {C.~E.}\ \bibnamefont {Thomas}}, \bibinfo {author} {\bibfnamefont {D.~J.}\ \bibnamefont {Wilson}}, \bibinfo {author} {\bibfnamefont {G.}~\bibnamefont {Moir}}, \bibinfo {author} {\bibfnamefont {M.}~\bibnamefont {Peardon}}, \ and\ \bibinfo {author} {\bibfnamefont {S.~M.}\ \bibnamefont {Ryan}} (\bibinfo {collaboration} {Hadron Spectrum}),\ }\href {\doibase 10.1007/JHEP02(2021)100} {\bibfield  {journal} {\bibinfo  {journal} {JHEP}\ }\textbf {\bibinfo {volume} {02}},\ \bibinfo {pages} {100} (\bibinfo {year} {2021})},\ \Eprint {http://arxiv.org/abs/2008.06432} {arXiv:2008.06432 [hep-lat]} \BibitemShut {NoStop}%
\bibitem [{\citenamefont {Rendon}\ \emph {et~al.}(2020)\citenamefont {Rendon}, \citenamefont {Leskovec}, \citenamefont {Meinel}, \citenamefont {Negele}, \citenamefont {Paul}, \citenamefont {Petschlies}, \citenamefont {Pochinsky}, \citenamefont {Silvi},\ and\ \citenamefont {Syritsyn}}]{Rendon:2020rtw}%
  \BibitemOpen
  \bibfield  {author} {\bibinfo {author} {\bibfnamefont {G.}~\bibnamefont {Rendon}}, \bibinfo {author} {\bibfnamefont {L.}~\bibnamefont {Leskovec}}, \bibinfo {author} {\bibfnamefont {S.}~\bibnamefont {Meinel}}, \bibinfo {author} {\bibfnamefont {J.}~\bibnamefont {Negele}}, \bibinfo {author} {\bibfnamefont {S.}~\bibnamefont {Paul}}, \bibinfo {author} {\bibfnamefont {M.}~\bibnamefont {Petschlies}}, \bibinfo {author} {\bibfnamefont {A.}~\bibnamefont {Pochinsky}}, \bibinfo {author} {\bibfnamefont {G.}~\bibnamefont {Silvi}}, \ and\ \bibinfo {author} {\bibfnamefont {S.}~\bibnamefont {Syritsyn}},\ }\href {\doibase 10.1103/PhysRevD.102.114520} {\bibfield  {journal} {\bibinfo  {journal} {Phys. Rev. D}\ }\textbf {\bibinfo {volume} {102}},\ \bibinfo {pages} {114520} (\bibinfo {year} {2020})},\ \Eprint {http://arxiv.org/abs/2006.14035} {arXiv:2006.14035 [hep-lat]} \BibitemShut {NoStop}%
\bibitem [{\citenamefont {Woss}\ \emph {et~al.}(2021)\citenamefont {Woss}, \citenamefont {Dudek}, \citenamefont {Edwards}, \citenamefont {Thomas},\ and\ \citenamefont {Wilson}}]{Woss:2020ayi}%
  \BibitemOpen
  \bibfield  {author} {\bibinfo {author} {\bibfnamefont {A.~J.}\ \bibnamefont {Woss}}, \bibinfo {author} {\bibfnamefont {J.~J.}\ \bibnamefont {Dudek}}, \bibinfo {author} {\bibfnamefont {R.~G.}\ \bibnamefont {Edwards}}, \bibinfo {author} {\bibfnamefont {C.~E.}\ \bibnamefont {Thomas}}, \ and\ \bibinfo {author} {\bibfnamefont {D.~J.}\ \bibnamefont {Wilson}} (\bibinfo {collaboration} {Hadron Spectrum}),\ }\href {\doibase 10.1103/PhysRevD.103.054502} {\bibfield  {journal} {\bibinfo  {journal} {Phys. Rev. D}\ }\textbf {\bibinfo {volume} {103}},\ \bibinfo {pages} {054502} (\bibinfo {year} {2021})},\ \Eprint {http://arxiv.org/abs/2009.10034} {arXiv:2009.10034 [hep-lat]} \BibitemShut {NoStop}%
\bibitem [{\citenamefont {H\"orz}\ \emph {et~al.}(2021)\citenamefont {H\"orz} \emph {et~al.}}]{Horz:2020zvv}%
  \BibitemOpen
  \bibfield  {author} {\bibinfo {author} {\bibfnamefont {B.}~\bibnamefont {H\"orz}} \emph {et~al.},\ }\href {\doibase 10.1103/PhysRevC.103.014003} {\bibfield  {journal} {\bibinfo  {journal} {Phys. Rev. C}\ }\textbf {\bibinfo {volume} {103}},\ \bibinfo {pages} {014003} (\bibinfo {year} {2021})},\ \Eprint {http://arxiv.org/abs/2009.11825} {arXiv:2009.11825 [hep-lat]} \BibitemShut {NoStop}%
\bibitem [{\citenamefont {Dawid}\ \emph {et~al.}(2025)\citenamefont {Dawid}, \citenamefont {Draper}, \citenamefont {Hanlon}, \citenamefont {H{\"o}rz}, \citenamefont {Morningstar}, \citenamefont {Romero-L{\'o}pez}, \citenamefont {Sharpe},\ and\ \citenamefont {Skinner}}]{Dawid:2025doq}%
  \BibitemOpen
  \bibfield  {author} {\bibinfo {author} {\bibfnamefont {S.~M.}\ \bibnamefont {Dawid}}, \bibinfo {author} {\bibfnamefont {Z.~T.}\ \bibnamefont {Draper}}, \bibinfo {author} {\bibfnamefont {A.~D.}\ \bibnamefont {Hanlon}}, \bibinfo {author} {\bibfnamefont {B.}~\bibnamefont {H{\"o}rz}}, \bibinfo {author} {\bibfnamefont {C.}~\bibnamefont {Morningstar}}, \bibinfo {author} {\bibfnamefont {F.}~\bibnamefont {Romero-L{\'o}pez}}, \bibinfo {author} {\bibfnamefont {S.~R.}\ \bibnamefont {Sharpe}}, \ and\ \bibinfo {author} {\bibfnamefont {S.}~\bibnamefont {Skinner}},\ }\href@noop {} {\  (\bibinfo {year} {2025})},\ \Eprint {http://arxiv.org/abs/2502.17976} {arXiv:2502.17976 [hep-lat]} \BibitemShut {NoStop}%
\bibitem [{\citenamefont {Brice\~no}\ \emph {et~al.}(2025)\citenamefont {Brice\~no}, \citenamefont {Costa},\ and\ \citenamefont {Jackura}}]{Briceno:2024ehy}%
  \BibitemOpen
  \bibfield  {author} {\bibinfo {author} {\bibfnamefont {R.~A.}\ \bibnamefont {Brice\~no}}, \bibinfo {author} {\bibfnamefont {C.~S.~R.}\ \bibnamefont {Costa}}, \ and\ \bibinfo {author} {\bibfnamefont {A.~W.}\ \bibnamefont {Jackura}},\ }\href {\doibase 10.1103/PhysRevD.111.036029} {\bibfield  {journal} {\bibinfo  {journal} {Phys. Rev. D}\ }\textbf {\bibinfo {volume} {111}},\ \bibinfo {pages} {036029} (\bibinfo {year} {2025})},\ \Eprint {http://arxiv.org/abs/2409.15577} {arXiv:2409.15577 [hep-ph]} \BibitemShut {NoStop}%
\bibitem [{\citenamefont {Varshalovich}\ \emph {et~al.}(1988)\citenamefont {Varshalovich}, \citenamefont {Moskalev},\ and\ \citenamefont {Khersonskii}}]{VMK}%
  \BibitemOpen
  \bibfield  {author} {\bibinfo {author} {\bibfnamefont {D.~A.}\ \bibnamefont {Varshalovich}}, \bibinfo {author} {\bibfnamefont {A.~N.}\ \bibnamefont {Moskalev}}, \ and\ \bibinfo {author} {\bibfnamefont {V.~K.}\ \bibnamefont {Khersonskii}},\ }\href@noop {} {\emph {\bibinfo {title} {Quantum Theory of Angular Momentum}}}\ (\bibinfo  {publisher} {World Scientific},\ \bibinfo {address} {Singapore},\ \bibinfo {year} {1988})\BibitemShut {NoStop}%
\bibitem [{\citenamefont {Byckling}\ and\ \citenamefont {Kajantie}(1973)}]{byckling1973particle}%
  \BibitemOpen
  \bibfield  {author} {\bibinfo {author} {\bibfnamefont {E.}~\bibnamefont {Byckling}}\ and\ \bibinfo {author} {\bibfnamefont {K.}~\bibnamefont {Kajantie}},\ }\href {https://books.google.com/books?id=d_cNAQAAIAAJ} {\emph {\bibinfo {title} {Particle Kinematics}}},\ A Wiley-Interscience publication\ (\bibinfo  {publisher} {Wiley},\ \bibinfo {year} {1973})\BibitemShut {NoStop}%
\bibitem [{\citenamefont {Kibble}(1960)}]{Kibble:1960zz}%
  \BibitemOpen
  \bibfield  {author} {\bibinfo {author} {\bibfnamefont {T.~W.~B.}\ \bibnamefont {Kibble}},\ }\href {\doibase 10.1103/PhysRev.117.1159} {\bibfield  {journal} {\bibinfo  {journal} {Phys. Rev.}\ }\textbf {\bibinfo {volume} {117}},\ \bibinfo {pages} {1159} (\bibinfo {year} {1960})}\BibitemShut {NoStop}%
\bibitem [{\citenamefont {Berman}\ and\ \citenamefont {Jacob}(1965)}]{Berman:1965gi}%
  \BibitemOpen
  \bibfield  {author} {\bibinfo {author} {\bibfnamefont {S.~M.}\ \bibnamefont {Berman}}\ and\ \bibinfo {author} {\bibfnamefont {M.}~\bibnamefont {Jacob}},\ }\href {\doibase 10.1103/PhysRev.139.B1023} {\bibfield  {journal} {\bibinfo  {journal} {Phys. Rev.}\ }\textbf {\bibinfo {volume} {139}},\ \bibinfo {pages} {B1023} (\bibinfo {year} {1965})}\BibitemShut {NoStop}%
\bibitem [{\citenamefont {Mikhasenko}\ \emph {et~al.}(2020)\citenamefont {Mikhasenko} \emph {et~al.}}]{JPAC:2019ufm}%
  \BibitemOpen
  \bibfield  {author} {\bibinfo {author} {\bibfnamefont {M.}~\bibnamefont {Mikhasenko}} \emph {et~al.} (\bibinfo {collaboration} {JPAC}),\ }\href {\doibase 10.1103/PhysRevD.101.034033} {\bibfield  {journal} {\bibinfo  {journal} {Phys. Rev. D}\ }\textbf {\bibinfo {volume} {101}},\ \bibinfo {pages} {034033} (\bibinfo {year} {2020})},\ \Eprint {http://arxiv.org/abs/1910.04566} {arXiv:1910.04566 [hep-ph]} \BibitemShut {NoStop}%
\bibitem [{\citenamefont {Jacob}\ and\ \citenamefont {Wick}(1959)}]{Jacob:1959at}%
  \BibitemOpen
  \bibfield  {author} {\bibinfo {author} {\bibfnamefont {M.}~\bibnamefont {Jacob}}\ and\ \bibinfo {author} {\bibfnamefont {G.~C.}\ \bibnamefont {Wick}},\ }\href {\doibase 10.1016/0003-4916(59)90051-X} {\bibfield  {journal} {\bibinfo  {journal} {Annals Phys.}\ }\textbf {\bibinfo {volume} {7}},\ \bibinfo {pages} {404} (\bibinfo {year} {1959})}\BibitemShut {NoStop}%
\bibitem [{\citenamefont {Brice\~no}\ \emph {et~al.}(2024)\citenamefont {Brice\~no}, \citenamefont {Jackura}, \citenamefont {Pefkou},\ and\ \citenamefont {Romero-L\'opez}}]{Briceno:2024txg}%
  \BibitemOpen
  \bibfield  {author} {\bibinfo {author} {\bibfnamefont {R.~A.}\ \bibnamefont {Brice\~no}}, \bibinfo {author} {\bibfnamefont {A.~W.}\ \bibnamefont {Jackura}}, \bibinfo {author} {\bibfnamefont {D.~A.}\ \bibnamefont {Pefkou}}, \ and\ \bibinfo {author} {\bibfnamefont {F.}~\bibnamefont {Romero-L\'opez}},\ }\href {\doibase 10.1007/JHEP05(2024)279} {\bibfield  {journal} {\bibinfo  {journal} {JHEP}\ }\textbf {\bibinfo {volume} {05}},\ \bibinfo {pages} {279} (\bibinfo {year} {2024})},\ \Eprint {http://arxiv.org/abs/2402.12167} {arXiv:2402.12167 [hep-lat]} \BibitemShut {NoStop}%
\bibitem [{\citenamefont {Weinberg}(2005)}]{weinberg:1995mt}%
  \BibitemOpen
  \bibfield  {author} {\bibinfo {author} {\bibfnamefont {S.}~\bibnamefont {Weinberg}},\ }\href@noop {} {\emph {\bibinfo {title} {{The Quantum Theory of Fields. Vol. 1: Foundations}}}}\ (\bibinfo  {publisher} {Cambridge University Press},\ \bibinfo {year} {2005})\BibitemShut {NoStop}%
\bibitem [{\citenamefont {Potapov}\ and\ \citenamefont {Taylor}(1977{\natexlab{a}})}]{Potapov:1977ux}%
  \BibitemOpen
  \bibfield  {author} {\bibinfo {author} {\bibfnamefont {V.~S.}\ \bibnamefont {Potapov}}\ and\ \bibinfo {author} {\bibfnamefont {J.~R.}\ \bibnamefont {Taylor}},\ }\href@noop {} {\  (\bibinfo {year} {1977}{\natexlab{a}})}\BibitemShut {NoStop}%
\bibitem [{\citenamefont {Potapov}\ and\ \citenamefont {Taylor}(1977{\natexlab{b}})}]{Potapov:1977sr}%
  \BibitemOpen
  \bibfield  {author} {\bibinfo {author} {\bibfnamefont {V.~S.}\ \bibnamefont {Potapov}}\ and\ \bibinfo {author} {\bibfnamefont {J.~R.}\ \bibnamefont {Taylor}},\ }\href@noop {} {\  (\bibinfo {year} {1977}{\natexlab{b}})}\BibitemShut {NoStop}%
\bibitem [{\citenamefont {Martin}\ and\ \citenamefont {Spearman}(1970)}]{Martin:102663}%
  \BibitemOpen
  \bibfield  {author} {\bibinfo {author} {\bibfnamefont {A.~D.}\ \bibnamefont {Martin}}\ and\ \bibinfo {author} {\bibfnamefont {T.~D.}\ \bibnamefont {Spearman}},\ }\href {https://cds.cern.ch/record/102663} {\emph {\bibinfo {title} {{Elementary-particle theory}}}}\ (\bibinfo  {publisher} {North-Holland},\ \bibinfo {address} {Amsterdam},\ \bibinfo {year} {1970})\BibitemShut {NoStop}%
\bibitem [{\citenamefont {Von~Hippel}\ and\ \citenamefont {Quigg}(1972)}]{VonHippel:1972fg}%
  \BibitemOpen
  \bibfield  {author} {\bibinfo {author} {\bibfnamefont {F.}~\bibnamefont {Von~Hippel}}\ and\ \bibinfo {author} {\bibfnamefont {C.}~\bibnamefont {Quigg}},\ }\href {\doibase 10.1103/PhysRevD.5.624} {\bibfield  {journal} {\bibinfo  {journal} {Phys. Rev. D}\ }\textbf {\bibinfo {volume} {5}},\ \bibinfo {pages} {624} (\bibinfo {year} {1972})}\BibitemShut {NoStop}%
\bibitem [{\citenamefont {Estabrooks}\ and\ \citenamefont {Martin}(1975)}]{Estabrooks:1975cy}%
  \BibitemOpen
  \bibfield  {author} {\bibinfo {author} {\bibfnamefont {P.}~\bibnamefont {Estabrooks}}\ and\ \bibinfo {author} {\bibfnamefont {A.~D.}\ \bibnamefont {Martin}},\ }\href {\doibase 10.1016/0550-3213(75)90048-6} {\bibfield  {journal} {\bibinfo  {journal} {Nucl. Phys. B}\ }\textbf {\bibinfo {volume} {95}},\ \bibinfo {pages} {322} (\bibinfo {year} {1975})}\BibitemShut {NoStop}%
\bibitem [{\citenamefont {Protopopescu}\ \emph {et~al.}(1973)\citenamefont {Protopopescu}, \citenamefont {Alston-Garnjost}, \citenamefont {Barbaro-Galtieri}, \citenamefont {Flatte}, \citenamefont {Friedman}, \citenamefont {Lasinski}, \citenamefont {Lynch}, \citenamefont {Rabin},\ and\ \citenamefont {Solmitz}}]{Protopopescu:1973sh}%
  \BibitemOpen
  \bibfield  {author} {\bibinfo {author} {\bibfnamefont {S.~D.}\ \bibnamefont {Protopopescu}}, \bibinfo {author} {\bibfnamefont {M.}~\bibnamefont {Alston-Garnjost}}, \bibinfo {author} {\bibfnamefont {A.}~\bibnamefont {Barbaro-Galtieri}}, \bibinfo {author} {\bibfnamefont {S.~M.}\ \bibnamefont {Flatte}}, \bibinfo {author} {\bibfnamefont {J.~H.}\ \bibnamefont {Friedman}}, \bibinfo {author} {\bibfnamefont {T.~A.}\ \bibnamefont {Lasinski}}, \bibinfo {author} {\bibfnamefont {G.~R.}\ \bibnamefont {Lynch}}, \bibinfo {author} {\bibfnamefont {M.~S.}\ \bibnamefont {Rabin}}, \ and\ \bibinfo {author} {\bibfnamefont {F.~T.}\ \bibnamefont {Solmitz}},\ }\href {\doibase 10.1103/PhysRevD.7.1279} {\bibfield  {journal} {\bibinfo  {journal} {Phys. Rev. D}\ }\textbf {\bibinfo {volume} {7}},\ \bibinfo {pages} {1279} (\bibinfo {year} {1973})}\BibitemShut {NoStop}%
\bibitem [{\citenamefont {Pelaez}\ and\ \citenamefont {Yndurain}(2005)}]{Pelaez:2004vs}%
  \BibitemOpen
  \bibfield  {author} {\bibinfo {author} {\bibfnamefont {J.~R.}\ \bibnamefont {Pelaez}}\ and\ \bibinfo {author} {\bibfnamefont {F.~J.}\ \bibnamefont {Yndurain}},\ }\href {\doibase 10.1103/PhysRevD.71.074016} {\bibfield  {journal} {\bibinfo  {journal} {Phys. Rev. D}\ }\textbf {\bibinfo {volume} {71}},\ \bibinfo {pages} {074016} (\bibinfo {year} {2005})},\ \Eprint {http://arxiv.org/abs/hep-ph/0411334} {arXiv:hep-ph/0411334} \BibitemShut {NoStop}%
\bibitem [{\citenamefont {Dudek}\ \emph {et~al.}(2012)\citenamefont {Dudek}, \citenamefont {Edwards},\ and\ \citenamefont {Thomas}}]{Dudek:2012gj}%
  \BibitemOpen
  \bibfield  {author} {\bibinfo {author} {\bibfnamefont {J.~J.}\ \bibnamefont {Dudek}}, \bibinfo {author} {\bibfnamefont {R.~G.}\ \bibnamefont {Edwards}}, \ and\ \bibinfo {author} {\bibfnamefont {C.~E.}\ \bibnamefont {Thomas}},\ }\href {\doibase 10.1103/PhysRevD.86.034031} {\bibfield  {journal} {\bibinfo  {journal} {Phys. Rev.}\ }\textbf {\bibinfo {volume} {D86}},\ \bibinfo {pages} {034031} (\bibinfo {year} {2012})},\ \Eprint {http://arxiv.org/abs/1203.6041} {arXiv:1203.6041 [hep-ph]} \BibitemShut {NoStop}%
\bibitem [{\citenamefont {Zemach}(1964)}]{Zemach:1963bc}%
  \BibitemOpen
  \bibfield  {author} {\bibinfo {author} {\bibfnamefont {C.}~\bibnamefont {Zemach}},\ }\href {\doibase 10.1103/PhysRev.133.B1201} {\bibfield  {journal} {\bibinfo  {journal} {Phys. Rev.}\ }\textbf {\bibinfo {volume} {133}},\ \bibinfo {pages} {B1201} (\bibinfo {year} {1964})}\BibitemShut {NoStop}%
\bibitem [{\citenamefont {Draper}\ \emph {et~al.}(2023{\natexlab{b}})\citenamefont {Draper}, \citenamefont {Hansen}, \citenamefont {Romero-L\'opez},\ and\ \citenamefont {Sharpe}}]{Draper:2023xvu}%
  \BibitemOpen
  \bibfield  {author} {\bibinfo {author} {\bibfnamefont {Z.~T.}\ \bibnamefont {Draper}}, \bibinfo {author} {\bibfnamefont {M.~T.}\ \bibnamefont {Hansen}}, \bibinfo {author} {\bibfnamefont {F.}~\bibnamefont {Romero-L\'opez}}, \ and\ \bibinfo {author} {\bibfnamefont {S.~R.}\ \bibnamefont {Sharpe}},\ }\href {\doibase 10.1007/JHEP07(2023)226} {\bibfield  {journal} {\bibinfo  {journal} {JHEP}\ }\textbf {\bibinfo {volume} {07}},\ \bibinfo {pages} {226} (\bibinfo {year} {2023}{\natexlab{b}})},\ \Eprint {http://arxiv.org/abs/2303.10219} {arXiv:2303.10219 [hep-lat]} \BibitemShut {NoStop}%
\bibitem [{\citenamefont {Adolph}\ \emph {et~al.}(2017)\citenamefont {Adolph} \emph {et~al.}}]{COMPASS:2015gxz}%
  \BibitemOpen
  \bibfield  {author} {\bibinfo {author} {\bibfnamefont {C.}~\bibnamefont {Adolph}} \emph {et~al.} (\bibinfo {collaboration} {COMPASS}),\ }\href {\doibase 10.1103/PhysRevD.95.032004} {\bibfield  {journal} {\bibinfo  {journal} {Phys. Rev. D}\ }\textbf {\bibinfo {volume} {95}},\ \bibinfo {pages} {032004} (\bibinfo {year} {2017})},\ \Eprint {http://arxiv.org/abs/1509.00992} {arXiv:1509.00992 [hep-ex]} \BibitemShut {NoStop}%
\bibitem [{\citenamefont {Aghasyan}\ \emph {et~al.}(2018)\citenamefont {Aghasyan} \emph {et~al.}}]{COMPASS:2018uzl}%
  \BibitemOpen
  \bibfield  {author} {\bibinfo {author} {\bibfnamefont {M.}~\bibnamefont {Aghasyan}} \emph {et~al.} (\bibinfo {collaboration} {COMPASS}),\ }\href {\doibase 10.1103/PhysRevD.98.092003} {\bibfield  {journal} {\bibinfo  {journal} {Phys. Rev. D}\ }\textbf {\bibinfo {volume} {98}},\ \bibinfo {pages} {092003} (\bibinfo {year} {2018})},\ \Eprint {http://arxiv.org/abs/1802.05913} {arXiv:1802.05913 [hep-ex]} \BibitemShut {NoStop}%
\end{thebibliography}%

\end{document}